\documentclass[12pt]{article}

\usepackage[titletoc]{appendix}
\usepackage[margin={1.0in}]{geometry}
\usepackage{rotating}
\usepackage{lscape}
\usepackage{graphicx}
\usepackage{color}
\usepackage{subfig}
\usepackage{comment}
\usepackage{threeparttable}
\usepackage{booktabs}
\usepackage{longtable}
\usepackage{threeparttablex}
\usepackage{tabularx}
\usepackage{pdflscape} 
\usepackage{natbib}

\usepackage{authblk}
\usepackage{mathtools}
\usepackage{dsfont}
\usepackage{lmodern} 
\usepackage{anyfontsize} 
\usepackage[sc]{mathpazo}
\usepackage{courier}
\usepackage[utf8]{inputenc}
\usepackage[T1]{fontenc}
\usepackage{setspace}
\usepackage{amssymb}
\usepackage{amsmath}
\usepackage{mathrsfs}
\usepackage{amsthm}
\usepackage{bbm}
\usepackage{microtype}
\usepackage{float} 
\usepackage{multirow}
\usepackage{enumerate}
\usepackage{hyperref} 

\usepackage{tikz}
\usepackage[nameinlink]{cleveref} 
\usepackage{afterpage}

\usepackage{array}
\usepackage{adjustbox}

\usepackage{xfrac}

\ExplSyntaxOn
\cs_new:Npn \expandableinput #1
  { \use:c { @@input } { \file_full_name:n {#1} } }
\ExplSyntaxOff

\hypersetup{
	colorlinks=true, 
	linktoc=all,        
	linkcolor=black,  
	urlcolor  = blue,
	citecolor = black,
    pdftitle={The Debt-Inflation Channel of the German (Hyper-)Inflation},
    pdfauthor={Markus Brunnermeier, Sergio Correia, Stephan Luck, Emil Verner, Tom Zimmermann}
}

\usepackage[nottoc,notlot,notlof]{tocbibind}

\hypersetup{pdfstartview={Fit}} 
\RequirePackage[font=small,format=plain,labelfont=bf,textfont=it]{caption}

\def\eatcell#1\unskip{}
\newcolumntype{E}{>{\eatcell}c@{}}
\newcolumntype{H}{>{\lrbox0}c<{\endlrbox}@{}}

\newcommand{\E}{\mathbb{E}}

\begin{document}

\title{\vspace{0cm} \LARGE \bf The Debt-Inflation Channel of the \\ German (Hyper-)Inflation}

\author{\vspace{.5cm} Markus Brunnermeier, Sergio Correia, Stephan Luck, \\ \vspace{-.15cm} Emil Verner, and Tom Zimmermann\textsuperscript{*} }

\date{ \today } %

\pagenumbering{gobble}
\maketitle

\begin{abstract}
\fontsize{12.0pt}{15.0pt}\selectfont
\noindent This paper studies how a large increase in the price level is transmitted to the real economy through firm balance sheets. Using newly digitized macro- and micro-level data from the German inflation of 1919-1923, we show that inflation led to a large reduction in real debt burdens and bankruptcies. Firms with higher nominal liabilities at the onset of inflation experienced a larger decline in interest expenses, a relative increase in their equity values, and higher employment during the inflation. The results are consistent with real effects of a \textit{debt-inflation} channel that operates even when prices and wages are flexible.

\end{abstract}

\let\oldthefootnote\thefootnote
\renewcommand{\thefootnote}{\fnsymbol{footnote}}
\footnotetext[1]{Brunnermeier: Princeton University, \href{mailto:markus@princeton.edu}{markus@princeton.edu}; Correia: Board of Governors of the Federal Reserve System, \href{mailto:sergio.a.correia@frb.gov}{sergio.a.correia@frb.gov}; Luck: Federal Reserve Bank of New York, \href{mailto:stephan.luck@ny.frb.org}{stephan.luck@ny.frb.org};
Verner: MIT Sloan, \href{mailto:everner@mit.edu}{everner@mit.edu}; Zimmermann: University of Cologne \href{mailto:tom.zimmermann@uni-koeln.de}{tom.zimmermann@uni-koeln.de}.
We would like to thank our discussants Fernando Alvarez, Juliane Begenau, Francesco D'Acunto, Hanno Lustig, Carolin Pflueger, and Bjoern Richter for valuable feedback.
We would also like to thank Joseph Abadi, Martin Beraja, Michael Bordo, Gabriel Chodorow-Reich, Marco Del Negro, Keshav Dogra, Yuriy Gorodnichenko, Carl-Ludwig Holtfrerich, Harold James, David Lopez-Salido, David Thesmar, Ivan Werning, Christian Wolf, and seminar participants at Babson College, Banca d'Italia, Barcelona Summer Forum, Bocconi-BAFFI CAREFIN-CEPR Workshop, CESifo Summer Institute, Harvard-MIT Financial Economics Workshop, IESE, MIT, National Bank of Hungary, NBER Monetary Economics, NBER Corporate Finance, NBER Development of the American Economy, Norges Bank, Northwestern University, Red Rock Finance, SAFE, SED, Toulouse School of Economics, UC Berkeley Haas, UPF, UT Austin, Seminario MAP, SITE 2023, 12th Tepper-LAEF conference, and University of Georgia for useful comments. We thank Viktoria Brueck, Edwin Diebold, Natalia Fischl-Lanzoni, Lara Fronert, Linn Görnig, Liz Hay, Jonathan Jablonski, Tom Jarosch, Vinnie Le, Lev Luskin, Christian Meyer, Jenna Wang and Lena Will for outstanding research assistance. We also thank Universit{\"a}tsbibliothek Mannheim for access to their digitized book library and Stefan Weil for helping navigate the digitized book library. The project has received support from the Deutsche Forschungsgemeinschaft (DFG) under Germany’s Excellence Strategy EXC2126/139083886, Princeton University's Bendheim Center for Finance, and the MIT Research Support Committee's Ferry Fund. The opinions expressed in this paper do not necessarily reflect those of the Federal Reserve Bank of New York or the Board of Governors of the Federal Reserve System.}

\let\thefootnote\oldthefootnote

\onehalfspacing

\clearpage
\pagenumbering{arabic}
    \clearpage

\begin{quote}
    ``The Germany of the inflation was paradise for anyone who owed money.'' 
    
    \hfill -- Frederick Taylor, \textit{The Downfall of Money} (2013).
\end{quote}

\section{Introduction}

In the presence of nominal debt contracts, unexpected inflation redistributes wealth from creditors to debtors \citep{Keynes1923tract}. If levered firms are financially constrained, such wealth redistribution can affect \textit{real} economic activity and potentially have aggregate effects \citep{Fisher1933,Tobin1982asset}. In theory,  unexpected inflation can thus increase economic activity via a \textit{debt-inflation} channel that can operate even when prices and wages are flexible. While the transmission of inflation to the real economy through financial frictions is well understood theoretically \citep{Gomes2016}, there is limited evidence supporting the empirical relevance of such a financial channel. 

In this paper, we use Germany's inflation of 1919--1923 as a laboratory to study how a large inflationary shock transmits to the real economy through the debt-inflation channel. The German inflation is a key event in monetary history. It has been studied by generations of economists to understand the causes and consequences of high inflation \citep[e.g.,][]{BrescianiTurroniCostantino1937TEoI,Cagan1956,Sargent1982}.\footnote{Throughout the paper, we refer to the entire postwar inflation from November 1918 to November 1923 as the ``German inflation.'' As we discuss below, we reserve the term ``hyperinflation'' for the phase from July 1922 to November 1923 when monthly inflation exceeded 50\%.} We revisit this important episode through the lens of several newly digitized sources at both the macro- and micro-level. The setting is particularly appealing for studying the debt-inflation channel because the inflation was to a large extent unanticipated. Moreover, the enormous  increase in the price level puts the effects of inflation on balance sheets into sharp relief, allowing us to study a channel that may be harder to identify empirically during times of moderate inflation.

We find that the debt-inflation channel was an important factor for the  transmission of the German inflation to the real economy. In the aggregate, the inflation was associated with a large reduction in real debt burdens for levered nonfinancial firms, which resulted in a large decline in bankruptcies. In the cross-section, firms with higher leverage at the onset of the inflation saw a larger decline in their interest expenses, along with a larger relative increase in their book and market equity values. The reduction in real debt burdens had real effects, as high-leverage firms saw the largest increase in employment during the inflation. Altogether, our findings highlight the role of nominal rigidity in debt contracts and financial frictions in the transmission of a large inflationary shock to the real economy.

We begin by describing the macroeconomic environment surrounding the German inflation. The postwar German inflation can be divided into two broad phases. The first phase occurs from the end of WWI in November 1918 to June 1922.\footnote{Inflation begins with the outset of WWI following the abandonment of the gold standard. The price level in Germany increased by a factor of 2.4 during WWI. We focus mainly on the post-WWI period because inflation was higher during this period and economic activity and inflation during the war were heavily influenced by wartime policies and controls.} In this phase, the price level increased by a factor of 30. Economic growth was strong and, unlike most other major economies at the time, Germany avoided a large economic downturn following WWI. The root cause of the inflation was deficit-financed war spending, massive WWI reparations, and a lack of political will to adjust to the burden of higher debt through reduced spending and increased taxation in an attempt to maintain social peace  \citep{Graham1931,Sargent1982,kindleberger1985,Feldman1997}. Moreover, there was no central bank policy response to high inflation; instead, the Reichsbank accommodated large deficits by discounting government securities. Evidence from forward exchange premium data and narrative accounts suggest that the inflation was initially unanticipated.

The second phase of the inflation is the hyperinflation from July 1922 to the stabilization in November 1923. This phase begins after political turmoil over WWI reparations and the assassination of the prominent foreign minister Rathenau. In this phase, the price level spirals out of control. In line with inflation expectations becoming unanchored and a flight from the mark, the forward exchange premium turns to a large discount from July 1922 onward. The hyperinflation coincides with deteriorating economic conditions and rising unemployment. The economic downturn is exacerbated by the invasion of the Ruhr and resulting collapse in industrial production. 


To study the impact of the inflation on the real economy through firm balance sheets, we construct a new firm-level database by digitizing a financial manual with firm-level information on balance sheets, income statements, employment, and outstanding bonds for about 700 joint-stock firms in Germany. We merge these data with newly digitized stock returns. Information on employment and stock returns is particularly valuable, as accounting statements potentially provide a distorted representation of firm financial conditions during the hyperinflation phase, especially in 1923.\footnote{Accounting statements become more reliable again starting in January 1924, when firms were required to draw up revalued ``Goldmark'' balance sheets, as we discuss in section \ref{sec:data}.} We supplement these data with a range of other newly digitized datasets, including data on bankruptcies, retail prices, and wages.


What are the macro-financial implications of the inflation? Using time-series variation, we find a striking negative relation between inflation and firm bankruptcies. Bankruptcies consistently declined with rising inflation and remained at historically low levels even with the economic turmoil of 1923. Moreover, the relation is convex. For low to moderate inflation, additional inflation is associated with sharply lower firm bankruptcies. At levels of annual inflation above 500\%, additional inflation only weakly reduced bankruptcies. Intuitively, once the price level has doubled several times within a few years, debts have already been wiped out, making bankruptcy increasingly unlikely. To better understand the negative relation between inflation and bankruptcies, we document that inflation is associated with a massive fall in financial leverage. We illustrate that debts outstanding before the inflation were wiped out by 1922. Interest expenses as a share of total expenses also fall by 10 percentage points from 1918 through 1923. 

The aggregate decline in real debt burdens and bankruptcies during the inflation raises several intriguing questions. What is the extent of redistribution toward levered firms from the inflation? Does the erosion of the real value of nominal debt have real effects in terms of firms' employment? Do equity-holders benefit from leverage, and how is this reflected in equity prices? 

To answer these questions, we examine the impact of inflation in the cross-section of firms. We estimate a difference-in-differences specification, sorting firms by their leverage---measured as the liabilities-to-assets ratio---before the onset of the inflation. Firms with higher initial leverage have significantly stronger employment growth relative to firms with lower leverage. In terms of magnitudes, a one-standard-deviation increase in leverage is associated with 7.5\% higher employment during the inflation. We estimate that the erosion of real debt from inflation increases employment after about one year. As a result, the expansionary effect of the erosion in debts outstanding before the inflation is strongest in the first phase of inflation up to 1922. The hyperinflation phase of 1922--1923 had limited additional real effects on high-leverage firms.

We provide several pieces of evidence to support the interpretation that the real effects on employment are driven by the debt-inflation channel. First, high-leverage firms see larger reductions in the share of interest expenses to total expenses. This pattern is consistent with inflation lowering real debt payments, relaxing cash flow constraints, and thereby allowing firms to increase spending on production inputs. 

Second, the debt-inflation channel should especially benefit firms with a higher proportion of long-term debt, as these firms are less exposed to a repricing of debt as expected inflation rises. Many firms in our sample relied on fixed-rate long-term debt financing. Exploiting additional details on the maturity structure of firms' liabilities, we estimate our main regression model separately for firms with high and low reliance on long-term debt finance. We find that the decline in interest expenses and the increase in employment are strongest for highly levered firms with a high proportion of long-term debt to total liabilities. 

Third, high-leverage firms see a larger increase in the real value of book equity. Firms do not use the windfall reduction in interest expenses to increase dividends, and instead keep the money inside the firm.

Finally, while equities perform very poorly during the inflation, high-leverage firms experience relatively higher stock market returns. On average, high-leverage firms have about 10\% higher annual risk-adjusted returns relative to low-leverage firms during the inflation. The equity market, however, incorporates the benefits of leverage relatively late in the inflation period, perhaps when these benefits were most salient to investors and when there was consensus about the permanent nature of the inflationary shock, in line with the expectations implied by the foreign exchange market. Overall, firm-level evidence suggests that the largely unexpected inflation redistributed wealth from debt holders to shareholders of levered firms, relaxing financing constraints and allowing these firms to expand employment and production.

We reinforce these results and rule out several potentially confounding explanations with a series of robustness tests. First, to alleviate the concern that firms strategically adjust their leverage in anticipation of inflation, we sort firms by their leverage in 1917. We thus proxy a firm's exposure to the debt-inflation channel using variation from a point in time when the outcome of WWI and the subsequent inflation dynamics could not have been anticipated. Further, to address the concern that firm leverage toward the end of the war is shaped by exposure to the war economy and developments during the war, we also show that our findings are robust to instrumenting leverage at the end of the war with leverage from before WWI. Second, we show that our estimates are robust to a range of firm-level controls such as size, the share of fixed assets, profitability, free cash flow relative to assets, and Tobin's Q before the inflation. Third, to address concerns that high-leverage firms might have differential exposure to credit supply shifts, political connections, or export revenues, we show that our findings are robust to controlling for proxies of these potential confounders and to restricting the sample to firms where these confounders are unlikely to be important. Fourth, we also show that our results are unlikely to be driven by differential cyclicality, as firms with high leverage at the onset of the inflation do not have higher business cycle exposure in other periods. Finally, higher employment growth for highly levered firms is unlikely to be explained by differential exposure to wage rigidity. Wages were mostly set at the industry level by union bargaining, and all the results are robust to the inclusion of detailed industry-by-time fixed effects.

The debt-inflation channel relies on nominal rigidity through debt contracts, rather than nominal rigidity in wages and prices, as traditionally assumed in New Keynesian models \citep[e.g.,][]{gali2015monetary}. There is an analogy between nominal debt rigidity and nominal wage rigidity, as both can reduce firm expenses following an inflationary shock. However, a difference is that existing nominal debt contracts cannot be renegotiated after a large inflationary shock, whereas prices and wages can potentially be reset. Consistent with this, we document that wages adjust with increasing frequency in response to rising inflation. For example, wages are adjusted every 9 months when annual inflation is close to 0\% but every 60 days once inflation exceeds 100\% and every 30 days or less once inflation exceeds 200\%. We find a similar pattern for retail prices. This evidence is consistent with menu cost models \citep{Alvarez2019}. While a quantification of the relevant roles of nominal rigidity through debt versus prices and wages is beyond the scope of this paper, the evidence is consistent with the view that the debt-inflation channel can operate even for large inflationary shocks when nominal rigidity in wages and prices becomes less relevant.

Our paper relates to three strands of literature. First, we contribute to the literature on the importance of financing frictions for the response of real outcomes to inflation and monetary policy shocks \citep[see, e.g.,][]{Ottonello2020}. Much of the literature on monetary policy transmission focuses on the real effects of inflation through nominal rigidities in prices or wages. A debt-inflation channel through nominal rigidities from nominal long-term debt is absent from most standard models. Important exceptions are the theoretical models of {\citet{Bhamra2011} and \citet{Gomes2016}}. Moreover, there is limited empirical evidence on this channel. \cite{Kroszner1999} studies the abrogation of gold clauses in the U.S. during the Great Depression and finds that firms with higher leverage experienced higher returns around the announcement, consistent with these firms benefiting from a larger devaluation of their debts. \cite{HausmanRhodeWieland2019} document that the departure from the Gold Standard during the Great Depression led to an increase in crop prices in the U.S., which boosted spending in regions with high farm debt. An empirical literature focusing on the 1970s inflation in the U.S. finds mixed evidence for the hypothesis that equity valuations of firms with leverage benefit from inflation \citep[see, for example,][]{summers1981inflation,FrenchRubackSchwert1983,Ritter2002}. More recently, \cite{KangPflueger2015} find that inflation risk is priced in corporate bond yields, as unexpectedly low inflation increases a firm's real liabilities. \cite{BhamraWeber2021} find that, in the aggregate, high inflation is associated with low equity market valuations and low default rates in U.S. data since 1970, similar to our evidence based on aggregate data for Germany's inflation.\footnote{\cite{BhamraWeber2021} also analyze the relation between expected inflation and equity valuations for firms with high versus low leverage. They find that the negative relation between inflation and equity valuations is weaker for high-leverage firms, consistent with our findings.} We complement these papers by examining whether inflation has real effects at the firm level during an extreme inflationary shock.

Second, our findings contribute to the literature on firm financing frictions and the transmission of shocks to firms' net worth. Empirical studies find that firms adjust investment and employment in response to exogenous shocks to their net worth  \citep[e.g.,][]{Blanchard1994,KaplanZingales1997,Rauh2006,Benmelech2019} or changes in leverage following debt restructurings \citep{Giroud2012}.  Our analysis shows that an unexpected inflationary shock can generate substantial real effects on firms' real activity through its impact on firm balance sheets.

Finally, our paper contributes to the large literature on the German inflation and big inflations more broadly.\footnote{Prominent studies of the German inflation include \cite{schacht1927stabilization}, \cite{BrescianiTurroniCostantino1937TEoI}, \cite{Graham1931}, \citet{Holtfrerich1986}, Chapter 5 of \citet{Eichengreen1995}, \cite{Feldman1997} and \citet{Hetzel2002}. Important studies of hyperinflations, including the German inflation, include \cite{Cagan1956}, \cite{Sargent1982}, \cite{Dornbusch1986}, and \cite{LopezMitchener2020}. \Cref{sec:aggregate} and Appendix \ref{app:historical_background} provide an overview of the historical context and reviews existing historical studies on Germany's inflation.} Few episodes have attracted as much attention from economists. The extensive work on the German hyperinflation, however, has almost entirely relied on aggregate time-series data.\footnote{A recent exception is \cite{braggion2023inflation}, who analyze security holdings of clients of a major bank. They find that investors exposed to higher local inflation have lower demand for stocks, likely due to money illusion.} We provide several new insights based on novel industry-  and firm-level data. Previous studies have discussed that inflation eroded public debt \citep{Dornbusch1985} and private debt \citep{Graham1931}, but we are the first to document that the fall in bankruptcies lines up closely with the preceding rise in the price level. Moreover, we are the first to quantify the impact of the debt-inflation channel of the inflation on firm balance sheets, stock market valuations, and real firm-level outcomes.

The remainder of the paper is structured as follows. Section \ref{sec:conceptual_framework} provides a conceptual framework to motivate the empirical analysis. Section \ref{sec:data} describes our newly digitized macro- and firm-level data. Section \ref{sec:aggregate} presents aggregate evidence on the debt-inflation and nominal rigidity channels of inflation. Section \ref{sec:firm_level} presents evidence of the financial channel in the cross-section of firms, and section \ref{sec:conclusion} concludes.


\section{Conceptual Framework}
\label{sec:conceptual_framework}
To fix ideas, this section lays out a conceptual framework summarizing our hypotheses for how a large increase in the price level can affect the real economy through financial channels. The discussion in this section is based on theoretical contributions to the macro-finance literature and a simple model presented in Appendix \ref{sec:model}. 

In theory, the equity owners of a firm financed with nominal debt will benefit from unanticipated inflation, as inflation reduces real debt burdens. This reduction in real debt burdens will increase the net worth and market value of equity of a levered firm. Absent financial frictions, the change in the capital structure does not affect real outcomes \citep{ModiglianiMiller}. However, in a world with financial frictions, the reduction in real debt burdens that benefits equity owners affects real economic outcomes through what we refer to as the \textit{debt-inflation channel}. The debt-inflation channel is the inverse case of Irving Fisher’s famous debt-deflation channel  \citep{Fisher1933}.

The \textit{debt-inflation channel} can operate through several related mechanisms. By reducing real debt burdens, inflation can reduce the likelihood of bankruptcy. The decline in the probability of bankruptcy lowers firms' interest costs, as lenders reduce their required interest for expected default cost \citep{Townsend1979,BGG1999}, as well as risk and liquidity premia \citep{Holmstrom1997}. Note that these channels are active only if debt contracts are nominal and at a fixed interest rate, as was the case in Weimar Germany. If instead debt is floating, indexed to inflation, or denominated in foreign currency, then inflation does not necessarily increase firms' net worth.

\textit{\textbf{Hypothesis I: Debt-Inflation, Firm Bankruptcies, and Financing Costs.} When firms have nominal debt and can default, unexpected inflation increases firms' net worth and lowers interest burdens, leading to a decline in bankruptcy rates.}

The increase in net worth and reduction in borrowing costs also enable a previously financially constrained firm to issue new debt to finance investment and production costs. Moreover, the decline in real interest cost directly frees up cash flows that can be used to purchase labor or capital. In the case of collateralized borrowing, the value of the collateral asset might rise. This relaxes borrowing constraints and boosts production further \citep{Kiyotaki1997,CordobaRipoll2004}. Furthermore, inflation can also reduce debt overhang that would otherwise distort borrowers' investment decisions, as the returns from investment would primarily benefit debt holders \citep{Myers1977,Giroud2012}. For explicit models of the debt-inflation channel, see \cite{Bhamra2011}, \cite{Gomes2016}, and \cite{Brunnermeier2016} as well as the model in \Cref{sec:model}.

\textit{\textbf{Hypothesis II: The Debt-Inflation Channel and Firm Activity.} 
Unexpected inflation leads to an expansion in real economic activities at highly indebted firms relative to less indebted firms.}

If firms face financing frictions or if financial distress is costly, the reduction in real debt burdens through inflation can boost real activity. Our empirical analysis focuses on cross-sectional implications primarily between highly and less indebted firms. There are also important general equilibrium effects whose study requires a structural macroeconomic model. For example, unanticipated inflation leads not only to gains for debtors but also to losses for nominal debt holders. If these debt holders are households, consumption can contract, reducing employment through a demand channel. Alternatively, lower household wealth might expand labor supply, further increasing employment. This latter channel can be more important if prices and wages are flexible so that output is not demand-determined (see Appendix \ref{sec:model}). Furthermore, if inflation reduces the net worth of banks holding the debt, then the reduction in credit supply can depress firm activity.\footnote{While our main focus is on the debt-inflation channel operating through nonfinancial firms' balance sheets, we discuss the impact of inflation on bank credit supply in section \ref{sec:firm_level}. For recent evidence that inflation can impair the intermediation of credit, see \cite{Drechsler2022} and \cite{AgarwalBaron2022}.}

Note that the debt-inflation channel due to financial frictions does not require the nominal rigidities traditionally assumed in macroeconomic models of inflation \citep[e.g.,][]{gali2015monetary}. In our model presented in Appendix \ref{sec:model}, we consider nominal rigidity due to a fixed (menu) cost of wage adjustment. Nominal rigidities further enhance the effects of moderate inflation, as they lead to a fall in real wages, boosting labor demand, employment, and output. In a very high inflationary environment, menu costs do not stop price and wage adjustment, effectively resulting in a flexible price economy with limited rigidity effects \citep{golosov2007menu}. High inflation could nonetheless have substantial real effects through the debt-inflation channel, as existing nominal debt contracts cannot be renegotiated following an inflationary shock, in contrast to wages and prices. Only after debt is effectively wiped out does the debt-inflation channel lose its power.


\section{Data}
\label{sec:data}

\subsection{Firm-Level Data}

We construct a firm-level dataset with annual information on balance sheets, income statements, and employment from \textit{Saling's B\"{o}rsen-Jahrbuch}, an investor manual. The data from \textit{Saling's} are available for over 700 firms each year.  In this paper, we focus on the sample of nonfinancial firms and exclude banks and insurance companies. We focus most of our data collection on the period 1916--1926, covering the entirety of the postwar inflation period (1919--1923). We also collect information on selected variables such as leverage and employment for some earlier years. Furthermore, \textit{Saling's} also allows us to collect additional information on firms' outstanding bonds and exporter status, as well as details on their executives and supervisory board members.

We digitize the balance sheets and income statements in \textit{Saling's} using optical character recognition (OCR), applying the methods discussed in \citet{Correia2022}. All OCR output is reviewed manually, with particular attention to cases where accounting identities fail to hold. Lastly, we standardize the financial statement items in order to obtain harmonized balance sheets and income statements. Other data from \textit{Saling's} such as employment and information on outstanding bonds are hand-collected. Appendix \ref{app:data} provides a detailed discussion on how these data were collected, validated, and standardized. As a concrete example of the underlying data source, we include an annotated snippet in Appendix \Cref{fig:salings-snippet}, which contains balance sheet, income statement, and employment information for \emph{H. Berthold AG}---the largest type foundry in the world at the time---as well as the firm name and location for \emph{Bing Werke}---another iconic German firm and one of the world's largest toy manufacturers. \Cref{tab:summary_statistics} provides summary statistics for key firm-level variables. 

An important point to note when using data from \textit{Saling's B\"{o}rsen-Jahrbuch} is that balance sheets provide a misleading account of firms' financial situations during the hyperinflation, especially in 1923.\footnote{Income statements can also be distorted due to the addition of nominal values at different price levels, but the distortions are most severe for balance sheets. For income statements, we rely on ratios such as the share of interest expenses to total expenses, which are less likely to be distorted by inflation as long as different expenses occur at similar times in the year. See \cite{Sweeney1934} for details on how inflation distorted balance sheets during the 1920s hyperinflations in Europe.} ``Inflation accounting'' did not exist at the time, and dealing with inflation was a major challenge for firms' accounting. In particular, real items such as fixed assets become significantly undervalued relative to nominal items such as cash holdings. In response to the distortion of paper mark balance sheets caused by the hyperinflation, the government passed the regulation on Goldmark accounts (\textit{Verordnung über Goldbilanzen}) in December 1923. The regulation required firms to prepare new opening balance sheets for financial years beginning on or after January 1, 1924 in Goldmarks.\footnote{The Goldmark was not an actual currency in circulation but was used for accounting purposes and equivalent to the new Rentenmark, which had an exchange rate of 4.2 per U.S. dollar.} This preparation required a full revaluation of all assets and liabilities \citep{Sommerfeld1924}. In Appendix \ref{app:data}, we provide a more detailed discussion of these measurement issues and the revalued Goldmark balance sheets.

Our analysis uses two approaches to overcome the inflation-induced measurement challenge. First, we rely on balance sheet variables before the hyperinflation (before 1922) and from the more accurate Goldmark balance sheets (usually from January 1, 1924), avoiding use of the 1923 balance sheet values. This allows us to see what happened to key balance sheet variables from before to after the inflation, but it has the obvious drawback of not being informative about the timing of the effects on balance sheet outcomes. Moreover, our main analysis sorts firms based on their balance sheet exposure to inflation before the start of the inflation.

Second, we examine variables that are not subject to accounting issues throughout the inflation---employment and stock prices---as well as ratios based on income statements, where measurement error is less severe. Employment is reported for about one-third of the firms in \textit{Saling's}. Aggregating employment growth across firms in \textit{Saling's} captures the aggregate fluctuations in employment reasonably well (see \Cref{fig:saling_emp_vs_unemp}). There are gaps in the firm-level employment variable for some firms, as some firms do not report employment in all years. In our baseline analysis, we use the raw employment data reported by firms, but in the appendix we show the results are robust to requiring that firms report in all years.

Finally, monthly stock prices for nonfinancial firms and banks are hand-collected from \textit{Berliner Börsen-Zeitung} (BBZ), a financial newspaper. We also collect dividends to construct total returns. Appendix \Cref{fig:bbz_example} provides an example of data from \textit{Berliner Börsen-Zeitung}. The equal-weighted stock price indexes constructed from the BBZ data closely track corresponding indexes published by a contemporary statistical publication (\textit{Wirtschaft und Statistik}), providing a reassuring validation of our hand-collected stock market data (see Appendix \Cref{fig:bbz_validation}).

\begin{table}[ht!]
  \centering
  \caption{Summary Statistics: Firm-Level Dataset.  }\label{tab:summary_statistics}
        \begin{minipage}{1.0\textwidth}
        \begin{center}
        \scalebox{0.9}{
        {
\def\sym#1{\ifmmode^{#1}\else\(^{#1}\)\fi}
\begin{tabular}{l*{1}{ccccc}}
\toprule
                    &           N&        Mean&   Std. dev.&        10th&        90th\\
\midrule
\( \text{Liabilities/Assets}_{i,1917} \)&         724&        0.43&        0.18&        0.20&        0.65\\
\( \text{Debt/Assets}_{i,1917} \)&         724&        0.32&        0.18&        0.09&        0.56\\
\( \ln(\text{Assets})_{i,1917} \)&         724&       15.52&        1.22&       14.08&       17.24\\
\( \text{Fixed Assets/Assets}_{i,1917} \)&         724&        0.40&        0.25&        0.11&        0.79\\
\( \text{FCF/Assets}_{i,1917} \)&         678&        0.01&        0.14&       -0.17&        0.14\\
\( \text{EBIT margin}_{i,1917} \)&         670&        0.12&        2.35&        0.14&        0.72\\
\( \text{Tobin's Q}_{i,1918} \) &         567&        1.11&        0.22&        0.85&        1.41\\
\(\Delta_{1919-1924} \text{Liabilities/Assets}_{i} \)&         662&       -0.26&        0.19&       -0.48&       -0.02\\
\( \text{Employment}_{i,t} \)&       2,355&    4,121.65&   10,435.41&      400.00&    8,745.00\\
\( \Delta \ln( \text{Employment})_{i,t} \)&       1,937&        0.03&        0.19&       -0.02&        0.18\\
\bottomrule
\end{tabular}
}
}
        \end{center}
        {\footnotesize \textit{Notes}: This table reports summary statistics for the firm-level dataset based on \textit{Saling's B\"{o}rsen-Jahrbuch}. $\text{Employment}_{i,t}$ is the level of employment for the 1914--1923 sample.  $\Delta \ln(\text{Employment})$ is the annual change in log firm employment (times 100) from 1915 through 1923.}
        \end{minipage}
\end{table}%

\subsection{Aggregate, Industry-level, and Regional Data}

We supplement our new firm-level dataset with aggregate-, industry-, and regional-level data by digitizing contemporary publications of various government agencies. Our main source is a publication called \textit{Zahlen zur Geldentwertung in Deutschland von 1914 bis 1923}, published by the government agency for statistical analysis, the \textit{Reichsamt f\"{u}r Statistik}, in 1925. This publication provides data on the daily mark-per-dollar exchange rate, cost-of-living and wholesale prices indexes by month, wages by industry, weekly prices for consumption goods in Berlin during 1923, as well as stock market indices by broad industry categories. Additional data on wages by industry and monthly prices for consumption goods at the city level are obtained from the contemporary publication \textit{Wirtschaft and Statistik}, which was published at a monthly frequency starting in January 1921, and at a bi-monthly frequency from January 1922 onwards. Further, we obtain information on firm bankruptcies and liquidations by industry and prices for wholesale goods from the \textit{Vierteljahrshefte zur Statistik des Deutschen Reichs Herausgegeben vom Statistischen Reichsamt} and the annual \textit{Statistisches Jahrbuch  f\"{u}r das deutsche Reich}, which were published through 1919--1923 at the quarterly and annual frequency, respectively. Finally, we digitize parts of the appendix of the \textit{Reichsarbeitsblatt}, published by the Ministry of Labor (\textit{Reichsarbeitsministerium}), which contains information on monthly unemployment.


\section{Aggregate Evidence on the Debt-Inflation Channel}
\label{sec:aggregate}
\subsection{Background on Weimar Germany's Inflation}

To set the stage, we start with a brief overview of Weimar Germany's inflation. Appendix \ref{app:historical_background} provides further historical background and a chronology of critical events.

\begin{figure}[!ht]
\begin{center}
\caption{The Price Level during Germany's Inflation. \label{fig:agg_price_level}}
\includegraphics[width=1.0\textwidth]{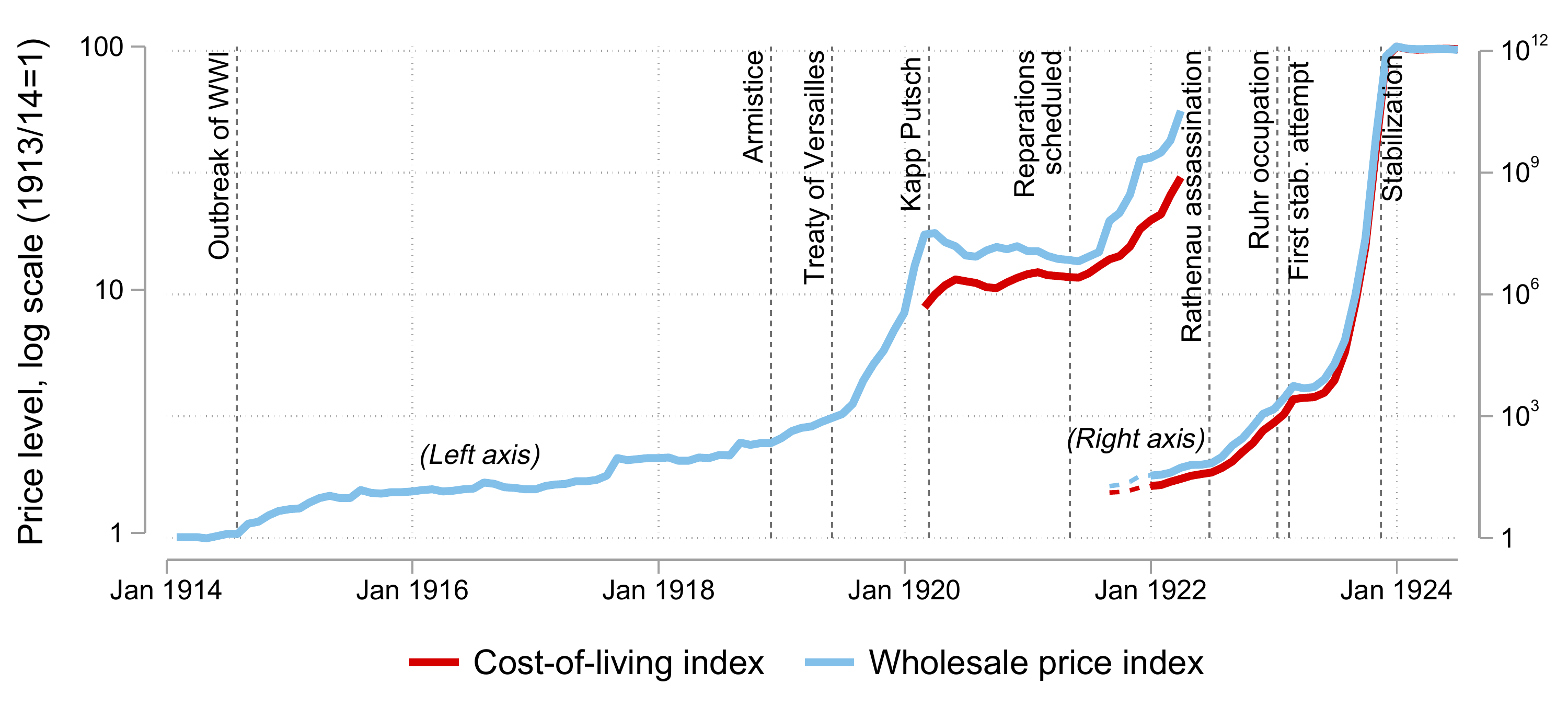}
\end{center}
\footnotesize{\vspace{-.5cm} \textit{Notes}: This figure shows the evolution of the price level in Germany between January 1914 and June 1924. The cost-of-living index only available from  February 1920 onwards. Due to the extreme changes in price levels, prices are reported in logarithms and over two axes, with the first phase of the inflation on the left axis and the second phase on the right axis. The source for the wholesale price index (Gesamtindex der Grosshandelspreise) and cost-of-living index (Lebenshaltung insgesamt) is \textit{Zahlen zur Geldentwertung in Deutschland von 1914 bis 1923}. }
\vspace{-.3cm}
\end{figure}

\paragraph{Two phases of the inflation.} \Cref{fig:agg_price_level} plots the time series of the wholesale price index and the cost-of-living index. Germany's inflation has its roots in WWI, when the gold standard was abandoned and the government increasingly financed the deficit by discounting government securities at the Reichsbank \citep{Feldman1997}. During WWI, the wholesale price index increased by a factor of 2.45. This rate of inflation was slightly higher than in the U.K., where prices increased by a factor of 2.3, but lower than in France, where prices increased by a factor of 3.3. At the end of the war, German inflation and public finances were not in significantly worse condition than France's \citep{Graham1931}.

There are two broad phases of the postwar inflation. The first phase is from the WWI Armistice in November 1918 to the summer of 1922. Panel (a) of \Cref{fig:agg_price_level} shows that inflation accelerated in the second half of 1919, after the signing of the Treaty of Versailles in June 1919. The Treaty assigned sole blame for the war on the Central Powers (the ``War Guilt Clause'') and imposed large and uncertain reparations on Germany. 
From the end of 1918 to the end of 1919, the wholesale price index increased by a factor of 3.3. The price level stabilized during 1920 with the Erzberger fiscal reforms and the successful suppression of the  ``Kapp Putsch,'' a right-wing coup attempt, in March 1920. The price level then rose again in May 1921 after the Reparations Commission determined Germany's exact reparations bill and imposed an ultimatum of a substantial upfront payment in 1921.

The second phase of the inflation runs from July 1922 to stabilization in November 1923, shown in panel (b) of \Cref{fig:agg_price_level}. This phase was ushered in by three events that undermined confidence in Germany's ability to meet reparations \citep{kindleberger1985}. First, in early June 1922, the French government insisted on the original reparations schedule, rather than a reduced schedule. 
Second, hopes of an international loan to stabilize the mark were disappointed by the Reparations Commissions Banker's Committee. Third, on June 24, the foreign minister Walther Rathenau was assassinated by an ultra-nationalist terrorist group. That day the mark depreciated by 7\% against the dollar. Germany then suspended and formally demanded a 2.5 year postponement of reparations in July 1922. From July 1922, monthly inflation exceeded 50\%, marking the start of the hyperinflation phase. Inflation rose further after the invasion of the Ruhr by France and Belgium in January 1923 in response to arrears on the delivery of reparations in kind. The occupation was met with passive resistance, which the government financed by discounting treasury bills, further fuelling inflation. Inflation was finally stabilized in November 1923 through a combination of monetary and fiscal reforms.

\begin{figure}[ht]
	\begin{center}
		\caption{Inflation Expectations Implied by Forward Exchange Rates. \label{fig:eizig}}

        \includegraphics[width=1.0\textwidth]{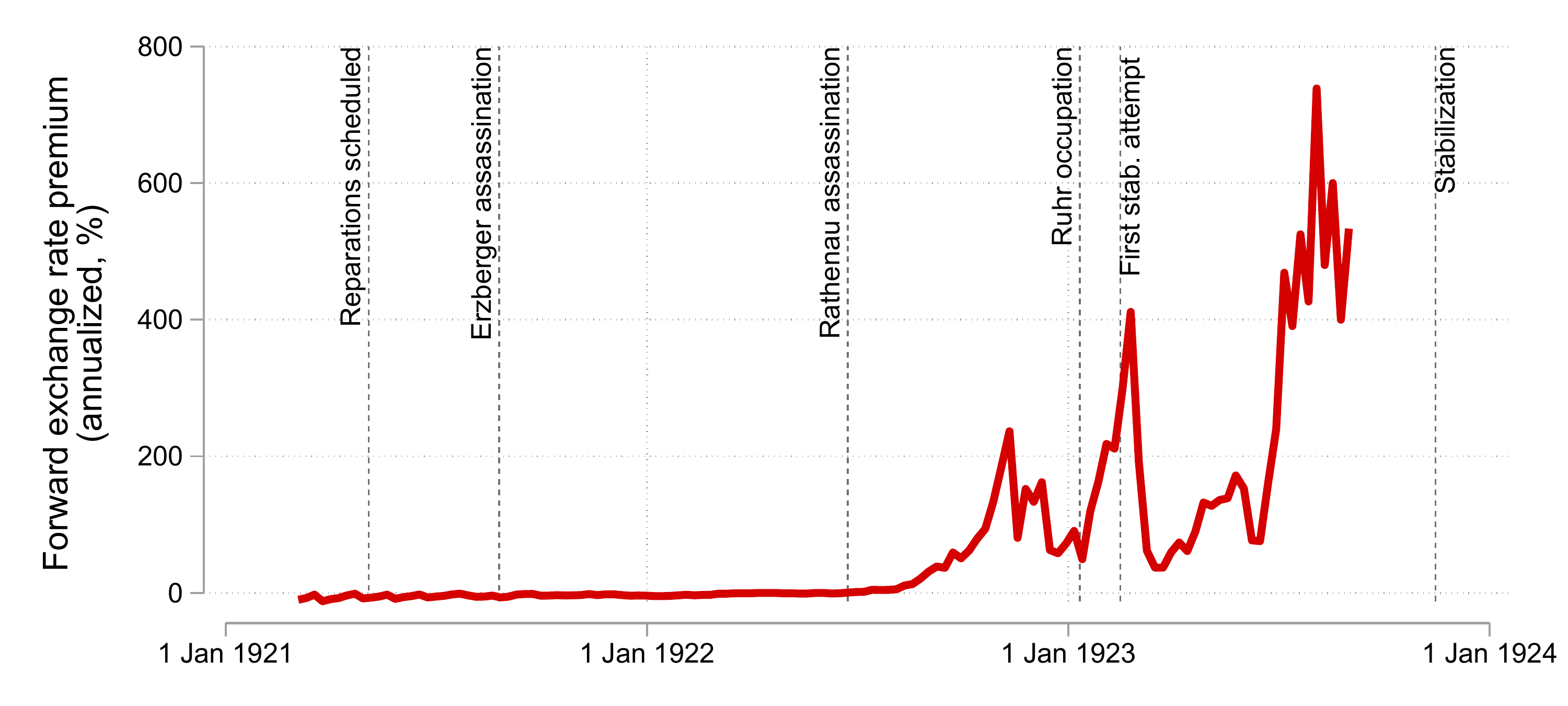}

    \end{center}
    \footnotesize{\textit{Notes}: Forward and spot exchange rate data are from \cite{Einzig1937}, which records the data from the weekly circular of the Anglo-Portuguese Colonial and Overseas Bank, Ltd. The annualized forward premium is based on one-month forward contracts and is calculated as $12 \frac{F-S}{S},$ where $S_t$ is the spot exchange rate in marks per sterling and $F_t$ is the forward exchange rate. A higher value implies that the forward price of sterling (in terms of marks) is at a premium compared to the spot price.}
\end{figure}

\paragraph{Inflation expectations.} Narrative accounts suggest inflation expectations were anchored only increased gradually in the first phase of the inflation \citep{kindleberger1985}.\footnote{\cite{Ferguson1995keynes} reports that Keynes lost \textsterling20,000 speculating on the mark (about \textsterling500,000 today), as foreign speculators boosted the value of the mark relative to his expectations of a larger depreciation. In September 1922, he wrote that ``everyone in Europe and America bought mark notes... the argument... was the same... Germany is a great and strong country; some day she will recover; when that happens the mark will recover also, which will bring a very large profit. So little do bankers and servant girls understand of history and economics'' \citep[][Vol. 18, p. 50]{Keynes_1978}.} Expectations of inflation then shifted decisively in the summer of 1922, and expectations of further depreciation became widespread \citep{Feldman1997}.

This narrative is supported by the behavior of the forward exchange rate. \Cref{fig:eizig} plots the annualized mark/sterling forward premium based on one-month forward contracts. From covered-interest parity and abstracting from a risk-premium, the forward premium is given by $\frac{F_t}{S_t} = \frac{1+i_t}{1+i_t^*}$, where $i_t$ is the mark interest rate and $i^*_t$ is the pound sterling interest rate. Taking an approximation and applying the Fisher equation yields $\frac{F_t-S_t}{S_t} \approx i_t - i_t^* = \E_t \pi_{t+1}- \E_t \pi^*_{t+1} +  r_t  - r^*_t$, where $\pi_t$ is the rate of inflation and $r_t$ the real interest rate. Assuming that most variation in interest rates is due to expectations about inflation in Germany ($\E_t \pi_{t+1}$), the forward premium provides a proxy of inflation expectations.\footnote{\cite{Frenkel1977} also makes this assumption and uses the same forward-premium data to construct a measure of inflation expectations to estimate money demand during the German hyperinflation.} 

Before June 1922, the mark forward rate was at a premium relative to the spot rate, suggesting that the forward market did not imply a large increase in Germany's inflation. Therefore, high inflation came as a surprise relative to expectations in the first phase of the inflation.\footnote{Another piece of evidence for anchored inflation expectations is that firms were able to issue long-term fixed-rate bonds in 1919--1920, as we discuss below (see \Cref{fig:bonds_maturity}). Moreover, in 1922 many borrowers repaid loans early to take advantage of the depreciated mark because they (wrongly) expected the mark to strengthen again \citep{Hughes1988paying}.} While this is perhaps surprising \textit{ex post}, inflation had been low from Germany's unification in 1871 until WWI, at an average rate of 0.7\% per year.\footnote{Inflation had been low in all major economies for multiple decades under the Gold Standard before WWI \cite{Eichengreen1995}.} The last hyperinflation occurred over a century earlier during the French Revolution \citep{Sargent1995}. 

The forward price of sterling moved decisively from a discount to a premium the week of the Rathenau assassination in late June 1922, suggesting that inflation expectations became unanchored \citep{Holtfrerich1986}. After the political turmoil during the summer of 1922, the forward traded at an increasingly large premium.

\begin{figure}[!ht]
    \caption{Real GDP in Germany and Other Major Economies, 1918--27.}
        \begin{center}
	    \includegraphics[width=0.7\textwidth]{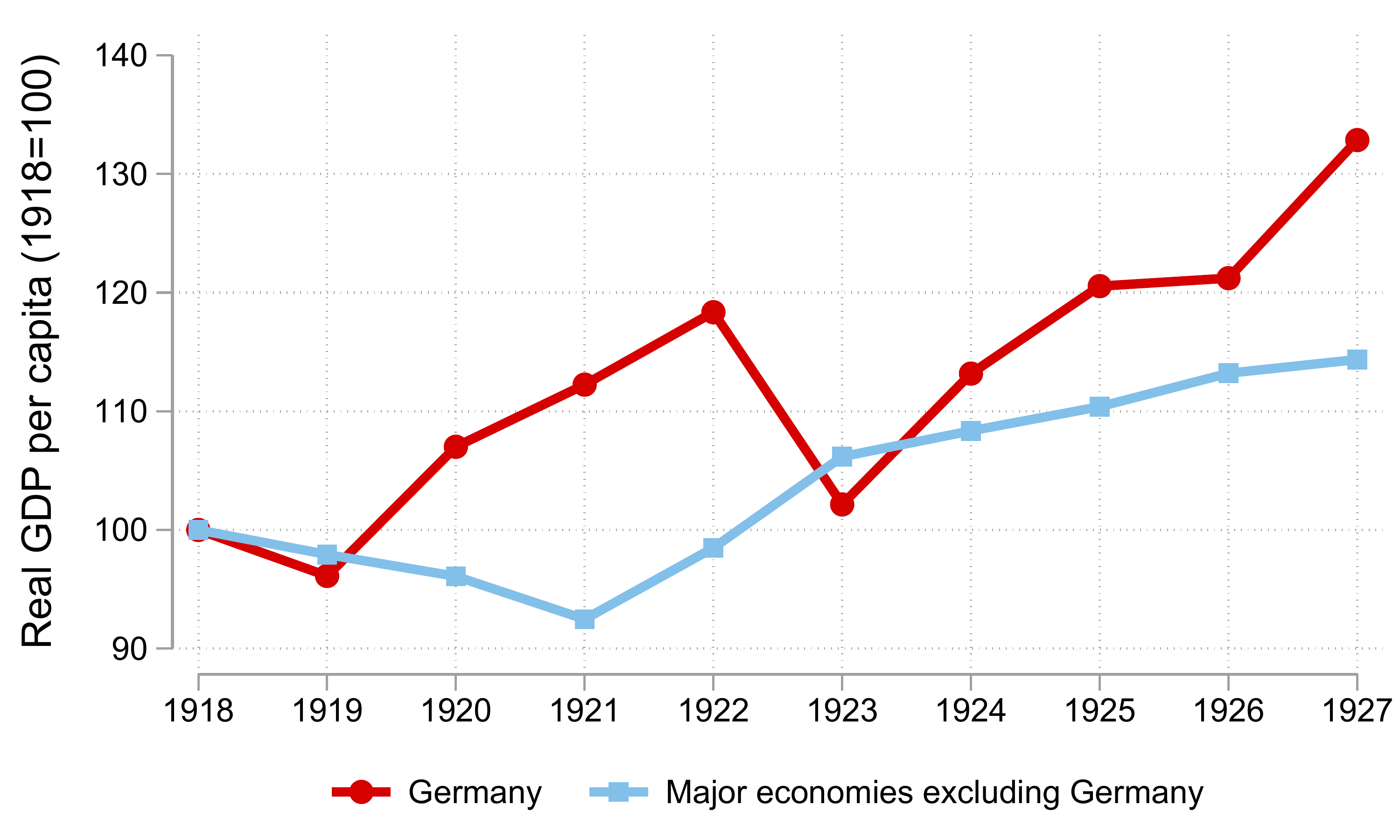}
        \end{center}
        \footnotesize{\vspace{-.5cm} \textit{Notes}: This figure presents real GDP per capita for Germany and an index of other major economies. The series are indexed to 100 in 1918. The data are from \cite{JST2017} and \cite{BarroUrsua}. ``Other major economies excluding Germany'' is an index of average real GDP growth per capita, weighted by lagged nominal GDP in U.S. dollars. The index is constructed using 15 countries with continuous coverage in the \cite{JST2017} database between 1914 and 1927 (Australia, Belgium, Canada, Denmark, Finland, France, Italy, Japan, Norway, Portugal, Spain, Sweden, Switzerland, the U.K., and the U.S.).}
	 \label{fig:real_GDP}
        \vspace{-.1cm} 
\end{figure}

\paragraph{Real activity during the inflation.} The postwar inflation was associated with a booming economy through the third quarter of 1922, followed by a severe bust starting at the end of 1922.\footnote{As  emphasized by \cite{Graham1931}: ``that business in Germany was booming during most of the inflation period is a universally admitted fact.'' \cite{Graham1931} further argued that inflation contributed to the boom.} \Cref{fig:real_GDP} plots an index of real GDP per capita for Germany starting in 1918. For comparison, we also plot an index of average real GDP per capita growth for other major industrial economies, weighted by GDP. While the U.S., U.K., and other industrial economies underwent deflation and declining output to maintain or return to pre-war gold parities, Germany's real GDP per capita rose by 20\% from 1919 to 1922. Further, unemployment was low from the end of WWI until the last months of 1922 (see \Cref{fig:unemp_def}). Germany's inflationary boom slows with the hyperinflation in the second half of 1922 and decisively reverses in early 1923, following the invasion of the Ruhr and the resulting passive resistance. In 1923, Germany saw a large fall in real GDP, and unemployment rose to nearly 30\% at the height of the hyperinflation and in the run-up to the stabilization.

\subsection{Aggregate Evidence on Firm Bankruptcies and Leverage} 

\begin{figure}[!ht]
\begin{center}
\caption{Inflation and Firm Bankruptcies. \label{fig:fin_PC}}
\includegraphics[width=0.7\textwidth]{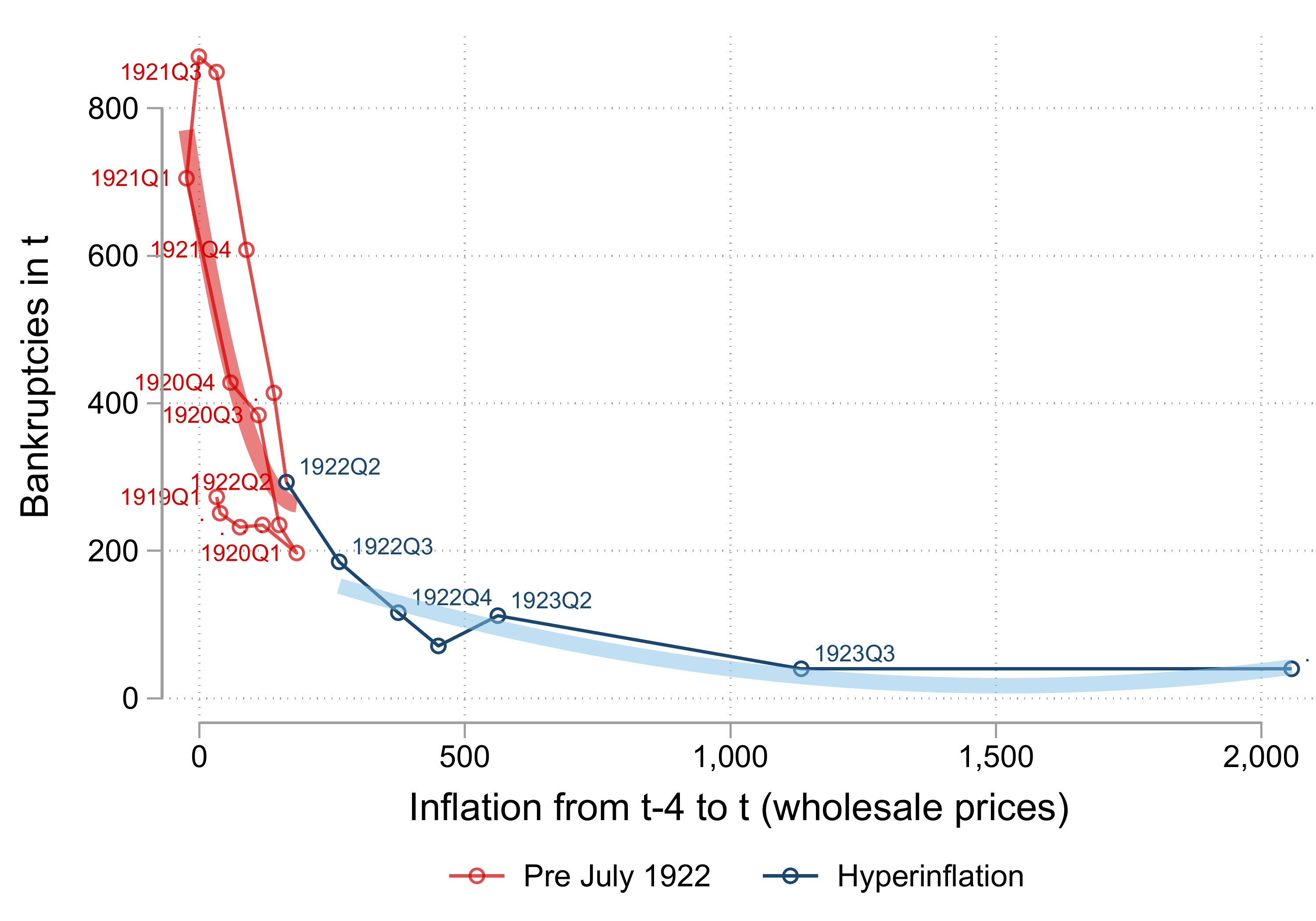}
\end{center}
\footnotesize{\textit{Notes}: This figure plots the number of firm bankruptcies in quarter $t$ against realized inflation over the past four quarters from $t-4$ to $t$. Inflation is calculated as the log change (times 100). Quarterly counts of firm bankruptcies are obtained from the \textit{Vierteljahrshefte zur Statistik des Deutschen Reichs Herausgegeben vom Statistischen Reichsamt}. Inflation of wholesale prices as reported in \textit{Zahlen zur Geldentwertung}. The thick lines represent quadratic fits, computed separately for each of the two stages of inflation.}
\vspace{-.1cm} 
\end{figure}

\paragraph{Inflation and firm bankruptcies.} In the presence of nominal debt contracts, unexpected inflation will increase the net worth of levered nonfinancial firms and reduce the likelihood of bankruptcy. \Cref{fig:fin_PC} plots the relation between the number of firm bankruptcies in a quarter and inflation over the past four quarters. There is a striking negative relation between inflation and firm bankruptcies. Bankruptcies fall with rising inflation in 1919, then rise with falling inflation in 1920, before falling as inflation rises from the second half of 1921. The relation is also convex. It is steep during the first phase of the postwar inflation for 1919Q1 to 1922Q2, but flatter in the hyperinflation phase from 1922Q3 to 1923Q4.\footnote{See \Cref{tab:financial_pc} in the appendix for the corresponding regression estimates. Further, while \Cref{fig:fin_PC}  uses the level of annual inflation on the x-axis,  \Cref{fig:fin_PC_accel} in the Appendix reveals a similar pattern when using  the ``accelerationist'' version that uses the change in inflation.} Once annual inflation is high, further increases in inflation are not associated with additional defaults. It is worth noting that bankruptcies remained low even in 1923, when output declined significantly.

\begin{figure}[!ht]
    \caption{Hyperinflation Led to Debt-Inflation and a Decline in Interest Expenses.}
        \begin{center}
        \subfloat[$\text{\textit{Debt-inflation}}_{i,t} $.]{\hspace*{-0.3in}\includegraphics[width=0.52\textwidth]{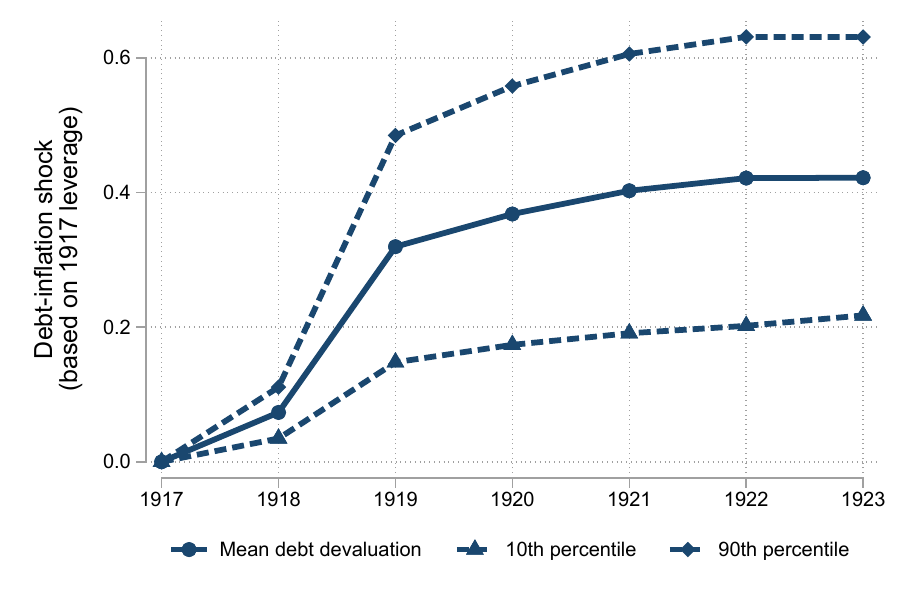}} 
        \subfloat[Evolution of expense shares over time. \label{fig:cost_shares}]{\includegraphics[width=0.52\textwidth]{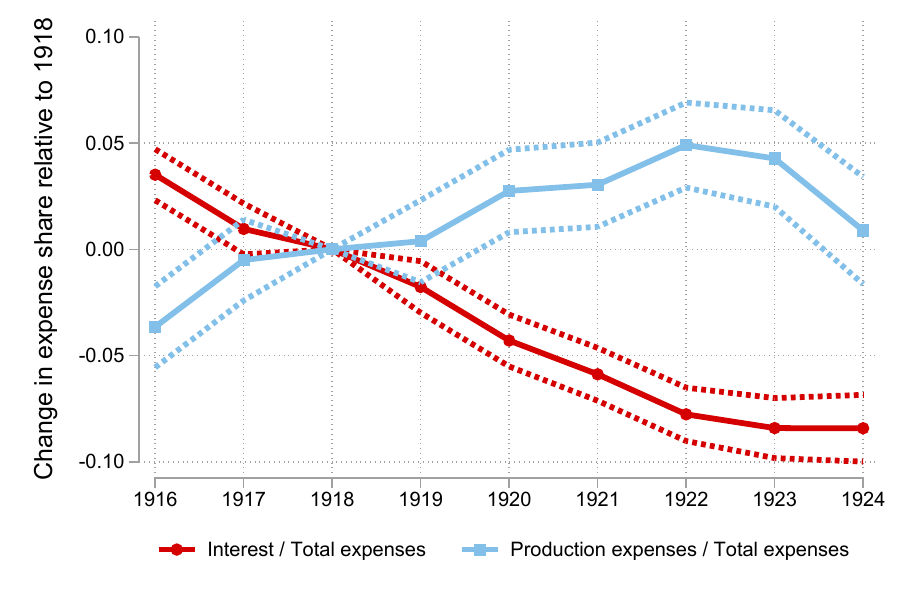}}
\end{center}
\vspace{-.7cm}
\footnotesize{\textit{Notes}: Panel (a) plots the average, 10\textsuperscript{th} percentile, and 90\textsuperscript{th} percentile of $\text{\textit{Debt-inflation}}_{i,t} $ over time. $\text{\textit{Debt-inflation}}_{i,t} $ is defined in \eqref{DV}. Panel (b) plots the evolution of interest expenses and production expenses, both as a share of total expenses, during the inflation. Specifically, it plots the sequence of estimated year fixed effects $\{\hat \gamma_t\}$ from firm-level two-way fixed effects regression regression of the form: $\text{Expense Share}_{it} = \alpha_i + \gamma_t + \epsilon_{it}, $ for interest expenses to total expenses and production expenses to total expenses as the dependent variables. Expenses are from the previous twelve months for fiscal year ending in a reported year. This regression captures the change in the expense share within firm. Errors bands represent 95\% confidence intervals based on standard errors clustered at the firm level. 
        }
	 \label{fig:leverage_hist}
\end{figure}

\paragraph{Inflation eroded real firm debt burdens.} Our hypothesis is that the negative relation between inflation and bankruptcies emerges because inflation erodes firms' nominal liabilities. \Cref{fig:leverage_hist} provides direct evidence that the inflation reduced firm leverage. In panel (a), we compute the realized cumulative \textit{debt-inflation} shock. This measures the reduction in initial book leverage induced by the inflation over time, relative to initial total assets. We base the calculation on leverage outstanding as of 1917, before the inflation. We then calculate the difference between debt at the price level in an initial year (normalized to one, $P_{1917}=1$) and leverage at subsequent price levels $P_t$, normalized by initial assets:
\begin{align}
\text{\textit{Debt-inflation}}_{i,t} =  \frac{D_{i,1917}}{D_{i,1917} + E_{i,1917} } - \frac{(D_{i,1917} / P_t) }{D_{i,1917} + E_{i,1917} }=\frac{D_{i,1917} }{D_{i,1917} + E_{i,1917} } \frac{\pi_t}{1+\pi_t},  \label{DV}
\end{align}
where $D_{i,1917}$ and $E_{i,1917}$ are the liabilities and equity of firm $i$ in $1917$, and $\pi_{t} $ is inflation from $1917$ to $t$.

Panel (a) of \Cref{fig:leverage_hist} plots the average of $\text{\textit{Debt-inflation}}_{i,t} $ over time. The figure shows that the real value of debts outstanding in 1917 was eroded throughout the inflation, particularly when inflation first accelerated in 1919. By 1922 most debts outstanding in 1917 were already wiped out, and the average of $\text{\textit{Debt-inflation}}_{i,t}$ reaches its maximum potential value of 43\%, the average initial leverage ratio. Panel (a) of \Cref{fig:leverage_hist} also plots the 10\textsuperscript{th} and 90\textsuperscript{th} percentiles of $\text{\textit{Debt-inflation}}_{i,t}$, showing that there is considerable variation across firms in the extent of erosion in real debt burdens. We exploit this variation in our cross-sectional analysis below.

\Cref{fig:lev_hist} in the appendix further compares the distribution of firm book leverage in 1917 and 1924 in the \textit{Saling's} firm-level data. 
We choose 1924 as the endpoint to utilize the more reliable Goldmark balance sheets, which firms were required to report by January 1924, shortly after the stabilization. \Cref{fig:lev_hist} shows that the distribution of leverage shifts significantly to the left during the German inflation. The liabilities-to-assets ratio falls by 20 percentage points for the average firm in the sample. At the same time, the fact that leverage does not collapse to zero indicates firms were partly able to re-lever during the inflation.

Panel (b) of \Cref{fig:leverage_hist} analyzes firm leverage from the perspective of interest expenses. Specifically, it shows the evolution in the share of interest expenses in firms' total non-depreciation expenses. We calculate the within-firm change using the estimated year fixed effects from a firm-level regression with firm and year fixed effects. Firms do not always break out interest expenses from other expenses in the income statements reported in the \textit{Saling's} data, so this measure is a lower bound on interest expenses. \Cref{fig:leverage_hist}(b) shows a clear decline in the share of interest expenses to total expenses during the inflation. The interest expense share declines by about 10 percentage points. The erosion of real debt thus directly boosted firms' interest coverage ratios and benefited their liquidity positions. 

\Cref{fig:leverage_hist}(b) also plots the evolution of production expenses to total expenses. Production expenses are defined as revenue minus EBITDA. Production expenses as a share of total expenses rise during the inflation. This is consistent with firms reallocating expenses from interest payments to payments to materials and employment to boost production.

\subsection{Price and Wage Flexibility during the Inflation}

Inflation reduced real debt burdens from fixed nominal debt contracts. How did wage and price setting respond? Real wages, an implicit nonfinancial liability for firms, may also have declined if wages were sticky and slow to respond to inflation, further benefiting firms. At the same time, wages can be renegotiated \textit{ex post} and can therefore have significantly shorter effective maturity than debt contracts. Similarly, prices, which matter for firm revenues, can also be adjusted after a shock to inflation.

We first analyze the frequency of wage adjustment during the inflation. The ``Stinnes-Legien'' agreement from November 1918 enshrined a set of workers' rights long coveted by the German labor movement, including the recognition of trade unions as the official representatives of the workforce. This allowed for industry-level union bargaining. We therefore collect industry-level wages for seven industries from 1920 through 1923. 

\begin{figure}[!ht]
	\begin{center}
	\caption{Interval between Price Adjustment Falls during the Inflation: Evidence from Wages and  Cost-of-Living Index Prices.}
   \subfloat[Frequency of wage adjustments and inflation.]{\includegraphics[width=0.49\textwidth]{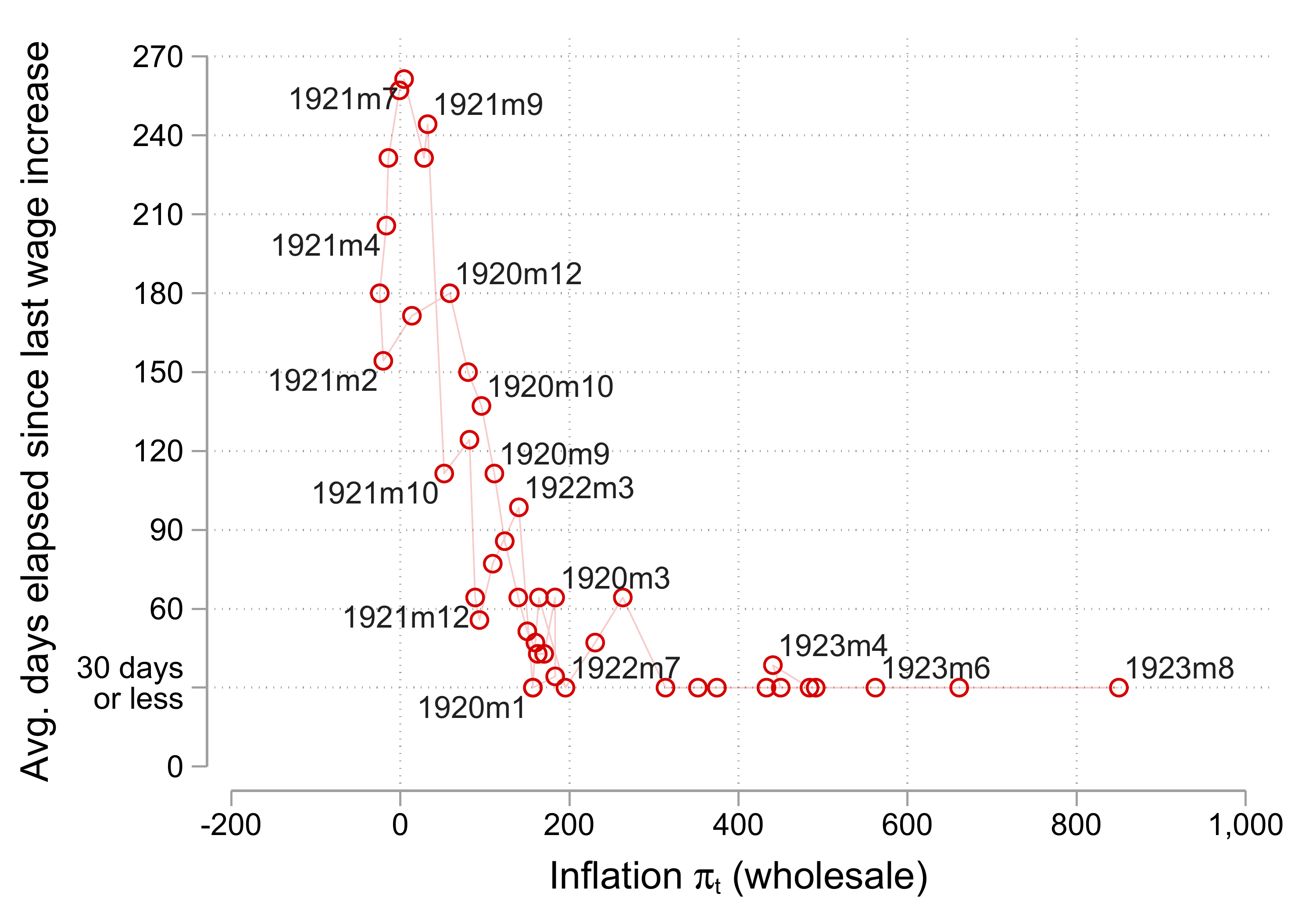}}
   \subfloat[Frequency of price adjustments and inflation.]{\includegraphics[width=0.49\textwidth]{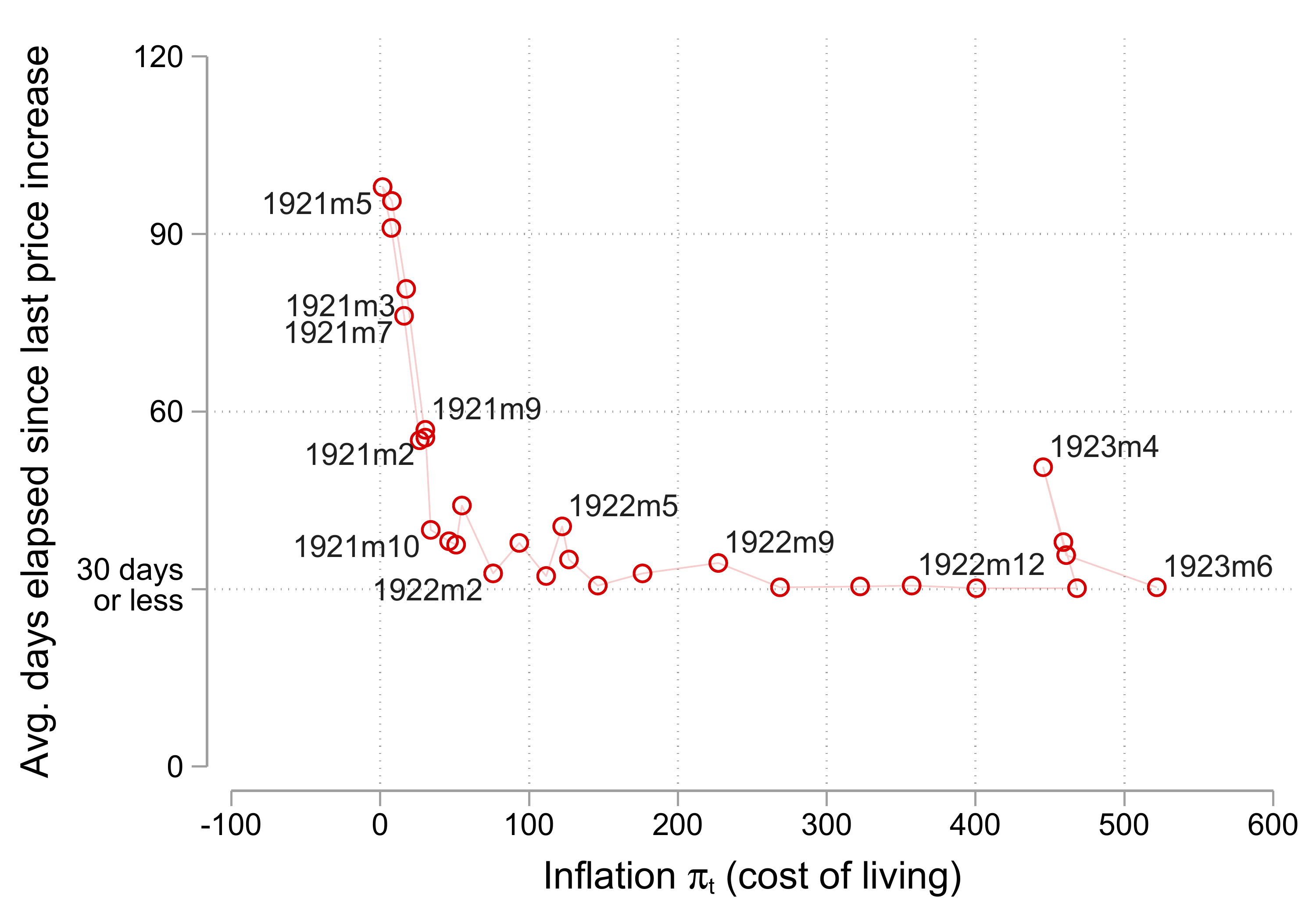}}
      \label{fig:price_interval}
\end{center}
	
    \footnotesize{\vspace{-.5cm} \textit{Notes}: This figure plots the duration of unchanged wages and of prices of products underlying the cost-of-living index against inflation. Wages and retail prices are as reported in \textit{Zahlen zur Geldentwertung in Deutschland von 1914 bis 1923} and \textit{Wirtschaft und Statistik} (various issues). Inflation is defined as the difference between the log of the price level in month $t$ and month $t-12$, times 100. See \Cref{fig:price_interval_additional_wages} and \Cref{fig:price_interval_additional} in the appendix for additional details.}
\end{figure}

Panel (a) in \Cref{fig:price_interval} examines the frequency of wage adjustment during the inflation.  It plots the number of days since wages were last increased, averaged across the seven industries, against wholesale price inflation.\footnote{\Cref{fig:price_interval_additional_wages} plots the average number of days elapsed since wages were last increased separately for each of the seven industries.} As inflation accelerated, wages were adjusted with increasing frequency. At low levels of inflation, wages were adjusted, on average, every 9 months. This frequency increased to every 60 days or less once inflation exceeded 100\% and every 30 days or less during hyperinflation.\footnote{Narrative evidence suggests that wages were adjusted weekly once the inflation reached hyperinflation \citep{Feldman1997}. We obtain union-bargained wages only at the monthly frequency and thus can only provide an upper bound on the time elapsed since the last wage increase.}

The increasing flexibility of wages informs the evolution of real wages. Relative to pre-war levels, real wages declined during the inflation. 
The decline in real wages is more pronounced for higher-skilled salaried workers (see Appendix \Cref{fig:real_wages} for state employees from 1914 through 1923). However, most of the decline in real wages had already occurred by the end of 1920. During 1920--1923, real wages fluctuate substantially, but there is no clear evidence of a downward trend. This is consistent with wages becoming flexible and racing to keep up with prices once inflation becomes high. 

Inflation also led to a shortening in the duration of retail prices. \Cref{fig:price_interval} panel (b) presents a similar analysis for the frequency in the adjustment of goods prices. The analysis is based on the prices of goods underlying the cost-of-living index. The data are at the city-product-month level for 18 major cities and 12 retail products such as sugar, pork, and milk. The price quoted for each city is the average of 4 samples taken over the course of a month for each city. An important caveat is that because data are not quoted prices but averages of quoted prices, they may overstate the frequency at which individual quoted prices were adjusted upwards.\footnote{Note that this caveat does not apply to wages. Further, to partially address this concern, we also hand-collected the daily prices of two widely circulated newspapers. These represent actual quoted prices, but only for two goods. We find similar results on the frequency of price adjustment for these newspaper prices, as shown in \Cref{fig:price_interval_newspapers}.}

\Cref{fig:price_interval}(b) relates the average days elapsed since the last price increase against the level of inflation. There is a negative relation between inflation and the duration of price adjustment. For cost-of-living inflation around 0\%, prices are adjusted every 80 to 100 days. However, once inflation exceeds 50\%, product prices are adjusted at least once every 30 to 40 days. By early 1922, before the hyperinflation phase,  prices of most goods in all cities are adjusted upwards every 30 to 60 days. By the hyperinflation, prices are adjusted even every 7 days or less (see Appendix \Cref{fig:price_interval_additional}). 

Overall, \Cref{fig:price_interval} indicates that prices and wages became increasingly flexible with rising inflation. The reduction in the time elapsed between price and wage adjustments with rising inflation is consistent with menu cost models \citep{Gagnon2009,Nakamura2018,Alvarez2019}. There was thus likely limited stickiness in the aggregate price and wage levels once the inflationary shocks became large \citep{golosov2007menu,caballero2007price}.\footnote{Nevertheless, menu costs could have induced large economic costs through price dispersion, as calibrated by \cite{Alvarez2019} using micro-data on prices in Argentina's 1989-90 hyperinflation.} Revenues and wage expenses of nonfinancial firms increased at a similar pace as overall inflation once inflation became high. In contrast, the debt-inflation channel may still be active even for large shocks to the price level, as long-term financial contracts cannot be renegotiated.


\section{Firm Level Evidence on the Debt-Inflation Channel}
\label{sec:firm_level}
The analysis of aggregate data in \cref{sec:aggregate} shows that the increase in the price level reduced the real value of nominal debt claims and interest payments, which in turn led to a drastic decline in firm bankruptcies. The aggregate data thus suggest that inflation relaxed firms' financial constraints and thus potentially stimulated economic activity through the debt-inflation channel. A concern with interpreting patterns in aggregate data, however, is the possibility of a spurious relationship due to unobserved aggregate shocks correlated with both inflation and bankruptcy rates. Moreover, while the inflation may have redistributed from debt-holders to equity-holders, this redistribution may not necessarily have affected real economic activity. 

In this section, we analyze cross-sectional variation using firm-level data to test whether firms with higher leverage experience differential outcomes during the inflation. We exploit that firms with relatively higher leverage naturally have a higher nominal balance sheet exposure to unexpected increases in the price level. This analysis allows us to tighten the empirical link between inflation and real economic outcomes through the debt-inflation channel.

\subsection{Leverage and Firm-level Employment}

\paragraph{Empirical specification and identification.} We test whether firms with high leverage expand their economic activity relative to firms with low leverage during the inflation. We proxy a firm's real economic activity by its total number of employees. Studying employment as the outcome variable is especially informative, as balance sheets become less reliable during the hyperinflation. Hence, the reported number of employees dominates alternative indicators of real economic activity constructed from financial statements, such as capital expenditure. 

We examine the evolution of firm outcomes across firms with high and low leverage through the lens of a standard dynamic difference-in-differences model. We estimate variants of the following specification:  
\begin{align}
    \ln(Employment_{it}) = \alpha_i + \delta_{st} + \sum_{y\neq 1918} \beta_y Leverage_{i} \mathbf{1}_{y=t}  + \sum_{y\neq 1918}  X_{i} \Gamma_y \mathbf{1}_{y=t} + \epsilon_{it}  \label{eq:dynamic}
\end{align}
where $Employment_{it}$ is the number of employees at firm $i$ in year $t$.  $Leverage_{i}$ is a measure of firm $i$'s leverage at the start of the inflation. Further, $\alpha_i$ is a firm fixed effect, $\delta_{st}$ is an industry-time fixed effect, and $X_{i}$ is a set of firm-level control variables that are interacted with time fixed effects.\footnote{The industry classification corresponds approximately to two-digit SIC industries.}  The sequence of estimates $\{\hat \beta_y\}$ captures the evolution of employment for high-leverage firms, relative to low-leverage firms, with 1918 as the benchmark year. Our main estimation period is 1914 to 1923.
 
We measure a firm's balance sheet exposure to inflation by its leverage ratio before the onset of the inflation. Intuitively, firms with relatively more nominal liabilities---to the extent these are not short-term and constantly repriced, floating rate, or indexed to the price level---benefit more from unexpected increases in the price level. In a frictionless economy in which the Modigliani-Miller theorem holds, equity-holders would benefit from unexpected inflation, but the resulting changes in a firm's capital structure would have no impact on firm real investment or employment decisions. However, if firms are financing constrained, unanticipated inflation can relax financing constraints for levered firms, leading them to increase investment and employment. In this context, it is important to emphasize that theory suggests that debt inflation can affect real outcomes through several related channels, see \cref{sec:conceptual_framework}.  We note that it goes beyond the objective of our paper to identify which exact frictions are ultimately relaxed by the inflation.

We define leverage either as the ratio of a total liabilities to total assets or a financial debt to total assets. The latter measure incorporates both short- and long-term financial debt such as trade credit, bank debt, and bonds.  The former measure also includes other nominal liabilities such as accrued wages, unpaid taxes, and pensions. Most debt contracts, especially long-term debt contracts, were fixed-rate.\footnote{While no detailed information on the contractual structure of bank debt is available, it is technically possible that short-term bank debt was  floating-rate and indexed to the Reichsbank's discount rate. However, the Reichsbank's discount rate remained unchanged at 5\% between summer 1914 and July 1922. The lack of a policy response hence implied that any floating-rate debt claims were effectively fixed-rate. The Reichsbank eventually gradually  increased its discount rate to 18 percent by April 1923 and then to 90\% in summer 1923.} Long-term debt was common, and firms often issued fixed-rate bonds with a maturity from 10 to 50 years. We discuss and exploit this additional variation further below.
 
The identifying assumption behind our empirical strategy is that of parallel trends: In the absence of differences in leverage, firms with high and low leverage would have evolved in parallel during the inflation. Identification does not require that leverage be randomly assigned, but it assumes that leverage at the onset of the inflation is uncorrelated with other shocks to employment during the inflation, conditional on controls. This identifying assumption would be violated if highly levered firms faced better investment opportunities. We take a series of steps to alleviate concerns about this threat to identification.

A key identification concern for our empirical strategy is that sophisticated firms could have strategically adjusted their leverage in anticipation of inflation. This behavior could, in turn, be correlated with subsequent investment opportunities and firm growth. To alleviate this concern, we calculate $Leverage_i$ using balance sheet data from 1917, more than a year before the acceleration of inflation in the middle of 1919. Recall that in 1917, the outcome of WWI, the Treaty of Versailles, and the subsequent inflation would have been extremely difficult to predict.  Hence, it is unlikely that firms chose their debt finance in anticipation of the postwar inflation dynamics.\footnote{Leverage is highly persistent. Therefore, it is unsurprising that our results are robust to using leverage in other proximate years such as 1918. Moreover, below we show that results are also robust to instrumenting leverage at the end of the war with pre-WWI leverage. Our findings support historical accounts which suggest that firms generally did not borrow in anticipation of the inflation \citep{Lindenlaub1985,Balderston1991}.}

We also study what factors explain the variation in firm leverage. \Cref{tab:lev_correlates}  shows that the main factors in explaining cross-firm variation in leverage are firm size and industry. The positive correlation between leverage and size is consistent with a size-dependent borrowing constraint, perhaps due to a lower probability of default for large firms \citep{RajanZingales1995,Gopinath2017}. Tobin's Q is also positively correlated with leverage but explains only a small share of its variation. After controlling for these variables, leverage is uncorrelated with EBIT margin and the share of fixed assets in total assets. Moreover, there is considerable unexplained variation, consistent with the evidence that debt ratios often vary widely even for similar firms \citep{Myers1984,GrahamLeary2011}. 

We further exploit that the cross-section provides a stronger test than the aggregate time series, as we can control for aggregate and industry-specific shocks impacting firms. Industry-year fixed effects, $\delta_{st}$, absorb aggregate shocks such as the Ruhr invasion as well as industry-specific shocks that might be correlated with leverage. For example, inflation led to a flight from the mark toward durable assets, which disproportionately benefited firms in industries producing these assets \citep{Graham1931}. Further, price and wage rigidity may vary across industries and be correlated with leverage. For example, \cite{DAcunto2018flexible} find that U.S. public firms with more flexible prices have higher financial leverage. In this context, union-bargained wage setting was common but industry-specific. Industry-year fixed effects absorb the effects of industry-specific rigidity in wages or prices. 

Finally, we include a range of controls to capture differences in firms' investment opportunities that may be correlated with leverage.  Our baseline analysis controls for size, fixed assets to total assets, free cash flow to total assets, and profit margin. We measure all of these variables in 1917, following the timing of our measure of $Leverage_i$. To account for expected differences in investment opportunities, we also control for Tobin's Q in 1918.\footnote{We construct Tobin's Q for 1918 based on balance sheet data from 1918 and stock prices from January 1919 (see Appendix \ref{variable-construction}).}  In addition, we explore additional controls for export status, political connections, exposure to the war economy, and credit supply shocks. To ensure that results are not driven by a differential cyclicality of high-leverage firms, we also present placebo tests by studying employment dynamics across high and low leverage firms in other business cycles. Finally, we study pre- and post-trends and show that the impact of leverage on employment is concentrated in the inflation period, with the largest effects occurring between 1919 and 1922.

\paragraph{Main Results.} 

\begin{figure}[!ht]
    \caption{Employment Dynamics across Low and High Leverage Firms. \label{fig:leverage_employment_dynamics}}
        \begin{center}
	\subfloat[Employment across high and low leverage firms.]{
        \includegraphics[width=0.49\textwidth]{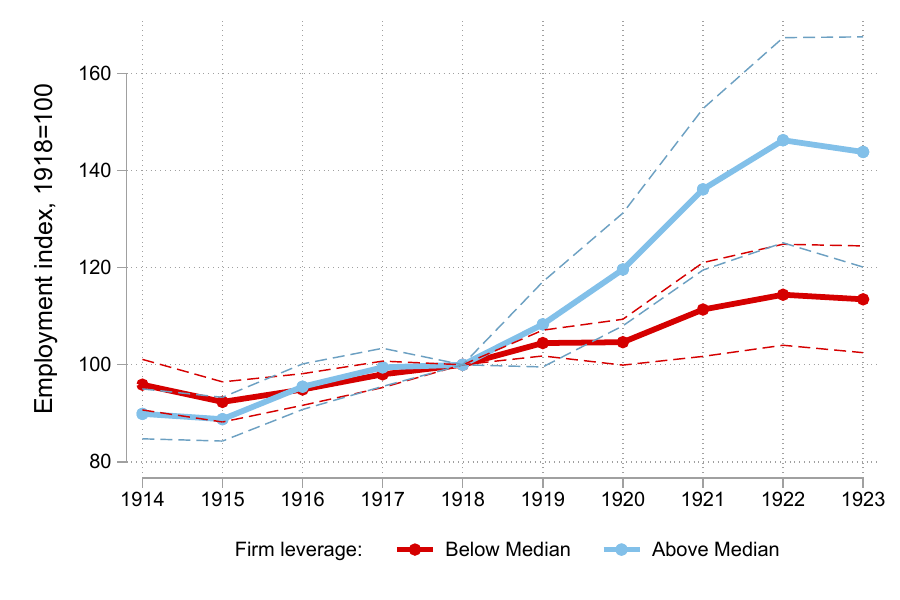}}  
        \hfill
     \subfloat[Results from estimating equation \eqref{eq:dynamic}.]{
        \includegraphics[width=0.49\textwidth]{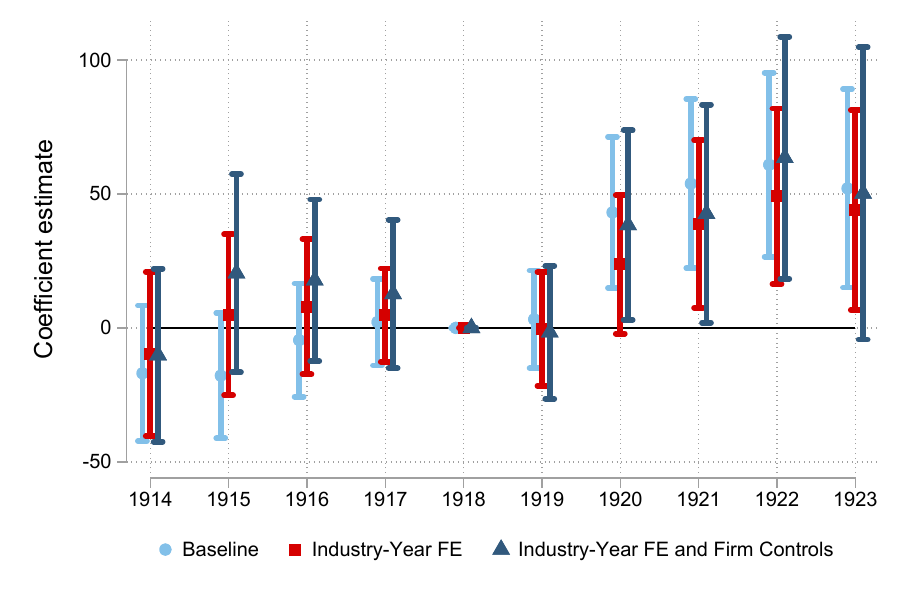}}  
                 
        \end{center}
        \footnotesize{ \vspace{-.5cm} \textit{Notes}: { Panel (a) presents the average employment growth for firms with above- and below-median leverage. Leverage is defined as the average ratio of liabilities to assets in 1917.  Employment is indexed to 100 in 1918 for each firm. Dashed lines represent 95\% confidence intervals. Panel (b) presents the sequence of coefficients  $\{\beta_y\}$ from estimating equation \eqref{eq:dynamic} with log employment (times 100) as the dependent variable. Firm-level control variables are firm size (log of assets), the share of fixed assets to total assets, free cash flow to assets, and profit margin (EBIT-to-revenue), all as of 1917, as well as Tobin's Q as of the end of 1918. Firm controls are interacted with year fixed effects. Errors bars represent 95\% confidence intervals based on standard errors clustered at the firm level.} 

}
\end{figure}

To illustrate our main results, we first plot the raw data. Panel (a) of \Cref{fig:leverage_employment_dynamics} presents the average evolution of employment for firms with above- and below-median leverage. Leverage is defined as liabilities-to-assets in 1917.\footnote{\Cref{fig:leverage_employment_dynamics_robustness} shows that the patterns documented in \Cref{fig:leverage_employment_dynamics}(a) are robust to computing employment growth with log changes, grouping firms by terciles of leverage, and sorting firms by debt-to-assets instead of liabilities-to-assets.} On average, firms see strong employment growth from 1918 through 1922. This is in line with evidence from \cref{sec:aggregate} that the inflation was associated with increased economic activity through 1922. Moreover, while high- and low-leverage firms expand their employment at similar rates from 1914 through 1918, the strongest employment gains during the inflation occur for firms with higher leverage.  Employment at firms with above-median leverage grows by approximately 45\% from 1918 through 1922. In contrast, firms in the lowest tercile experience only a 15\% increase in employment.

Panel (b) in \Cref{fig:leverage_employment_dynamics} presents the results from estimating \eqref{eq:dynamic} using $\sfrac{Liabilities}{Assets}_{i,1917}$ as the measure of leverage. Firms with higher leverage see stronger employment growth in 1919--1922 relative to firms with lower leverage. The estimates are similar when including industry-year fixed effects and firm-level controls interacted with year fixed effects. In line with the evidence in panel (a), panel (b) of \Cref{fig:leverage_employment_dynamics}  reveals that employment at high-leverage firms expands faster after the start of the inflation and remains elevated through 1923. 

Panels (a) and (b) in \Cref{fig:leverage_employment_dynamics} also reveal that there is limited systematic difference in employment growth across high and low leverage firms before the acceleration of inflation in 1919. Panel (a) suggests that high-leverage firms perhaps grew slightly faster between 1915 and 1916, possibly due to differential exposure to the massive labor market reallocation during WWI. However, the estimates in panel (b) suggest this pattern is not robust to the inclusion of industry-year fixed effects. Hence, whatever is driving the potential pre-trend between 1915 and 1916 seems to be driven by industry-specific factors which, in turn, cannot explain the postwar boom in employment in highly levered firms. 
Alternative specifications reported in \Cref{app:additional_results} reinforce the evidence on the absence of employment pre-trends; see \Cref{fig:leverage_employment_dynamics_robustness} and \Cref{fig:dd_emp_robust_all}.

\begin{figure}[!ht]
    \caption{Binned Bivariate Means of Employment Growth and Leverage. \label{fig:binsreg}}
        \begin{center}
        
        \includegraphics[width=0.8\textwidth]{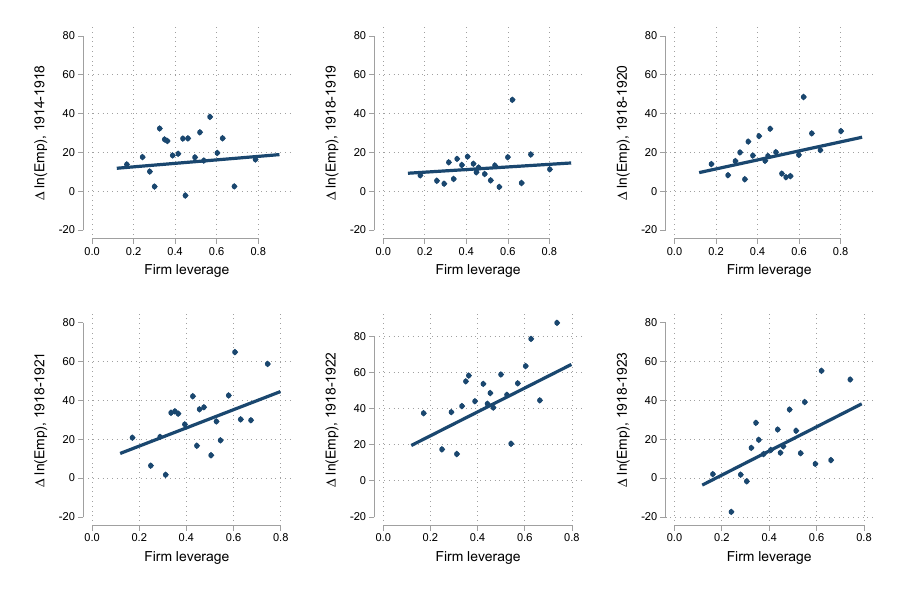}
                 
        \end{center}
        \footnotesize{ \vspace{-.5cm} \textit{Notes}: This figure plots binned bivariate means of leverage before the inflation against firm employment growth (log changes, times 100) across various years. Leverage is defined as liabilities-to-assets in 1917. The figure is constructed using the methodology of \citet{Cattaneo2019binscatter}, using 20 equally sized bins. It controls for industry fixed effects and the following firm-level variables: log of assets, fixed assets to total assets, free cash flow to assets, and profitability, all as of 1917, and Tobin's Q as of 1918. 
         
}
\end{figure}

As an alternative way to visualize the results, \Cref{fig:binsreg} plots binned scatterplots of employment growth for various years relative to 1918 against firm leverage in 1917. The figure shows that there is no relation between firm growth and leverage before the inflation (top-left panel). As inflation then accelerates after 1919, high-leverage firms outperform low-leverage firms. Altogether, we interpret these patterns as an indication that the debt-inflation channel played an important role in how the inflation starting in 1919 transmitted to the real economy.

\begin{table}[!ht]
  \centering
   \caption{Firm Leverage and Employment.}\label{tab:leverage_employment}
        \begin{minipage}{1.0\textwidth}
        \begin{center}
        \scalebox{0.85}{{   \def\sym#1{\ifmmode^{#1}\else\(^{#1}\)\fi}  \begin{tabular}{l*{8}{c}}   \toprule  Dependent Variable & \multicolumn{8}{c}{ln(Employment)} \\    \cmidrule(lr){2-9}       
                &\multicolumn{3}{c}{OLS}                                 &\multicolumn{1}{c}{2SLS}&\multicolumn{3}{c}{OLS}                                 &\multicolumn{1}{c}{2SLS}\\\cmidrule(lr){2-4}\cmidrule(lr){5-5}\cmidrule(lr){6-8}\cmidrule(lr){9-9}
                &\multicolumn{1}{c}{(1)}         &\multicolumn{1}{c}{(2)}         &\multicolumn{1}{c}{(3)}         &\multicolumn{1}{c}{(4)}         &\multicolumn{1}{c}{(5)}         &\multicolumn{1}{c}{(6)}         &\multicolumn{1}{c}{(7)}         &\multicolumn{1}{c}{(8)}         \\
\midrule
\( \text{Liab/Assets}_{i,1917} \times \mathbf{1}_{t\geq1920} \) &     57.3\sym{***}&     46.5\sym{**} &     41.6\sym{**} &                  &                  &                  &                  &                  \\
                &   (14.3)         &   (20.0)         &   (20.8)         &                  &                  &                  &                  &                  \\
\( \text{Liab/Assets}_{i,1918} \times \mathbf{1}_{t\geq1920} \) &                  &                  &                  &     52.1\sym{**} &                  &                  &                  &                  \\
                &                  &                  &                  &   (25.4)         &                  &                  &                  &                  \\
\( \text{Debt/Assets}_{i,1917} \times \mathbf{1}_{t\geq1920} \) &                  &                  &                  &                  &     56.5\sym{***}&     49.9\sym{**} &     49.4\sym{**} &                  \\
                &                  &                  &                  &                  &   (16.0)         &   (21.5)         &   (22.6)         &                  \\
\( \text{Debt/Assets}_{i,1918} \times \mathbf{1}_{t\geq1920} \) &                  &                  &                  &                  &                  &                  &                  &     63.8\sym{**} \\
                &                  &                  &                  &                  &                  &                  &                  &   (28.8)         \\
\midrule
Observations    &     2355         &     1910         &     1866         &     1866         &     2355         &     1910         &     1866         &     1866         \\
Number of Firms &      351         &      280         &      278         &      278         &      351         &      280         &      278         &      278         \\
\(R^2_{overall}\)&     0.97         &     0.97         &     0.97         &    0.047         &     0.97         &     0.97         &     0.98         &    0.065         \\
\(R^2_{within}\)&    0.033         &    0.087         &    0.049         &                  &    0.039         &    0.094         &    0.059         &                  \\
Year FE         &\checkmark         &\checkmark         &                  &                  &\checkmark         &\checkmark         &                  &                  \\
Firm FE         &\checkmark         &\checkmark         &\checkmark         &\checkmark         &\checkmark         &\checkmark         &\checkmark         &\checkmark         \\
Baseline Controls $\times$ $\mathbf{1}_{t\geq1920}$&                  &\checkmark         &\checkmark         &\checkmark         &                  &\checkmark         &\checkmark         &\checkmark         \\
Industry-Year-FE&                  &                  &\checkmark         &\checkmark         &                  &                  &\checkmark         &\checkmark         \\
IV              &                  &                  &                  &\( L/A_{i,1917} \)         &                  &                  &                  &\(D/A_{i,1917} \)         \\
First-stage F-stat&                  &                  &                  &    191.9         &                  &                  &                  &    230.2         \\
\bottomrule
\end{tabular}
}
}
        \end{center}
        {\footnotesize \textit{Notes}: This table reports results from estimating equation \eqref{eq:diff-in-diff} for log employment (times 100) as the dependent variable. Baseline controls are firm size (log of assets), the share of fixed assets to total assets, free cash flow to assets, and profit margin (EBIT-to-revenue), all as of 1917, as well as Tobin's Q as of the end of 1918. The estimation period is from 1914 through 1923. Standard errors in parentheses are clustered at the firm level. *,**, and *** indicate significance at the 10\%, 5\%, and 1\% level, respectively.
         }
        \end{minipage}
\end{table}%

\Cref{tab:leverage_employment} reports results for employment from standard a difference-in-differences specification of the following form:
\begin{align}    \ln(Employment_{it}) = \alpha_i + \delta_{st} + \beta (Leverage_{i} \, \mathbf{1}_{t\geq 1920} ) + X_{i} \,  \mathbf{1}_{t\geq 1920}   \, \Gamma  + \epsilon_{it}, 
    \label{eq:diff-in-diff}
\end{align}
where  $\mathbf{1}_{t\geq 1920}$ is an indicator variable that takes the value of one after 1919. Columns 1-3 present the estimate measuring $Leverage_i$ with the ratio of liabilities to assets in 1917. The estimated coefficient on $Leverage_{i}$ interacted with $\mathbf{1}_{t\geq 1920}$ is positive and statistically significant at the 5\% level or lower. Further, as above, the effect of leverage on employment during the inflation is robust to the inclusion of firm-level control variables and industry-year fixed effects. Column 4 presents the estimates from a specification that instruments liabilities-to-assets in 1918 with the same variable in 1917. We find that the results are quantitatively similar when instrumenting leverage after the end of the war with leverage from before the end of the war.\footnote{Appendix \Cref{tab:leverage_employment_1918} shows that the results are similar when estimating \eqref{eq:diff-in-diff} by least squares and measuring leverage with data from 1918.} Moreover, the estimates are also similar when measuring leverage with debt-to-assets (see columns 5-8).\footnote{Column 1 Appendix \Cref{tab:robust1_liab} and \Cref{tab:robust1_debt} report the coefficient estimates on the control variables. These tables reveal that firms with higher Tobin's Q at the onset of the inflation have higher employment growth during the inflation.} 

The magnitude of the effect is economically meaningful. Increasing a firm's leverage ratio by one standard deviation (18 percentage points) implies 7.5\% (=41.6$\times$0.18) higher employment between 1920 and 1923, compared to the period from 1916 through 1919, based on the estimate in column 3 of \Cref{tab:leverage_employment}. The magnitude of the employment effects is not easily benchmarked to other existing estimates due to the lack of studies on the links between debt and inflation. However, we note that the response of employment to the debt-inflation shock is broadly in line with findings from the macro-finance literature, which documents that balance sheet and financing shocks can have considerable real effects. \cite{Benmelech2019} find that a 10 percentage point higher debt-to-assets was associated with a 3-5\% larger decline in employment for large U.S.\ firms during the Great Depression. This estimate of the negative effect of leverage during a deflationary environment is quantitatively similar to our estimate of the positive effect of leverage during a large inflation. Our findings are also consistent with large estimated effects on investment and employment from revaluations of foreign currency-denominated debt in emerging markets  \citep{Aguiar2005,Salomao2022,Verner2020} and shocks to bank credit supply \citep{ChodorowReich2013,Huber2018}.

\paragraph{Timing of the debt-inflation and employment effects.}

We next study how the timing of the erosion in real debt burdens lines up with the expansion in employment. The increase and plateau in employment at high relative to low leverage firms documented in \Cref{fig:leverage_employment_dynamics} approximately correspond with the dynamics of $\text{\textit{Debt-inflation}}_{i,t}$ in \Cref{fig:leverage_hist}(a). The dynamics of debt-inflation can thus potentially explain why the strongest employment gains occur from 1919 through 1921.

To examine this connection more formally, \Cref{tab:DV_reg} presents results from estimating the following specification:
\begin{align}
 \ln(Employment_{it}) = \alpha_i + \delta_{st} + \beta \times \text{\textit{Debt-inflation}}_{i,t} + \Gamma X_{i,t} + \epsilon_{it}. \label{eq:dv}
\end{align}

Column 1 in \Cref{tab:DV_reg} shows that $\text{\textit{Debt-inflation}}_{i,t}$ predicts higher employment. Column 2 also includes the lag of the debt-inflation shock, $\text{\textit{Debt-inflation}}_{i,t-1}$, and shows that the lag is an even stronger predictor than the contemporaneous shock. This implies that, in response to a reduction in real debt burdens, firms expand employment the following year. Column 3 shows the results are robust to including our firm-level controls interacted with year fixed effects as well as industry-time fixed effects. 

We further also use the debt-inflation shock directly as an instrument for leverage. Specifically, we estimate 
\begin{align}
 \ln(Employment_{it}) = \alpha_i + \delta_{st} + \beta \times Leverage_{i,t} + \Gamma X_{i,t} + \epsilon_{it}, \label{eq:dv2}
\end{align}
and we instrument $Leverage_{i,t}$ with $(\text{\textit{Debt-inflation}}_{i,t},\text{\textit{Debt-inflation}}_{i,t-1})$. This exercise asks how firm employment responds to changes in leverage using only variation in leverage caused by the debt erosion from inflation. The estimates in columns 4 and 5 in \Cref{tab:DV_reg} confirm that reductions in leverage resulting from inflation lead to a significant increase in firm employment. Note that the magnitudes of the estimates in \Cref{tab:DV_reg} are similar to the estimates in \Cref{tab:leverage_employment} but with the opposite sign, as a larger debt-inflation implies a larger reduction in leverage. This is consistent with the effects we document above being driven by the debt-inflation channel. 

\begin{table}[!ht]
  \centering
   \caption{Time Varying Debt-Inflation Shock and Firm Employment.}\label{tab:DV_reg}
        \begin{minipage}{1.0\textwidth}
        \begin{center}
        \scalebox{0.8}{
        { \def\sym#1{\ifmmode^{#1}\else\(^{#1}\)\fi} \begin{tabular}{l*{5}{c}}   \toprule  Dependent Variable & \multicolumn{5}{c}{ln(Employment)} \\    \cmidrule(lr){2-6}       
                &\multicolumn{3}{c}{OLS}                                 &\multicolumn{2}{c}{2SLS}             \\\cmidrule(lr){2-4}\cmidrule(lr){5-6}
                &\multicolumn{1}{c}{(1)}         &\multicolumn{1}{c}{(2)}         &\multicolumn{1}{c}{(3)}         &\multicolumn{1}{c}{(4)}         &\multicolumn{1}{c}{(5)}         \\
\midrule
 \( \text{Debt-inflation}_{i,t} \) &     51.4\sym{***}&    -16.9         &    -32.9         &                  &                  \\
                &   (17.5)         &   (15.7)         &   (25.2)         &                  &                  \\
 \( \text{Debt-inflation}_{i,t-1} \) &                  &     71.2\sym{***}&     75.8\sym{**} &                  &                  \\
                &                  &   (20.5)         &   (29.7)         &                  &                  \\
 \( \text{Liabilities/Assets}_{i,t} \) &                  &                  &                  &    -79.2\sym{***}&    -61.0\sym{**} \\
                &                  &                  &                  &   (25.0)         &   (29.5)         \\
\midrule
Observations    &     1696         &     1696         &     1365         &     1597         &     1297         \\
Number of Firms &      328         &      328         &      262         &      324         &      259         \\
\(R^2_{overall}\)&     0.98         &     0.98         &     0.98         &    -0.25         &   -0.055         \\
\(R^2_{within}\)&    0.020         &    0.031         &    0.071         &                  &                  \\
Year FE         &\checkmark         &\checkmark         &                  &\checkmark         &                  \\
Firm FE         &\checkmark         &\checkmark         &\checkmark         &\checkmark         &\checkmark         \\
Baseline Controls $\times$ Year FE&                  &                  &\checkmark         &                  &\checkmark         \\
Industry-Year-FE&                  &                  &\checkmark         &                  &\checkmark         \\
P-value for joint sign.&                  &     .002         &      .04         &                  &                  \\
First-stage F-stat&                  &                  &                  &     58.3         &     36.8         \\
\bottomrule
\end{tabular}
}
}
        \end{center}
        {\footnotesize \textit{Notes}: This table reports results from estimating equation \eqref{eq:dv}. The estimation period is 1917 to 1923, as $\text{\textit{Debt-inflation}}_{i,t}$ is defined with reference to leverage in $1917$ as in equation \eqref{DV}. $X_{i}$ is a set of firm-level controls, which are interacted with year dummies. These control variables are: firm size (log of assets), the share of fixed assets to total assets, free cash flow to assets, and profit margin (EBIT-to-revenue), all as of 1917, as well as Tobin's Q as of the end of 1918. ``P-value for joint sign.'' refers  to the p-value from an F-test of joint significance of  $\text{\textit{Debt-inflation}}_{i,t}$ and $\text{\textit{Debt-inflation}}_{i,t-1}$.  Columns 5 and 6 present two-stage least squares estimates using $\text{\textit{Debt-inflation}}_{i,t}$ and $\text{\textit{Debt-inflation}}_{i,t-1}$ as instruments for $Liabilities/Assets_{i,t}$.  *,**, and *** indicate significance at the 10\%, 5\%, and 1\% level, respectively.
         }
        \end{minipage}
\end{table}%

\subsubsection{Additional Robustness}

Next, we present a series of additional tests to support the argument that high leverage firms see a relative expansion of their employment through a debt-inflation channel rather than through alternative, confounding channels.

\paragraph{State intervention and political connections.} Germany's inflation occurs in the aftermath of WWI. An important concern is that firm leverage in 1917 and employment growth after the war could be correlated with  state intervention in the economy during the war. For example, large, politically connected firms may have gained preferential access to debt financing during the war, which may also have governed their ability to expand during the postwar years. We present four pieces of evidence to alleviate this type of concern.

First, we construct proxy variables for firm political connections using information on each firm's executives and supervisory board members. \textit{Saling's} provides both the names and professions of individual executives and board members for each firm.\footnote{Appendix \ref{app:data-board}  describes the dataset on executives and board members and illustrates how firms interconnect through boards. For more background on the executives and board members in interwar Germany see also \citet{Ferguson2008} and \citet{Huber2021}.} This additional information allows us to proxy the distance of a firm through the board member network to the government, the Reichsbank, and Hugo Stinnes (a prominent and politically connected industrialist). We find that our baseline results are generally robust to the inclusion of these controls (see \Cref{tab:robust1_liab} and \Cref{tab:robust1_debt}), suggesting that the debt-inflation channel is not confounded by privileges stemming from political connections.

Second, firms in industries that were particularly relevant for the war effort would be most likely to be impacted by state intervention. However, we find that the effect of leverage on employment is, if anything, stronger when we exclude war-relevant industries, as defined by \cite{Kocka1984}; see columns 1-3 in \Cref{tab:war_exposure}. Third, we show that the results are robust to excluding the largest firms in the sample and to flexibly controlling for size using size deciles (see columns 4-7 in \Cref{tab:war_exposure}). 

Finally, we show that the results are robust to instrumenting leverage before the inflation in 1917 with firm leverage before WWI (see \Cref{tab:preWWI_IV}). This approach does not use variation in leverage that could have been determined by state intervention during the war. 

\paragraph{Export activity.} The inflation and depreciation of the mark benefited exporting firms \citep{Feldman1997}. Export activity could be correlated with leverage, leading to an upward-biased estimate of the debt-inflation channel. Industry-year fixed effects control for industry-level differences in exposure to the exchange rate channel, but there may be differences in exposure to export status within industries. We take two approaches to address this threat to identification.

First, we hand-collect information on firm export status based on business descriptions in \textit{Saling's}.\footnote{See \Cref{sec:exporter} for details.} Including firms' reported export status in our main specification \eqref{eq:diff-in-diff}, we indeed see that exporters have stronger employment growth during the inflation. Nevertheless, because export status is only weakly correlated with leverage, the estimated effect of leverage is similar when including this control (see \Cref{tab:robust1_liab}, \Cref{tab:robust1_debt}, and \Cref{fig:dd_emp_robust_all}). Second, we find broadly similar results when excluding firms in industries that are most active in exporting (see \Cref{tab:exporter_status}). 

\paragraph{Credit supply shifts.} Our hypothesis is that high-leverage firms benefit from the inflation through the erosion of their debt burdens. A related channel that may differentially affect high and low leverage firms is a shift in credit supply caused by inflation. Financing was available in much of the first phase of the inflation when additional inflation was not expected. 
However, credit supply contracted in 1922 with increased expectations of depreciation.\footnote{See Appendix \ref{app:historical_background} for a detailed discussion of the impact of the inflation on banks and credit conditions.} 

The bias introduced by the credit supply channel in the estimation of \eqref{eq:dynamic} could be either positive or negative, as it is not clear whether high or low leverage firms are most exposed to a credit supply contraction. Controlling for size, industry, and profit margin partially addresses this concern, as these are common proxies of exposure to credit supply \citep[e.g.,][]{gertler1994monetary}. Furthermore, our evidence below that the effect of leverage is strongest for firms with longer maturity debt also points to the importance of the firm balance sheet channel.

As additional robustness, we add controls that more directly proxy for differential exposure to shifts in bank credit supply. First, we collect information on firm-bank relationships in 1919. We follow \cite{Doerr2018} and measure firm-bank relationships based on information on the banks that paid out a firm's dividends (\textit{Zahlstellen}). With this information, we control for bank-time fixed effects, distinguishing between seven major banks, other banks, and firms without banking relationships. This essentially compares two firms with different leverage but connected to the same bank, thereby holding fixed bank-specific changes in credit supply.\footnote{We note that this only controls for credit-supply shocks that are bank specific. It would not capture credit-supply changes that affect firms differently based on their unobservables.} Second, we control for the distance to Berlin, as firms located closer to Berlin may have had better access to Reichsbank credit during the inflation. Third, we control for distance to commercial banks through the board member network. The estimated effect of leverage on employment is quantitatively similar with these controls (see \Cref{tab:robust1_liab}, \Cref{tab:robust1_debt}, and \Cref{fig:dd_emp_robust_all}).

\paragraph{Cyclicality.} A related concern is that firms with high leverage at the onset of the inflation could be more exposed to business cycle risk \citep[see, e.g.,][]{GiroudMueller2017}. To alleviate this concern, we study whether high-leverage firms also saw higher growth during Germany's post-inflation expansion over 1926-1928. This expansion saw a rapid fall in unemployment from 30\% in 1926 to less than 5\% in late 1927. \Cref{fig:cyclicality} shows that firms with higher leverage did not experience faster growth during this cycle. This finding holds when sorting firms by their leverage either in 1917---before the inflation---or in 1925---after the hyperinflation and subsequent stabilization but before the post-inflation expansion. The evidence that firms with high leverage did not have differential growth in the post-inflation boom also implies that the expansion of employment during the inflation was persistent.

\paragraph{Additional sample restrictions.} Another concern is that the baseline results may be affected by the unbalanced nature of the panel data on employment. \Cref{fig:dd_emp_robust_samples} addresses this concern by estimating \eqref{eq:dynamic} on full and balanced panels. The patterns are similar across these different during samples during the 1914--1923 period.

\subsection{Additional Evidence Supporting the Debt-Inflation Channel}

We next provide evidence on the mechanisms through which the inflation affected levered firms. First, we ask, do more levered firms pay less in interest expenses and thereby have more liquidity to spend on production expenses? Second, are the effects stronger for firms with a higher proportion of long-term debt? Third, do firms with higher leverage see a larger increase in their book equity values and higher valuations in the stock market?

\paragraph{Interest expenses.} First, we study the dynamics of interest expenses at the firm level. Recall from \Cref{fig:cost_shares} that the average share of interest expenses to total expenses falls during the inflation. We now ask whether this fall is more pronounced for high-leverage firms. 

Similar to our analysis of firm-level employment, we estimate equation \eqref{eq:diff-in-diff} using the ratio of interest expenses to total expenses as the dependent variable.\footnote{Appendix \Cref{fig:dd_expenses} shows the results from estimating the dynamic difference-in-differences specification \eqref{eq:dynamic} with interest expenses as the dependent variable. The figure shows that the effect is persistent and builds throughout the inflation.} \Cref{tab:leverage_interest} shows  results. Firms with a higher share of nominal liabilities relative to total assets at the onset of the inflation dedicate fewer resources to servicing their debt claims, as the real burden of debt service declined. A similar pattern arises when using financial debt-to-assets ratio as the measure of leverage. The reduction in interest expenses translates into a similar increase in production expenses as a share of total expenses (see \Cref{tab:s_prod_cost}). Altogether, these findings are in line with highly levered firms benefiting from the inflation by decreasing the amount of resources that need to be spent on interest payments, allowing these firms to hire more employees and spend more funds on salaries, raw materials, and other production inputs. As noted above, our data do not allow us to quantify the relative importance of these different mechanisms. 

Note that theory suggests that firms that have higher leverage before the inflation and thus experience a larger debt reduction should not necessarily experience a larger decrease in their interest expenses. Rather, if credit is available, these firms could choose to take on new debt after their original debt has been inflated away. Indeed, we find that the average liabilities-to-assets ratio, while falling considerably throughout the inflation, remains well above zero, see \Cref{fig:lev_hist} in the Appendix. This evidence suggests that the increase in employment is partially related to the increase in available cash flows from reduced interest expenses, as well as to the ability to raise additional debt.

\begin{table}[!ht]
  \centering
   \caption{Firm Leverage and Interest Expenses.}\label{tab:leverage_interest}
        \begin{minipage}{1.0\textwidth}
        \begin{center}
        \scalebox{0.8}{{   \def\sym#1{\ifmmode^{#1}\else\(^{#1}\)\fi}  \begin{tabular}{l*{4}{c}}   \toprule  Dependent Variable & \multicolumn{4}{c}{Interest Expenses/Tot. Expenses} \\    \cmidrule(lr){2-5}       
                &\multicolumn{1}{c}{(1)}         &\multicolumn{1}{c}{(2)}         &\multicolumn{1}{c}{(3)}         &\multicolumn{1}{c}{(4)}         \\
\midrule
Liab/Assets$_{i,1917}$ $\times$ $\mathbf{1}_{t\geq1920}$&    -0.10\sym{***}&    -0.14\sym{***}&                  &                  \\
                &  (0.037)         &  (0.050)         &                  &                  \\
Debt/Assets$_{i,1917}$ $\times$$\mathbf{1}_{t\geq1920}$&                  &                  &    -0.16\sym{***}&    -0.13\sym{***}\\
                &                  &                  &  (0.036)         &  (0.047)         \\
\midrule
Observations    &     3256         &     2321         &     3256         &     2321         \\
Number of Firms &      520         &      335         &      520         &      335         \\
\(R^2_{overall}\)&     0.65         &     0.68         &     0.65         &     0.68         \\
\(R^2_{within}\)&    0.010         &    0.041         &    0.025         &    0.042         \\
Year FE         &\checkmark         &                  &\checkmark         &                  \\
Firm FE         &\checkmark         &\checkmark         &\checkmark         &\checkmark         \\
Baseline Controls $\times$ $\mathbf{1}_{t\geq1920}$&                  &\checkmark         &                  &\checkmark         \\
Industry-Year-FE&                  &\checkmark         &                  &\checkmark         \\
\bottomrule
\end{tabular}
}
}

              \end{center}
        {\footnotesize \textit{Notes}: 
            This table reports results from estimating \eqref{eq:diff-in-diff}, with firm $i$'s share of interest expenses to total expenses as the dependent variable. $Leverage_{i}$ is firm $i$'s ratio of either total liabilities to assets (columns 1-2) or finance debt to assets in 1917 (columns 3-4).  Baseline controls are as defined in \Cref{tab:leverage_employment}. The estimation period is from 1916 through 1923. Standard errors in parentheses are clustered at the firm level. *,**, and *** indicate significance at the 10\%, 5\%, and 1\% level, respectively.
         }
        \end{minipage}
\end{table}%

\paragraph{Impact on firms with longer maturity debt.}  The debt-inflation channel should be especially strong for firms with a higher proportion of long-term debt \citep[see, for example,][]{Gomes2016}. As inflation rises and increases \textit{expected} inflation, firms that have secured long-term financing are less exposed to a repricing of new loans and a tightening of credit conditions. As a finer test, we therefore examine whether the benefits of leverage are strongest within firms with more long-term debt financing. 

Nonfinancial German firms relied extensively on fixed-rate long-term bond financing. We collect detailed information on the contractual loan terms of all bonds issued by firms in our sample as of 1918 and 1919. These loan terms are detailed in \Cref{tab:bond_terms} in the appendix. All bonds in the sample pay a fixed coupon, and the coupon rates are almost all between 400 and 500 basis points of par. None of the bonds we analyze have ``gold clauses.''\footnote{The absence of gold clauses for German corporate bonds is consistent with \cite{FlandreauSussman2005} and \cite{BordoMeissnerRedish2003}, who find that under the gold standard, German government bonds also did not carry gold or exchange clauses.} After origination, the loan terms typically specified an interest-only period lasting 5 years, on average, during which no amortization took place. Once repayment started, bonds would be typically amortized evenly until a specified final repayment date. Notably, bonds were of very long maturity. For instance, for  bonds outstanding in 1918 and 1919, the median origination year was 1906 and the median final maturity year was 1940. \Cref{fig:bonds_maturity} in the Appendix also shows that most bonds were originated before WWI, and a final maturity date after 1950 was not uncommon.

\begin{figure}[!htpb]
    \caption{Firm Leverage, Long-term Debt, Interest Expenses, and Employment.}
    \label{fig:LTdebt_interaction}
        \begin{center}
          \includegraphics[width=0.8\textwidth]{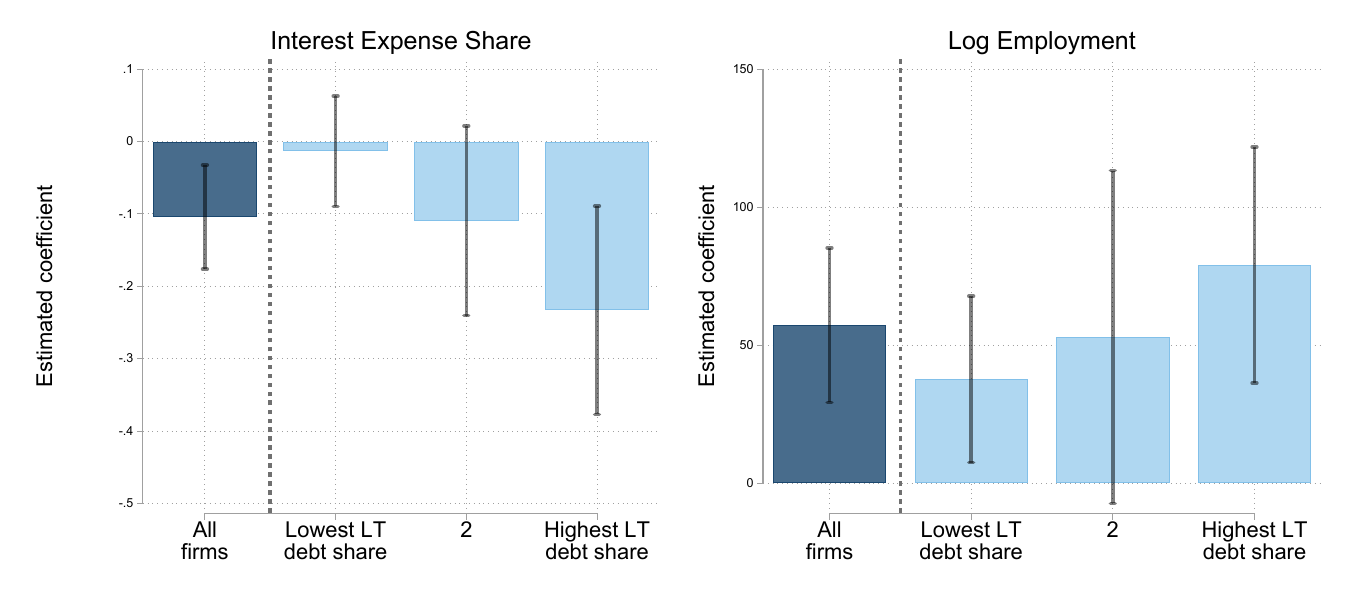}  
\end{center}
\vspace{-.7cm}
\footnotesize{\textit{Notes}: This figure reports results from estimating equation \eqref{eq:diff-in-diff} separately for firms in the lowest, middle, and highest tercile of the distribution of long-term debt to total debt in 1917. The dependent variable is either interest expenses as a share of total expenses or log employment (times 100).  Standard errors in parentheses are clustered at the firm level. *,**, and *** indicate significance at the 10\%, 5\%, and 1\% level, respectively.   }
\end{figure}

To test whether the impact of the debt-inflation channel is stronger for firms with relatively more long-term debt financing, we estimate equation \eqref{eq:diff-in-diff} separately for firms in the lowest, middle, and highest tercile of the distribution of long-term debt to total debt in 1917. The results are presented in \Cref{fig:LTdebt_interaction}.\footnote{\Cref{tab:leverage_financials-LTDebt} presents the table version from a triple-difference-in-differences specification where we interact leverage before the inflation with either the share of long-term debt in total debt or with indicator variables for the tercile of long-term debt.} Firms with a high share of long-term debt experienced the largest reduction in the share of interest payments to total expenses. The effect is close to zero for firms in the lowest tercile of the long-term debt and builds for firms in higher terciles. For example, a one-standard-deviation increase in the liabilities-to-assets ratio implies that firms in the second and third terciles of the long-term debt share distribution experience a reduction of interest expense shares of about 1.1 percentage points and 3.9 percentage points, respectively.

\Cref{fig:LTdebt_interaction} also compares the effect of leverage on employment across the long-term debt distribution. Although the results are less precise, we indeed find that not only do more highly levered firms hire more employees during the inflation, but the effect is strongest for firms with more long-term debt. A one-standard-deviation increase in leverage is associated with 6.6\% higher employment for firms in the lowest tercile of the long-term debt share and 13.1\% in the highest tercile. 

Further, observe that while the effect of leverage on interest expenses is negligible for firms in the lowest tercile of the long-term debt distribution, these firms do see a significant expansion in employment. This suggests that the erosion of nominal debt burdens likely operates---as discussed above---not only through reduced interest payments and relaxed cash flow constraints, but also through other channels such as a net worth channel that allows firms to take on additional debt.

\paragraph{Book equity.} The reduction in real debt burdens should increase the book equity values of levered firms.  On average, the real value of nonfinancial firms' book equity increased by 119\% during the postwar inflation from 1919 to 1924. \Cref{tab:lev_1919_1924} asks whether this increase was larger for highly levered firms. As discussed in section \ref{sec:data}, balance sheet outcomes during the hyperinflation (especially in 1923) are subject to systematic measurement error. For this analysis, we therefore focus on the change in real book equity from before the start of the postwar inflation in 1918 to the aftermath of the inflation in 1924, when the new and more accurate Goldmark balance sheets are available. We present estimates of the following long-difference firm-level regression for book equity to assets and real book equity as the outcome variables:
\begin{equation}
    \Delta_{18-24} \ln(E_i) = \alpha + \beta Leverage_{i,1917} + X_i \Gamma + \epsilon_i, \label{longdiff}
\end{equation}
where $ \Delta_{18-24} \ln(E_i)$ is the change in book equity deflated by wholesale prices from 1918 to 1924. 

Column 1 in \Cref{tab:lev_1919_1924} shows that firms with higher liabilities-to-assets before the inflation see a larger increase in the real value of book equity from the before the start of the inflation in 1918 to the immediate aftermath in 1924. Column 2 shows that the estimate is larger when including firm controls and industry fixed effects. In column 3, we further control for growth in real equity from 1916 to 1918 to account for any differential pre-trends in the dependent variable. In terms of magnitudes, the estimate in column 3 implies that a one-standard-deviation increase in liabilities-to-assets before the inflation is associated with a 33\% larger increase in the real book value of equity from 1918 to 1924. Columns 4-6 show that the results are robust to measuring leverage with debt-to-assets in 1917. As a placebo exercise, columns 7-8 show that firms with higher leverage in 1917 did not experience faster equity growth before the inflation from 1916 to 1918. Overall, this evidence implies that more levered firms---firms with a larger exposure to the erosion of real debt burdens---see a larger increase in real book equity values. Book equity measures thus suggest that the inflation led to redistribution from debt to equity holders of levered firms.

\begin{table}[!ht]
  \centering
  \caption{Firm Leverage and Growth in Real Book Equity during the Inflation.}\label{tab:lev_1919_1924}
        \begin{minipage}{1.0\textwidth}
        \begin{center}
        \scalebox{.8}{
        {   \def\sym#1{\ifmmode^{#1}\else\(^{#1}\)\fi}   \begin{tabular}{l*{8}{c}}   \toprule Dependent Variable & \multicolumn{6}{c}{\( \Delta_{1918-24} \ln(E) \) } & \multicolumn{2}{c}{\( \Delta_{1916-18} \ln(E) \) } \\ \cmidrule(lr){2-7} \cmidrule(lr){8-9}
                &\multicolumn{1}{c}{(1)}         &\multicolumn{1}{c}{(2)}         &\multicolumn{1}{c}{(3)}         &\multicolumn{1}{c}{(4)}         &\multicolumn{1}{c}{(5)}         &\multicolumn{1}{c}{(6)}         &\multicolumn{1}{c}{(7)}         &\multicolumn{1}{c}{(8)}         \\
\midrule
\( \text{Liabilities/Assets}_{i,1917} \)&     90.1\sym{***}&    181.1\sym{***}&    180.3\sym{***}&                  &                  &                  &    -2.35         &                  \\
                &   (24.0)         &   (64.0)         &   (63.6)         &                  &                  &                  &   (6.73)         &                  \\
\addlinespace
\( \text{Debt/Assets}_{i,1917} \)&                  &                  &                  &     38.9\sym{*}  &    113.8\sym{***}&    112.7\sym{***}&                  &    -5.79         \\
                &                  &                  &                  &   (21.5)         &   (43.7)         &   (43.2)         &                  &   (7.37)         \\
\midrule
Observations    &      595         &      475         &      475         &      595         &      475         &      475         &      531         &      531         \\
\(R^{2}\)       &    0.007         &    0.103         &    0.108         &    0.001         &    0.090         &    0.094         &    0.132         &    0.133         \\
\hline \hline Industry FE&                  &\checkmark         &\checkmark         &                  &\checkmark         &\checkmark         &\checkmark         &\checkmark         \\
Firm Controls   &                  &\checkmark         &\checkmark         &                  &\checkmark         &\checkmark         &\checkmark         &\checkmark         \\
Lagged dep. var., 1916-18&                  &                  &\checkmark         &                  &                  &\checkmark         &                  &                  \\
\bottomrule
\end{tabular}
}
}
        \end{center}
        {\footnotesize \textit{Notes}: This table presents estimates of equation \eqref{longdiff}. The dependent variable in columns 1-6 is the change in log deflated book equity (times 100) from 1918 to 1924, using the wholesale price index as the deflator. The dependent variable in columns 7-8 is the change in log deflated book equity (times 100) from 1916 to 1918. Firm controls are size (log of assets), the ratio of fixed assets to total assets, and free cash flow to assets, all as of 1917, as well as Tobin's Q as of the end of 1918. ``Lagged dep. var., 1916--18'' refers to a specification that controls for the change in log real book equity from 1916 to 1918. Robust standard errors in parentheses. *,**, and *** indicate significance at the 10\%, 5\%, and 1\% level, respectively.}
        \end{minipage}
\end{table}%

\paragraph{Market equity returns.} Do more levered firms see an increase in their stock market valuations? The decline in interest expenses and increase in real book equity values suggest that the inflation should increase the market equity returns of highly levered firms. Studying market equity values also has the benefit that they are not subject to accounting errors but instead reflect the equity values of firms as perceived by investors in real-time. At the same time, equity investors subject to ``money illusion'' may also misperceive the impact of inflation on firm value \citep{ModiglianiCohn1979,Ritter2002,CampbellVuolteenaho2004,CohenPolkVuolteenaho2005}, and the evidence on whether unexpected inflation boosts market equity values of more levered firms is mixed \citep{FrenchRubackSchwert1983,Modigliani1984,BhamraWeber2021}. 

\Cref{tab:lev_return} reports market equity returns across firms with high and low leverage during the inflation. We follow the standard asset pricing approach and sort firms into five portfolios by quintiles of lagged leverage each year of the postwar inflation, from 1919 to 1923.\footnote{The estimates are smaller and usually statistically insignificant when fixing leverage before the inflation (in 1917 or 1918). For the result that high-leverage firms have higher stock returns, resorting firms each year using lagged leverage is important, especially as most of the excess return for high-leverage firms occurs in 1922 and 1923. Note that we avoid using leverage based on balance sheets from 1923, which are subject to measurement error, as we discussed in \cref{sec:data}. Results are also similar if we use leverage from balance sheets only through 1921.}  We then compute the equal-weighted average log total return for each portfolio, as well as the difference between the high-minus-low portfolio. Returns are deflated by wholesale prices. Panel A reports results using the liabilities-to-assets ratio as a proxy for leverage, and Panel B reports results using the ratio of financial debt to assets.

\begin{table}[!ht]
  \centering
  \caption{Stock Returns across Portfolios Sorted by Leverage.}\label{tab:lev_return}
        \begin{minipage}{1.0\textwidth}
        \begin{center}
        \scalebox{.9}{
        \begin{tabular}{ccccc}
        \multicolumn{5}{c}{\textit{Panel A: Sorting by Liabilities-to-Assets.}} \\ \toprule
        \multicolumn{3}{c}{\( \text{Liabilities/Assets}_{i,t-1}\)} & \multicolumn{2}{c}{ \(\text{Return}_{t}\)}  \\  \cmidrule(l){1-3} \cmidrule(l){4-5}
        Quintile & Mean & S.E. & Mean & S.E. \\
        \midrule
        1 & 0.29 & (0.01) & -35.82 & (3.94)  \\
2 & 0.48 & (0.00) & -34.89 & (3.91)  \\
3 & 0.58 & (0.01) & -31.11 & (3.90)  \\
4 & 0.66 & (0.01) & -31.78 & (3.67)  \\
5 & 0.79 & (0.01) & -24.77 & (3.89) \\ \midrule
High minus low & 0.51 & (0.01) & 11.05 & (4.63)
 \\    \bottomrule   
        \multicolumn{5}{c}{} \\ 
        \multicolumn{5}{c}{\textit{Panel B: Sorting by Debt-to-Assets.}} \\ \toprule
        \multicolumn{3}{c}{\( \text{Debt/Assets}_{i,t-1}\)} & \multicolumn{2}{c}{ \( \text{Return}_{t}\)}  \\  \cmidrule(l){1-3} \cmidrule(l){4-5}
        Quintile & Mean & S.E. & Mean & S.E. \\
        \midrule
        1 & 0.35 & (0.01) & -34.96 & (4.00)  \\
2 & 0.48 & (0.01) & -35.58 & (3.81)  \\
3 & 0.57 & (0.01) & -32.64 & (3.74)  \\
4 & 0.65 & (0.01) & -27.09 & (3.88)  \\
5 & 0.77 & (0.01) & -27.91 & (3.86) \\ \midrule
High minus low & 0.44 & (0.01) & 7.05 & (4.62)
 \\ 
        \bottomrule
        \end{tabular}
        }
        \end{center}
        {\footnotesize \textit{Notes}: This table presents total returns on five portfolios of nonfinancial firms sorted by lagged leverage. Panel A uses firm liabilities-to-assets in year $t-1$ as the measure of leverage, while panel B uses debt-to-assets. Returns are defined as log total returns (times 100) and are deflated by the wholesale price index. The analysis uses annual data on returns from 1919 to 1923.
         }
        \end{minipage}
\end{table}%

\Cref{tab:lev_return} reveals two notable patterns. First, returns are, on average, negative during the inflation, despite the large increase in the book value of equity during this time. Prior research argues that the poor performance of the stock market during Germany's inflation is explained by lower expected future cash flows, including from high expected taxes on firms, high uncertainty, and money illusion on the part of investors. \cite{BrescianiTurroniCostantino1937TEoI} argues that investors mistook large nominal capital gains for large real capital gains. Consistent with this, \cite{braggion2023inflation} present evidence of money illusion among stock market investors during Germany's hyperinflation. 

Second, returns are increasing across the leverage portfolios. The difference between the high minus low portfolio is large (7-11\% annual return).\footnote{The magnitude of this effect is broadly in line with the effects documented by \citet{Kroszner1999} who studies the abrogation of gold clauses in the US.} The high-minus low leverage portfolio thus hedges about one-third of the overall decline in stock prices during the inflation. \Cref{fig:hml_portfolio} plots the cumulative return on the high-minus-low leverage portfolio over time. The portfolio has positive returns in the second half of 1919 and again from the start of 1922 to the end of the inflation in November 1923. Returns are especially high during 1922--1923, when annual inflation surpassed 1,000\%. The relatively poor performance in 1921 is puzzling given the high rate of inflation in that year (141\%). One possible explanation is that investors were slow to realize the benefits of high inflation for highly levered firms, as hypothesized by \citet{ModiglianiCohn1979}. Another explanation could be that investors anticipated some recovery in the mark that would partly reverse the erosion of long-term debt. This could help explain why the returns for high-leverage firms are especially elevated after inflation expectations became unanchored in 1922.

\Cref{tab:fama_macbeth} explores the robustness of the relation between leverage and subsequent returns. We estimate cross-sectional regressions of subsequent monthly returns on lagged leverage, lagged size (market cap), lagged market-to-book, and market beta. We then average the cross-sectional regressions, following \cite{FamaMacbeth1973}. This procedure allows us to control for other factors that might predict returns. We find that more leveraged firms, measured using either liabilities-to-assets or debt-to-assets, have higher returns during the inflation, even after accounting for standard risk factors.

The relative outperformance of high-leverage firms in the stock market could be driven by higher prices or payouts in the form of dividends, depending on whether firms return the unexpected windfall to shareholders or keep the money inside the firm. Interestingly, we find that dividend yields decline during the inflation and essentially fall to zero by 1923 (see \Cref{fig:dividend_yield}). Firms keep the money inside the firm, presumably because firms have better access to stores of value than investors. This result is also consistent with the increase in book equity for high-leverage firms. That firms reduce, rather than increase, payouts is also consistent with the substantial real effects of the reduction in real debt burdens.

Nonfinancial firms with higher leverage fare relatively better during the inflation. By definition, the gain must come at someone else's expense. Thus, we ask: who loses at the same time?  Nonfinancial firms' equity should have greater positive net nominal exposure compared to banks. Nonfinancial firm equity is a levered claim on real assets, while bank equity is a levered claim on nominal assets. Moreover, banks are exposed to duration mismatch. \Cref{fig:bbz_validation} in the Appendix compares the returns on nonfinancial firms and banks during the inflation. While inflation was associated with negative real returns on both indexes, we find that nonfinancial firms performed better than banks. Altogether, our findings suggest that the inflation redistributed from those that held debt of nonfinancial firms (such as banks) to equity holders, with equity investors in levered firms benefiting relatively more.


\section{Conclusion}
\label{sec:conclusion}

This paper examines how inflation transmits to the real economy through a \textit{debt-inflation} channel via firm balance sheets. We study Weimar Germany's big inflation from 1919 to 1923 using newly digitized macro and firm-level data. Inflation led to a substantial decline in nonfinancial firms' leverage and interest expenses, resulting in a downward sloping and convex relation between inflation and firm bankruptcies. Exploiting variation across firms in initial leverage, we find that high-leverage firms saw larger increases in employment, book equity, and market equity valuations during the inflation. These firms also saw larger declines in the share of interest expenses. At the same time, we find that prices and wages were adjusted at shorter and shorter frequencies with rising inflation, consistent with menu cost models. These results are consistent with the view that the inflation affected real activity through a debt-inflation channel that can be operative even in the absence of nominal rigidities in prices and wages.

Our analysis invites questions of external validity. Previous researchers have studied hyperinflations as extreme events that can provide insights into the workings of inflation more broadly \citep[e.g.,][]{Sargent1982}. The debt-inflation channel may be present during times of more moderate inflation if debt contracts are nominal, long-term, and denominated in domestic currency. However, with other debt contract structures, such as floating or foreign currency debt, inflation would be neutral or even negative for more levered firms. Moreover, the debt-inflation channel, while still present, may be dominated by other forces during times of lower inflation \citep[see, e.g.,][]{Ottonello2020}. For example, if monetary policy responds aggressively to rising inflation by raising interest rates and tightening financial conditions, this can offset the expansionary effects from the reduction in real debt burdens.

Abstracting from general equilibrium effects, our cross-sectional estimates imply that wiping out all debt from nonfinancial firms can account for a 18\% increase in employment.\footnote{Liabilities in 1917 were on average 43\% of total assets, see \Cref{tab:summary_statistics}. Wiping out the real value of these debts and applying the coefficients in column (3) of Table, the increase in total employment would total 18\% (0.43 $\times$ 41.6).} The magnitude of this effect is similar to the overall employment increase during the inflation. Our findings thus highlight that the debt-inflation channel may have played an important role in the overall transmission of the inflation to the real economy. However, it is important to emphasize that our study does not quantify the aggregate effect of the inflation. On the one hand, we focus on firm debt due to data availability, but the debt-inflation channel through household debt could also have been quantitatively important \citep{Doepke2006,Diamond2022}. On the other hand, firms with high leverage could have benefited at the expense of firms with low leverage, without much positive net impact. Moreover, we have not estimated the effect of inflation through bank balance sheets and credit supply. We also do not capture the reduction in spending by savers who lost out from the inflation or from the inefficiencies that stem from the necessity of bartering at the height of the hyperinflation. These mechanisms are likely to offset at least some of the expansionary effects of the debt-inflation channel.

    
    {\singlespacing
    \footnotesize
    \bibliographystyle{chicago}
    \bibliography{literature}
    }
    \doublespacing
    \begin{appendices}
   	\clearpage
   	
\appendix

\setcounter{footnote}{0}
\pagenumbering{arabic} \setcounter{page}{1}

$ $

\begin{center}
\LARGE { \singlespacing \bf The Debt-Inflation Channel of the \\ German Hyperinflation \\
Online Appendix}
 \\

\author{\vspace{1.0cm} \large Markus Brunnermeier, Sergio Correia, Stephan Luck, \\ \vspace{-.2cm} Emil Verner, and Tom Zimmermann\textsuperscript{*} \\ \vspace{0.7cm}}

\normalsize
\date{ \today } %
\end{center}
\let\oldthefootnote\thefootnote
\renewcommand{\thefootnote}{\fnsymbol{footnote}}
\footnotetext[1]{Brunnermeier: Princeton University, \href{mailto:markus@princeton.edu}{markus@princeton.edu}; Correia: Board of Governors of the Federal Reserve System, \href{mailto:sergio.a.correia@frb.gov}{sergio.a.correia@frb.gov}; Luck: Federal Reserve Bank of New York, \href{mailto:stephan.luck@ny.frb.org}{stephan.luck@ny.frb.org};
Verner: MIT Sloan, \href{mailto:everner@mit.edu}{everner@mit.edu}; Zimmermann: University of Cologne \href{mailto:tom.zimmermann@uni-koeln.de}{tom.zimmermann@uni-koeln.de}.}

\let\thefootnote\oldthefootnote

\singlespacing

\begin{itemize}
\item Appendix \ref{app:historical_background}: Historical Background
\item Appendix \ref{sec:model}: Model of the Debt-Inflation Channel
\item Appendix \ref{app:additional_results}: Supplementary Figures and  Tables
\item Appendix \ref{app:data}: Data Appendix
\end{itemize}

   	\clearpage
\section{Historical Background}
\label{app:historical_background}

In this section, we provide additional historical context for the German Hyperinflation. We first provide a chronology of the key events in the inflation. Then, we discuss existing economic history research on the economic impact of the inflation.

\subsection{Chronology of Key Events}

\paragraph{WWI and first phase of the postwar inflation.} The origins of the German inflation lie in WWI \citep{Feldman1997}.  Before the start of WWI, the mark was on the gold standard, and the exchange rate relative to the U.S. dollar stood at 4.2 marks per dollar. The Banking and Currency Laws of August 4, 1914 led to the abandonment of the gold standard, and the Reichsbank started to discount \emph{Reichsschatswechsel}, de facto moving to a fiat currency. See \Cref{fig:reichsbank_goldmarks} and \Cref{fig:reichsbank_papiermarks} for the evolution of the Reichbank's balance sheets throughout the inflation. Exchange controls and price controls were maintained during WWI, which suppressed inflation but led to distortions and black marketeering \citep{Feldman1997}.

Relative to the Allied Powers, Germany relied heavily on domestic loan issuance rather than on new taxes and foreign bond issuance to finance the war \citep{Graham1931,Feldman1997}. However, Germany's public finances were not in a significantly worse condition than France's in the immediate aftermath of the war. Prices also rose in the U.K., U.S., and France during the war.\footnote{By the end of WWI, the mark had depreciated to 8 marks per U.S. dollar.} The most important difference would be the reparations imposed at the Treaty of Versailles \citep{Graham1931,Sargent1982}.

The WWI Armistice was signed on November 11, 1918, leading to the end of WWI fighting. November 1918 marked the start of the German Revolution. In January 1919, a Constitutional congress was convened, and the new Weimar Constitution was adopted on August 11, 1919. Meanwhile, the Treaty of Versailles was signed on June 28, 1919, which included the War Guilt Clause. As a result, Germany owed staggering but initially uncertain reparations, negatively impacting Germany's public finances. \cite{LopezMitchener2020} provide evidence that economic policy uncertainty due to reparations was an important factor contributing the the rise in inflation in Germany. Under the treaty, Germany also lost 13\% of its land area and 10\% of its population. At the same time, the signing of the Treat of Versailles ended the Allied war blockade of Germany, which hamstrung Germany's economy and public finances \citep{Graham1931}. 

Following high inflation in the second half of 1919, inflation slowed in 1920. This was due to the Erzberger fiscal reforms of 1919 and 1920, which led to large tax increases, and the suppression of the Kapp Putsch in March 1920, which led to a strengthening of the mark. However, inflation accelerated again in the spring of 1921 after the Reparations Committee determined exact the reparations in May 1921, which amounted to about 6\% of GDP per year \citep{Dornbusch1985}. Moreover, the London Ultimatum required an up front payment of 1.5 billion gold marks, about half of tax revenues,  in 1921 \citep{Dornbusch1985}. Inflation continued to increase following the assassination of finance minister Mathias Erzberger on August 26, 1921. 


\paragraph{Second phase of the inflation.} The summer of 1922 was the turning point in the inflation when high inflation turned into hyperinflation. There were three important shocks that explain the transition. First, in early June, the French government decided that the Bankers Committee could not provide reparations relief to Germany by reconsidering the May 1921 reparations schedule. This was in part a response to the Treaty of Rapallo, signed on April 16, 1922 between Germany and the Soviet Union, which opened diplomatic relations between the two countries and involved a mutual cancellation of financial claims. The Treaty violated the Treaty of Versailles. Second, the Bankers Committee determined that Germany did not have the credit to warrant an international loan to stabilize the mark \citep{kindleberger1985}. Germany suspended all payments of reparations in June 1922, and Germany formally demanded postponement of reparations for 2.5 years on July 12, 1922.  \cite{Cagan1956} dates the start of the hyperinflation in July 1922, and \cite{Cagan1991} refers to summer 1922 as the start of a ``new regime'' of collapse in the confidence in the mark.

Third, conflict over reparations was compounded by the assassination of the highly capable foreign minister Walther Rathenau  by right-wing paramilitaries on June 24, 1922. \textit{The Economist} noted that the Rathenau assassination and political turmoil were followed by ``panic on the Berlin exchange bourse'' (July 1, 1922). The mark depreciated by 7\% on the day of the assassination. This led to a flight from the mark to foreign exchange, as markets began to expect additional depreciation, resulting in a large capital outflow.\footnote{The size of the capital outflow and the amount of German wealth held abroad is highly uncertain and was debated in the context of Germany's ability to meet reparations \citep{Feldman1997}.} The boost to international competitiveness from the depreciation also subsided by the second half of 1922, leading to a rising trade deficit, while firms pulled back on investment due to a credit crunch, shortage of working capital, and increased uncertainty \citep{Feldman1997}.

Economic performance and inflation took another turn for the worse with the occupation of the Ruhr by France and Belgium in January 1923, following arrears on German deliveries of reparations in kind. The occupation was met by passive resistance, which the Reichsbank financed by discounting of Treasuries. This led to a surge in the issuance of paper currency. During 1922 and 1923, the Reichsbank also discounted commercial bills to alleviate the credit shortage.

There was a pause in the inflation from mid February to mid April 1923, when the Reichsbank attempted a first stabilization of the mark by intervening in the foreign exchange market. This briefly led to falling prices and an appreciation of the mark. The intervention was abandoned due to a large loss of central bank foreign currency reserves, as exchange rate was unsustainable given the large deficit \citep{Dornbusch1986}. From May to October 1923, the price level spiraled out of control with increasingly higher rates of monthly inflation. 

The the height of the hyperinflation, the economy was in crisis. Food shortages became common, as farmers refused to accept marks for their products \citep{Feldman1997}. Worker-employer relations deteriorated, as workers demanded wage increased to keep pace with inflation. In July 1923, government employee wages became explicitly indexed to inflation. Economic distress led to rising left- and right-wing extremism.

\paragraph{Stabilization.} Consensus for the need for stabilization grew in the hyperinflation phase, as there was a realization that the costs of inflation began to exceed the benefits of the monetary stimulus \citep{Feldman1997}. The foundations for the stabilization were laid starting in August 1923. The Cuno government was replaced by a ``Great Coalition" government with Gustav Stresemann as chancellor and the SPD in the finance ministry. The new government introduced new tax measures with accelerated indexation and issued a 500 million Goldmark loan, which paved the way for a new monetary unit linked to the Goldmark. Passive resistance in the Ruhr was ended on September 26, 1923. At this stage, the economy was in crisis; worker-employer relations had broken down and farmers had stopped accepting marks for products, leading workers to raid farmers' fields for food \citep{Feldman1997}. In October 1923, the SPD left the finance ministry after cabinet reshuffle, and Hans Luther became the new finance minister. Inflation peaked at a monthly rate of 30,000\% (more than 20\% per day) in October, and exchange rate based pricing became widespread. The extremely rapid increase in prices led to a fall in real money balances \citep{Cagan1956}.

A monetary reform was introduced on October 15, 1923. The decree created a new currency unit called the Rentenmark, which was declared equivalent to 1 trillion ($10^{12}$) paper marks. The Rentenmark would be issued by a new bank, the Rentenbank, which would replace the Reichsbank's note issue function. The Rentenbank was backed by ``fictitious'' claims on industry and land and faced limits on the amount of loans it could make to the government and private sector, as well as limits on the maximum amount of Rentenmarks that could be issued. The legislation also prohibited the Reichsbank from discounting government bills. The Rentenbank came into operation and started issuing Rentenmarks on November 15, 1923. The exchange rate was fixed from November 20, 1923. There was a final depreciation of the currency from 1.26 trillion paper marks/USD to 4.2 trillion paper marks/USD between November 14 and November 20, leading to a large reduction in the real money supply. The Rentenmark was then stabilized at 4.2 Rentenmarks/USD, and the Rentenmark was then equivalent to one Goldmark.

There were several important factors behind the success of the stabilization.\footnote{In contrast to the stabilization of Austria and Hungary, the German stabilization did not involve foreign assistance.} \cite{Sargent1982} argues that fiscal stabilization in the form of both increased taxes and cuts in government spending were crucial for success of stabilization. Government spending was cut through a 25\% reduction in personnel over four months and by retiring civil servants over age 65. With this fiscal reform, Stresemann and Luther balanced the budget. In contrast to the fiscal contraction, \cite{Sargent1982} emphasizes that the stabilization coincided with strong money growth. In additional to fiscal reform, \cite{Dornbusch1985} emphasizes the importance of exchange rate stabilization, very high discount rates (at times around 90\% per year) in November and December 1923, discounting restraints on the Reichsbank and Rentenbank, political stabilization with the end of passive resistance, and the large decrease in real money balances from the final 330\% devaluation between November 14 and 20, 1923. The success of the stabilization was highly uncertain in the first few months.\footnote{The stabilization also coincided with the death of the Reichsbank President Havenstein, who was replaced by Hjalmar Schacht.}

In August 1924, the Dawes Plan substantially aided Germany's fiscal situation by providing reparations relief. Reparations payments were temporarily suspended, and the Dawes plan assigned Germany a more manageable schedule of payments. The plan involved a reorganization of the Reichsbank and the introduction of the Reichsmark to replace the Rentenmark. The Reichsmark (sign RM) was equal to one Rentenmark. Under the plan, France and Belgium agreed to withdraw from the Ruhr.

\subsection{Historical Accounts of the Economic Impact of the Inflation}

\paragraph{Aggregate effects of inflation and stabilization.} The German economy experienced high growth and low unemployment from the end of the war to the second half 1922, avoiding the ``Depression'' of 1920-21 in the US, UK, and France (see \Cref{fig:real_GDP}).\footnote{\cite{Graham1931} writes: ``That business in Germany was booming during most of the inflation period is a universally admitted fact'' (p. 278).} From the final months of 1922, inflation was associated with contraction, and 1923 saw a large decline in production due to a combination of the Ruhr occupation, hyperinflation, and stabilization. \cite{Graham1931} argues that much of the adverse real effects of the inflation were due to coincident factors such as the loss of productive capacity during the war and the invasion of the Ruhr, although both \cite{Graham1931} and \cite{Garber1982} suggest that inflation may have resulted in a distortionary reallocation of resources toward large capital goods producers. \cite{Feldman1997} argues that the hyperinflation itself contributed to economic crisis toward the second half of 1922, due to capital flight and a credit shortage, increased uncertainty that led firms to hold back production, large distortions in relative prices, breakdown in labor relations, a breakdown of trade, and rising social unrest.

The impact of the stabilization has also been the subject of debate. \cite{Sargent1982} argues that the stabilization was not associated with substantial negative effects and was actually expansionary based on annual industrial production data, though it is difficult to know how much of the increase in industrial production from 1923 to 1924 was due to the end of the Ruhr crisis. \cite{Garber1982} argues that aftermath of stabilization was associated with large transitional costs through a reallocation of resources away from industrial firms that benefited from the inflation.

\paragraph{Distributional effects of the inflation and the impact on firms.} The inflation had distributional effects through balance sheets, as we show in the paper. Debt-financed industrialists and landowners, especially those with mortgages, benefited from the inflation, while households on fixed income lost out.\footnote{\cite{Graham1931} notes that 40 billion marks of mortgage debt (one-sixth of German wealth) in 1913 was wiped out by the inflation. While urban landlords benefited from the erosion of their mortgage debt, strict regulation of rents made housing almost free for tenants during the hyperinflation. } \cite{Feldman1997} notes that this redistribution was well understood by contemporaries \citep[e.g., pp. 54, 514][]{Feldman1997}. 
This allowed industrial firms to self-finance a higher share of their activity, making them less reliant on banks \citep[][p. 276]{Feldman1997}. \cite{Graham1931} argues these redistributive effects were expansionary, but also notes that it caused over-investment and misallocation of resources to less productive users. Inflation also wiped out much of public debt, though lags between assessment and collection increased the deficit through the Tanzi effect \citep{Dornbusch1986}.

There were also distributional effects through wages and prices. Real wages declined up through 1920, especially for skilled workers in the middle class (see \Cref{fig:real_wages}). By the hyperinflation stage, wages raced to keep pace with rising prices.  The depreciation of the mark also disproportionately benefitted exporters, who were able to regain foreign markets \citep{Graham1931}.

Firms responded to the inflation by increasing consolidation, such as the Stinnes' Siemens Concern. Mergers were financed by cheap debt. Vertical integration allowed firms to reduce uncertainty about the cost of materials. Horizontal integration was ostensibly pursued to the diversify risk of volatile goods prices \citep{Graham1931}. This wave of consolidation was an acceleration on previous structural trends in the German economy \citep[][p. 272]{Feldman1997}. Some of the industrial concerns built up during the inflation collapsed during the stabilization.

\paragraph{Banking and credit conditions during the inflation} Banks saw large declines in the real value of their capital during the inflation. Based on data on 19 credit banks \cite{Goldschmidt1928}, finds that deflated bank capital declined by 54\% from 1918 to 1923, with most of the decline occurring in 1919. Bank credit was available for firms in the first phase of the inflation (1919-21), but the second phase of the inflation witnessed a ``credit crisis.'' This section provides further details on banking and credit conditions during the inflation.

There was a banking boom from 1919-21, as banks saw large inflows of mark-denominated foreign deposits from speculators betting on an appreciation of the mark. Banks also benefited from a widening deposit spread and from commissions on the high activity of stock and money market transactions. Banks were reported to be ``swimming in money'' during the first phase of the postwar inflation \citep{Feldman1997}. 

As a result, credit conditions were not particularly tight in the first phase of the postwar inflation, and banks continued lending to industrial firms in this period. \cite{BrescianiTurroniCostantino1937TEoI} refers to a report of the General Association of German Banks and Bankers for 1923, which reported that: ``Thanks to the aid of the banks, German industry and commerce were given the means not only to preserve their resources but to increase them in considerable measure. Industry rapidly recognized that it was economically more advantageous to incur the highest possible debts at the bank rather than to keep large deposits.''  \cite{BrescianiTurroniCostantino1937TEoI} argues that banks lost by providing cheap credit to firms, perhaps because they did not understand the implications of inflation. Banks were slow to raise interest rates due to various factors. \cite{BrescianiTurroniCostantino1937TEoI} argues they did not require high interest rates because they did not anticipate inflation and because the ``phenomenon of monetary depreciation had not yet been properly understood by the majority of bank directors.'' \cite{schacht1927stabilization} notes that interest rates on bank loans were usually set based on the Reichsbank's discount rate, which remained low for much of the inflation. 

While firms benefited from \textit{ex post} low real interest rates, \cite{Lindenlaub1985} argues that, before late 1921, businesses generally did not respond to inflation by maximizing borrowing in anticipation of low real interest rates. This is consistent with narrative evidence and the behavior of forward exchange rate, which both indicate that agents did not anticipate continued high inflation before the summer of 1922. Therefore, while some industrial firms benefited from high leverage at the expense of banks, it is not clear that this was a systematic policy of the nonfinancial corporate sector.

Between 1921 and 1922, there was a large decline in real deposits, as depositors sought assets that would provide a better hedge against depreciation and to avoid taxes \citep{Feldman1997}. During this phase, the term structure of deposits also shortened, leading to a shortening of loan terms. Banks also gradually raised interest rates, though never sufficiently to yield positive \textit{ex post} returns \citep{Graham1931}.

Credit conditions became tight during the hyperinflation phase, starting in summer 1922. In this phase, it became very difficult for firms to obtain credit and external financing almost disappeared \citep{Graham1931,Dornbusch1986, Feldman1997}.\footnote{\cite{Neumeyer1998} presents theory and evidence from Argentina that high inflation leads to a disappearance of nominal financial contracts due to high expected inflation with a low probability of inflation stabilization.} For example, in July 1922, \textit{The Economist} noted an ``an extreme shortness of money,'' due to contraction of supply and elevated demand for credit, as depreciation became widely anticipated.\footnote{In July 1922, \textit{The Economist} also reported that ``the instability of the standard of value is gradually killing long-period credit in Germany.'' } Narrative accounts refer to a credit ``famine'' or ``crisis'' \citep{Balderston1991,Feldman1997}. This is evidenced in rising interest rates on various types of credit, including money market interest rates \citep{Holtfrerich1986,Feldman1997}.

In the hyperinflation phase, many businesses were severely liquidity constrained with rising nominal input prices and wages. Banks could not keep up the supply of credit to finance firms' working capital.\footnote{Banks could not index advances to the price level \citep{Balderston1991}. More broadly, indexation of financial contracts was not widespread due to restrictions on foreign currency pricing, thought commodity-indexed bonds started to be issued in late 1922 \citep{Feldman1997}.} This led to the reintroduction of bills of exchange, which could be discounted at the Reichsbank. From the middle of 1922, the Reichsbank partially substituted for credit bank's credit by discounting bills of large firms, which made these firms less reliable on the banks \citep{Feldman1997}. The loan bureaus of the Reichsbank also became more active in granting credit. Large firms benefited from discounting at low real rates at the Reichsbank. Banks could also discount bills at the Reichsbank, and \cite{Graham1931} argues banks recouped some of their losses by discounting bills at low rates at the Reichsbank, which transferred losses from banks to all holders of currency. .\footnote{\cite{Graham1931} notes: ``It has indeed been suggested that the big industrial borrowers virtually stole the banks, but, insofar as this occurred, the commercial bank directorates largely recouped their losses at the expense of the Reichsbank.''}

The inflation was also characterized by an acceleration in banking sector consolidation through banking alliances.\footnote{A notable example was the merger between Darmstadt Bank für Handel und Industrie and Nationalbank für Deutschland into Danat Bank (Damstädter- und Nationalbank) in July 1922.} Over the period 1914-1925, the Berlin ``great banks'' absorbed many provincial and private banks \citep{Balderston1991}. For example, Deutsche Bank increased its number of branches from 15 in 1913 to 142 in 1924 \citep{Feldman1997}. \cite{Balderston1991} argues that it is not clear exactly why mergers accelerated. \cite{Feldman1997} argues it was because provincial banks traded at a discount relative to the big banks and because big banks partly saw acquiring the assets of smaller banks (at least the real assets such as buildings) as an inflation hedge.

\renewcommand\thefigure{\thesection.\arabic{figure}}
\renewcommand\thetable{\thesection.\arabic{table}}

\section{Model of the Debt-Inflation Channel of Inflation}
\label{sec:model}

This section lays out a simple model to illustrate the following mechanisms:
\begin{enumerate}
\item Inflation and Bankruptcies: When firms have nominal debt and can default, unexpected inflation increases firms' net worth, leading to a decline in default rates.
\item The Debt-Inflation Channel and Firm Activity: If firms are financing-constrained, unexpected inflation relaxes financing constraints and leads to an increase in labor demand, employment, and output. The debt-inflation channel is stronger for a higher initial level of leverage.
\item The Nominal Rigidity Channel: If unions face a menu cost in adjusting wages, small increases in inflation have a large effect on output through the nominal rigidity channel by reducing real wages. The nominal rigidity channel thus complements the debt-inflation channel. However, for high inflation, wages become flexible, and inflation only has real effects through the debt-inflation channel.
\end{enumerate}

We consider a static model with two subperiods: ``morning'' and ``evening.'' The economy is populated by a unit mass of entrepreneurs, who operate the productive technology, and workers, who monopolistically provide labor to firms.

\paragraph{Firms.} Firms are run by risk neutral entrepreneurs and with utility function $U(C)=C$. Firms have initial capital stock $K_0$ and owe nominal debt to workers $D_0$. Capital is subject to a real shock $Z_i \sim G$. In the morning, the entrepreneur decides whether to default or produce. The real value of an entrepreneur is the maximum of zero and
\begin{align*}
J = K_0 - Z_i - \frac{D_0}{P}  + V,
\end{align*}
where $V$ is the value of the firm to the entrepreneur from continuing production (defined below) and $P$ is the price level, which is assumed to be exogenous.\footnote{The price level can be endogenized by assuming that there is stock of money that is required for transactions and that is randomly adjusted by the monetary authority. In that case, the price level is determined by $M = P(K_0-K+Y),$ where $K_0$ is initial capital, $K$ is capital used in production, $Y$ is aggregate production.} Firms with negative value default. The cutoff value for $Z^*$ for default is defined by:
\begin{align}
Z^* =   K_0 - \frac{ D_0}{P} + V(Z^*). \label{zstar}
\end{align}
When a firm defaults, the entrepreneur gets zero consumption and exits the economy. The capital of the entrepreneur is then destroyed (i.e., it has a liquidation value of zero). The measure of active entrepreneurs is $G(Z^*)$. The value of nominal debt in the economy is $G(Z^*)D_0$.


Each firm $i$ operates a Cobb-Douglas technology using capital and labor $\{L_{ij}\}_{j \in [0,1]}$ from each worker,
$$ Y_i = F(K_i, \{L_{ij}\}) = A K_i^\alpha L_i^{1-\alpha} ,$$
where 
$$L_i = \bigg(\int\limits_0^1 (L_{ij})^{\frac{\epsilon-1}{\epsilon}} dj\bigg)^{\frac{\epsilon}{\epsilon-1}}, $$
is a CES aggregate of labor provided by each worker $j$ to entrepreneur $i$. 

Each firm with $Z\leq Z^*$ uses initial capital net of debt along with intra-period debt $D_i$ to invest in capital and pay labor in the morning. The flow of funds condition for entrepreneur $i$ is
\begin{align}
 D_i - D_{0} - P Z_i &= WL_i + P(K_i - K_0) \label{fof}
\end{align}
To introduce financing constraints, we assume that firms are subject to a working capital constraint on $D_i$, similar to \cite{jermann2012macroeconomic}. The constraint is given by:
\begin{align}
D_i \leq  \xi P A K_i^\alpha L_i^{1-\alpha}.  \label{constraint}
\end{align} 
Combining \eqref{fof} and \eqref{constraint} yields the following constraint:
\begin{align}
D_{0} + P Z_i + WL_i + P(K_i -K_0) \leq  \xi P A K_i^\alpha L_i^{1-\alpha} \label{constraint2}
\end{align}
The firm's problem is
\begin{align*}
P \cdot V = &\max_{K_i,L_i} P A K_i^\alpha L_i^{1-\alpha}  - W L_i - PK_i \quad  \text{s.t.} \;\; \eqref{constraint2}. \nonumber
\end{align*}
The first-order conditions are:
\begin{align*}
[K_i]: \quad  F_K - 1 - \lambda (1 - \xi  F_K) &= 0 \\
[L_i]: \quad P F_L - W - \lambda(W-\xi P F_L) &=0,
\end{align*}
where $\lambda$ is the Lagrange multiplier on the constraint \eqref{constraint2}.

There are two cases, depending on whether the financing constraint binds. In the following, for simplicity to illustrate the main points, we consider parametrizations of $G$ and $\xi$ such that all firms are constrained. Regardless of whether the constraint binds, the capital-labor ratio is
\begin{align}
\frac{K_i}{L_i} &= \frac{\alpha}{1-\alpha} \frac{W}{P}. \label{K_L}
\end{align}
When the constraint binds, we can solve for firm $i$'s labor demand by combining \eqref{constraint2} and \eqref{K_L}:
\begin{align}
L_i^d = \frac{ K_0 - \frac{ D_{0}}{P} - Z_i }{   \frac{1}{1-\alpha} \frac{W}{P} - \xi A \left( \frac{\alpha}{1-\alpha} \frac{W}{P} \right) ^\alpha  }.  \label{ld_2} 
\end{align}
Firm labor demand is an increasing function of its initial resources, $K_0 - \frac{D_{0}}{P} - Z_i$.  Again assuming all firms are constrained, aggregate labor demand is given by
\begin{align}
L^d =  G(Z^*) \frac{K_0 - \frac{D_0}{P} - \frac{ \int_{\underline Z}^{Z^*} Z dG(Z) }{G(Z^*)}}{  \frac{1}{1-\alpha} \frac{W}{P} - \xi A \left( \frac{\alpha}{1-\alpha} \frac{W}{P} \right)^\alpha  } \label{ld_aggregate}
\end{align}

For a constrained firm, the real value of production is,
\begin{align*}
V = \frac{ D_{0}}{P} + Z_i - K_0 + (1-\xi)A K_i^\alpha L_i^{1-\alpha}
\end{align*}
so the value of the firm from not defaulting is
\begin{align*}
J = (1-\xi)A K_i^\alpha L_i^{1-\alpha}=(1-\xi)A \left(\frac{\alpha}{1-\alpha} \right)^\alpha \left( \frac{W}{P} \right)^\alpha L^d_i.
\end{align*}
Thus, a firm defaults if it would choose a negative amount of labor, $L^d_i$. From \eqref{ld_2}, we see that the cutoff value of default is the value such that $L^d_i=0$, or 
\begin{align*}
Z^* = K_0 - \frac{D_0}{P},
\end{align*}
In other words, the firm defaults if the real value of its initial debt exceeds the value of capital. it has negative initial net worth. The share of defaulting firms is $1-G(Z^*)$, which provides our first result.
\newline

\noindent \textbf{Result 1: Debt-Inflation and Firm Bankruptcies.} The share of defaulting firms declines for higher levels of inflation, $P$: 
\begin{align*}
\frac{\partial (\text{Default share})}{\partial P} &= - \frac{D_0}{P^2} G'\left( K_0 - \frac{D_0}{P} \right) <0.
\end{align*}

\paragraph{Workers.} The household chooses its overall level of consumption and labor to maximize
\begin{align*}
 U(C,L) = \ln(C) - \chi \frac{L^{1+\varphi}}{1+\varphi}
\end{align*}
subject to the budget constraint: 
\begin{align*}
 C &= \frac{W}{P}L + \frac{G(Z^*) D_0}{P}
\end{align*} 
The budget constraint uses the assumption that workers hold a diversified portfolio of debt claims with aggregate nominal value $G(Z^*) D_0$. Each worker sets a wage $W_j$ at which they are willing to work. Given that the production technology aggregates different varieties of labor according to a CES function, the total units of labor demanded from a worker $j$ who sets wage $W_j$ will be
$$ L(W_j) = \bigg(\frac{W_j}{W}\bigg)^{-\epsilon} L^d , $$
where $L^d$ is the aggregate quantity of labor demanded by entrepreneurs. In equilibrium, all varieties of labor set the same wage $W_j=W$. Household aggregate labor supply is given by
\begin{align*}
\frac{W}{P} = \frac{\epsilon}{\epsilon -1} \chi L^\varphi C,
\end{align*}
which can be rewritten as
\begin{align}
\frac{W}{P} &= \frac{\frac{\epsilon}{\epsilon-1}  \chi L^\varphi}{1- \frac{\epsilon}{\epsilon-1}  \chi L^{1+\varphi}} \left( \frac{G(Z^*) D_0}{P} \right).  \label{Ls}
\end{align}
An increase in inflation $P$ lowers real debt held by households, raising labor supply through a wealth effect. Intuitively, households reduce consumption of leisure and increase labor as the inflation erodes their wealth.\footnote{If we instead assume that utility is quasi-linear, $U(C,L)=C- \chi \frac{L^{1+\varphi}}{1+\varphi}$, then labor supply would be $\frac{W}{P} = \frac{\epsilon}{\epsilon-1} \chi L^\varphi$, removing the wealth effect.}

\paragraph{Flexible wage equilibrium.} With flexible wages, the equilibrium in the labor market is given by the solution to \eqref{ld_aggregate} and \eqref{Ls}. Capital of non-defaulting firms can be consumed immediately or used for production, at which point it depreciates entirely. Hence, the aggregate resource constraint is
$$ \int_{\underline Z}^{Z^*} A K_i^\alpha L_i^{1-\alpha} dG(Z)  = \int_{\underline Z}^{Z^*} [ C_{ie} + K_i - K_0 ] dG(Z) + C_{w}, $$
where $C_e$ and $C_w$ denote the consumption of the entrepreneur and household.

\begin{figure}[!ht]
\caption{Labor Market Equilibrium with Flexible Wages Following an Inflation Shock \label{labor_market}}
\begin{center}
\subfloat[Labor market equilibrium]{
\includegraphics[scale=.4]{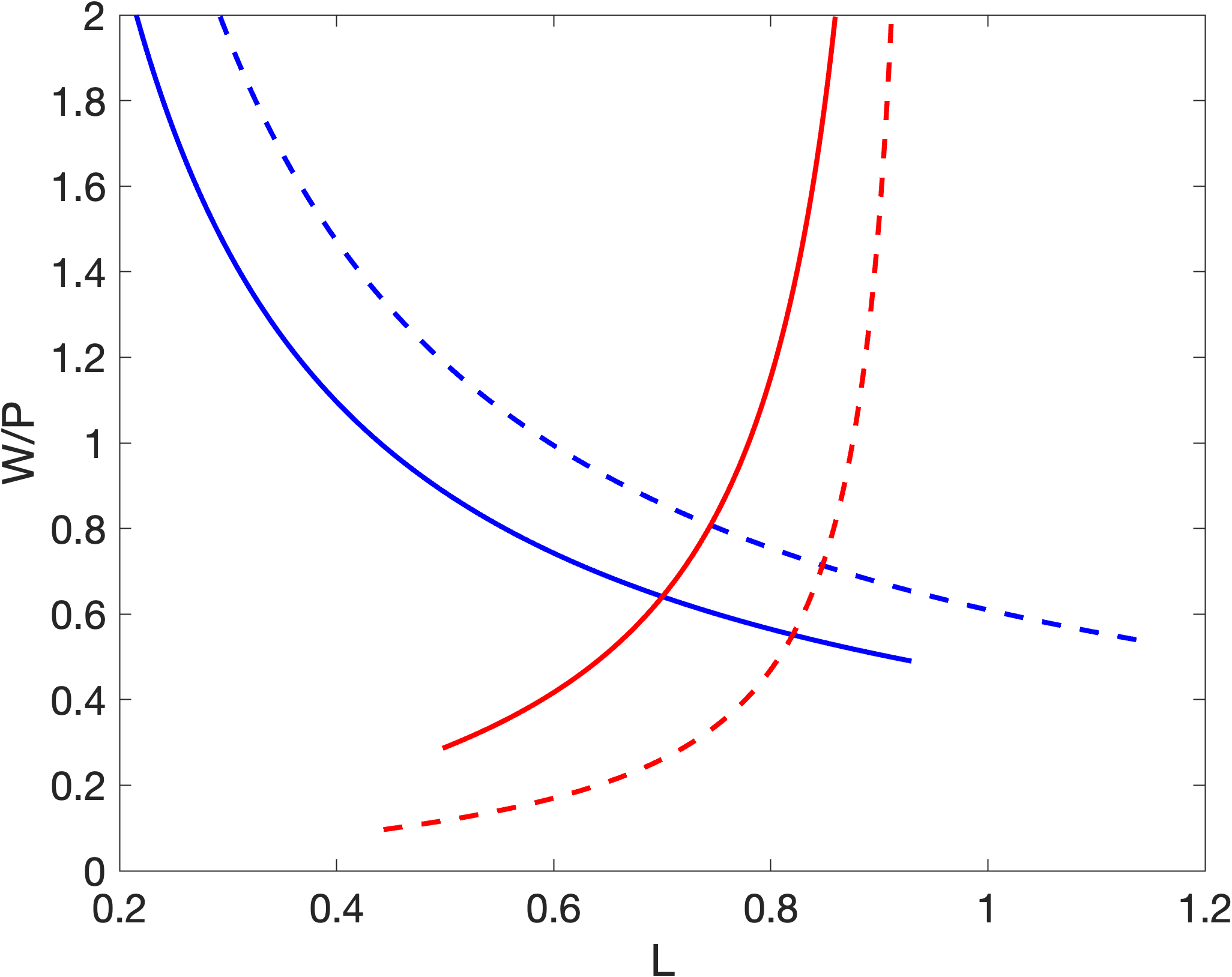}}

\subfloat[Outcomes as a function of inflation $P$]{
\includegraphics[scale=.4]{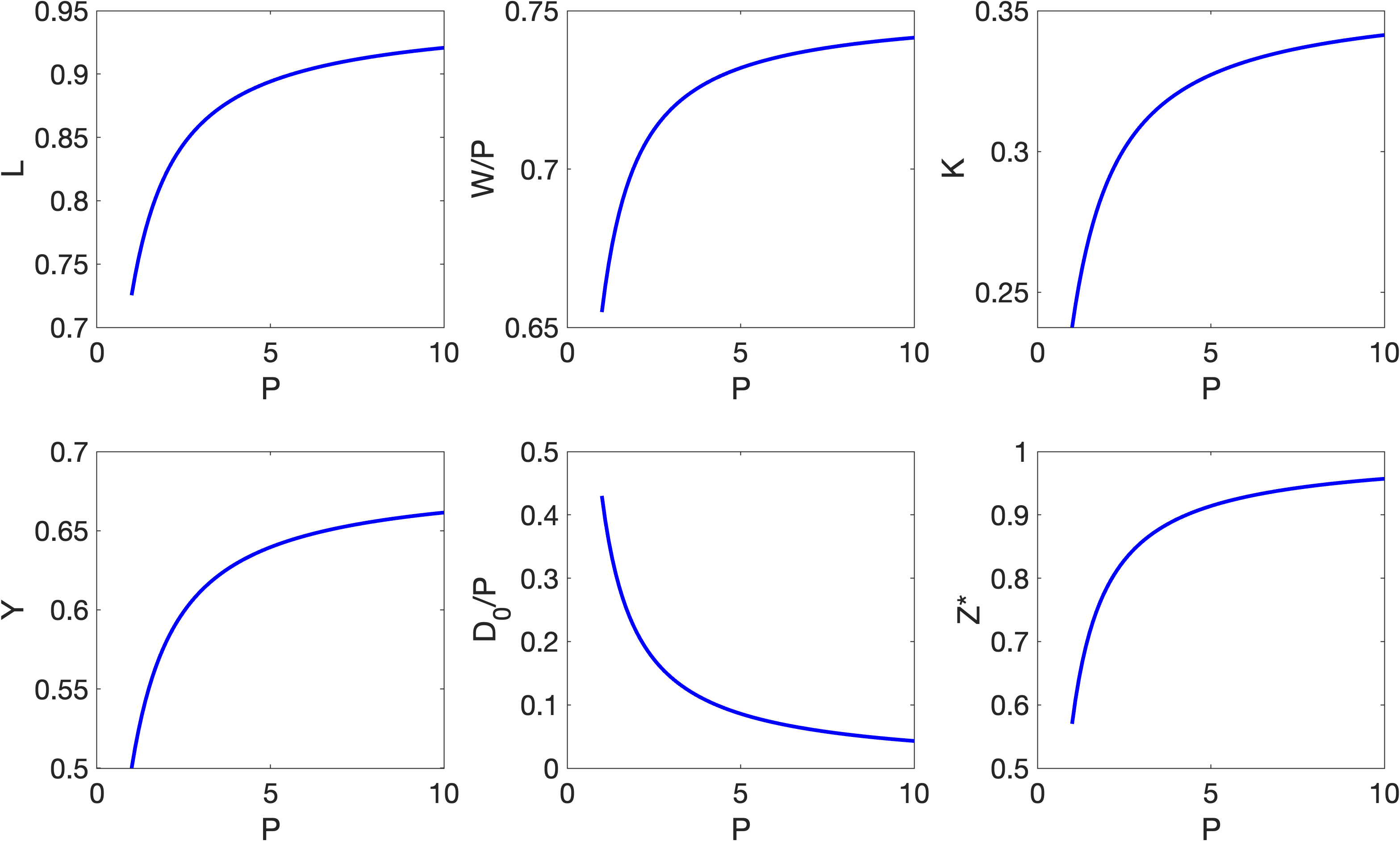}}

\end{center}
\footnotesize{\textit{Notes}: Panel (a) illustrates the labor market equilibrium for a low (solid curves) and high (dashed curves) of $P$. Panel (b) plots equilibrium outcomes from the model with flexible wages as a function of $P$.}
\end{figure}

The labor market equilibrium is illustrated in \Cref{labor_market}. In response to an increase in the price level $P$, labor demand shifts outward, as firms financing constraints are relaxed. Moreover, labor supply shifts outward due to the negative wealth effect. The increase in labor supply dampens the increase in the real wage, consistent with the fact that real wages did not rise and actually declined during the German inflation.
\newline

\noindent \textbf{Result 2: The Debt-Inflation Channel and Firm Activity.} If firms have nominal debt and are financing-constrained, inflation boosts labor demand \eqref{ld_aggregate}, increasing employment and output. The higher the level of initial debt $D_{0}$, the stronger the increase in labor demand and, thereby, the debt-inflation channel. The increase in the real wage is offset by the wealth effect on labor supply from the erosion of workers' real debt holdings.

\paragraph{Introducing nominal wage rigidity.} We introduce nominal rigidity by assuming that initially, the wage for all workers is set at $W_0$, which we assume is the equilibrium flexible wage with $P=1$. Workers can alter their wages $W_j$, but incur a cost by doing so. Specifically, there is a menu cost of altering the wage: a worker that changes $W_j$ from its baseline $W_0$ pays a cost $\psi \geq 0$ regardless of the final value of $W_j$. The presence of a menu cost will generate different behavior of the economy in times of normal inflation and times of hyperinflation, since workers will choose to change their wages only when inflation is at a high enough level. 

Denote $W^*$, $L^*$ denote equilibrium in the labor market if the wage is adjusted. $L^d(W)$ is labor demand for a given wage $W$. The wage is adjusted if utility from the flexible price equilibrium, net of the cost of adjustment, exceeds the utility from the allocation with $W=W_0$:
\begin{align*}
\ln\left( \frac{W^*}{P} L^* + G(Z^*)\frac{D_0}{P} \right)  - \chi \frac{(L^*)^{1+\varphi}}{1+\varphi}  - \psi \geq \ln\left( \frac{W_0}{P} L^d(W_0) + G(Z^*) \frac{D_0}{P} \right) - \chi \frac{L^d(W_0)^{1+\varphi}}{1+\varphi}
\end{align*}

\begin{figure}[!ht]
\caption{Labor Market Equilibria with Nominal Wage Rigidty for Different Levels of $P$ \label{fig:Leq_sticky_wage} }
\begin{center}
\includegraphics[scale=0.4]{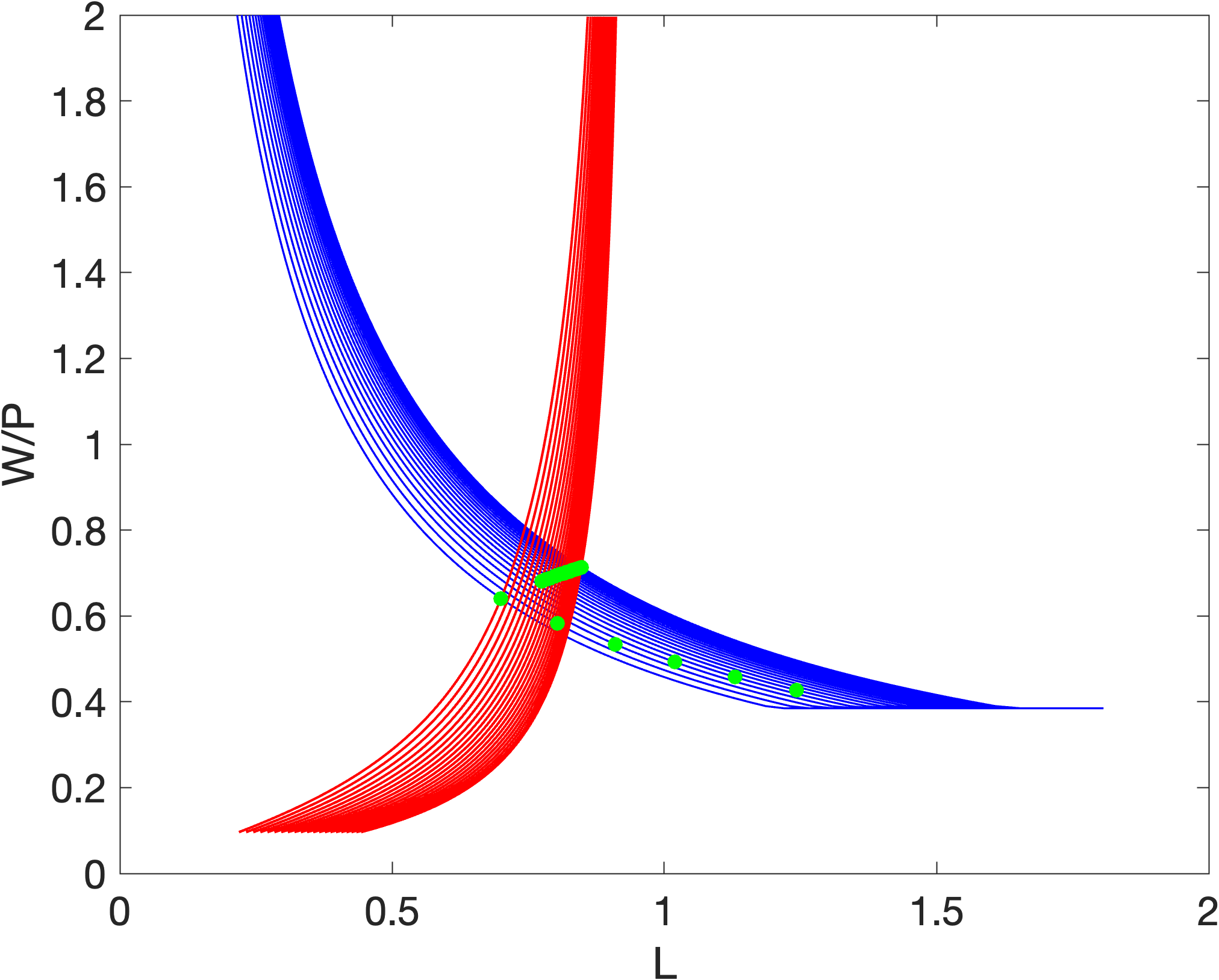}
\end{center}
\footnotesize{\textit{Notes}: This figure illustrates labor market equilibrium for increasing values of $P$ in the model with nominal wage rigidity. The downward sloping blue curves are labor demand curves for different levels of $P$, while the upward sloping red curves are labor supply curves. Green dots indicate the equilibrium, which depends on whether the nominal wage is adjusted.}
\end{figure}

\noindent \textbf{Result 3: Nominal Rigidity Channel of Inflation}: Labor market equilibrium in response to shock to $P$ depends on the size of the inflation shock. For a small inflation, the nominal wage is not updated, $W/P$ falls, and employment increases significantly through both the nominal rigidity and debt-inflation channels. For a large increase in the price level, the wage is updated, and inflation only affects real outcomes through the debt-inflation channel. This result is illustrated with an example in \Cref{fig:Leq_sticky_wage} for various levels of the price level $P$.

\clearpage
\renewcommand\thefigure{\thesection.\arabic{figure}}
\renewcommand\thetable{\thesection.\arabic{table}}
\setcounter{figure}{0}
\setcounter{table}{0}

\section{Supplementary Figures and Tables}
\label{app:additional_results}


\begin{figure}[htpb]
    \caption{Map of Firm Headquarter Locations.}
    \begin{center}
        \includegraphics[width=0.90\textwidth]{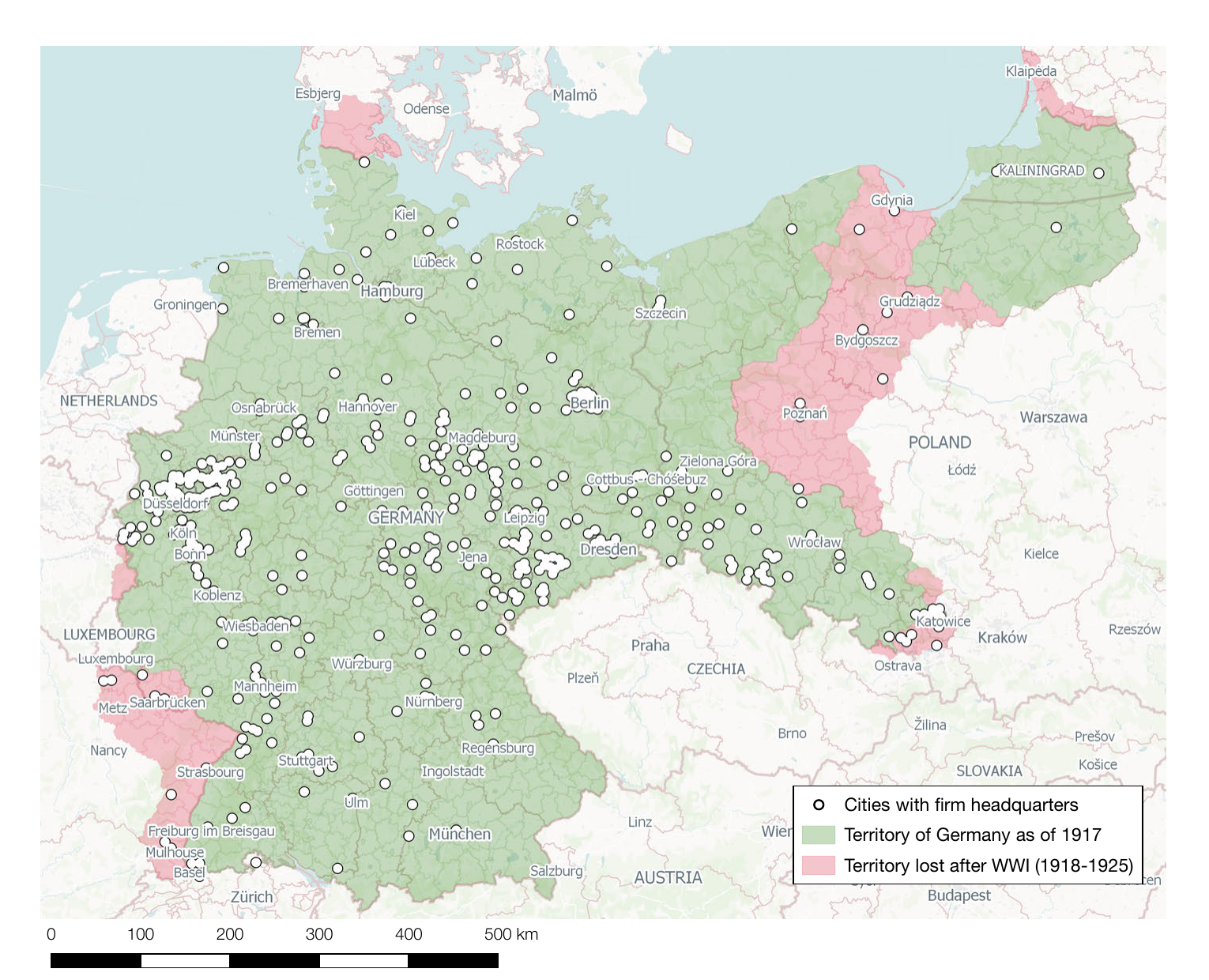}
        \end{center}
	 \label{fig:saling_map}
        \footnotesize{\textit{Notes}: This map portrays the headquarter locations of all firms in our \textit{Saling's} sample, alongside Germany's territorial extent as of 1917.}
\end{figure}

\begin{figure}[htpb]
    \caption{Balance Sheet Dynamics in Saling: Deflated Levels.}
    \begin{center}
        \subfloat[Evolution of median of liability components in paper marks, deflated.]{\includegraphics[width=0.5\textwidth]{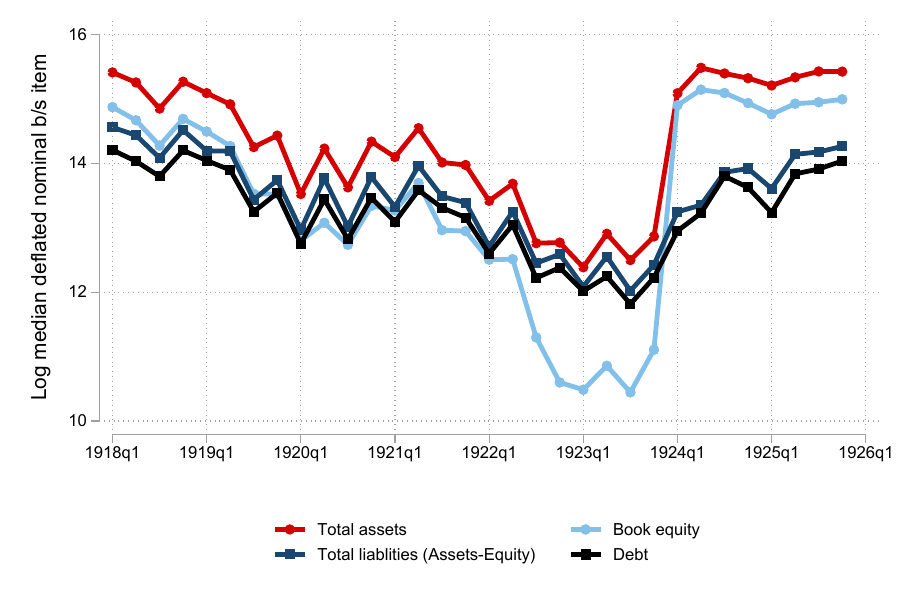}} \hfill 
        \subfloat[Evolution of median of asset components in paper marks, deflated.]{\includegraphics[width=0.5\textwidth]{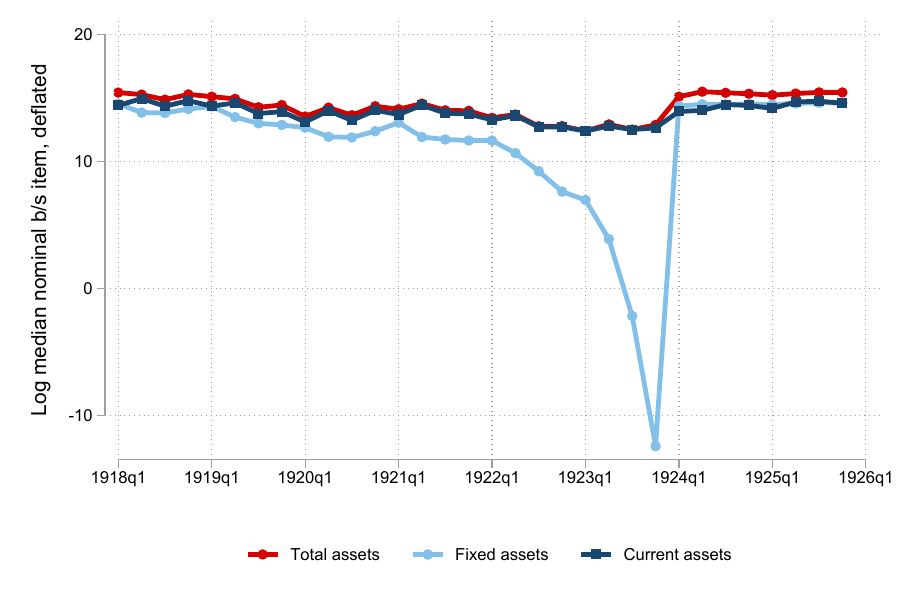}}

          \subfloat[Evolution of median of current asset components in paper marks, deflated.  \label{fig:saling_med_current_assets}]{\includegraphics[width=0.5\textwidth]{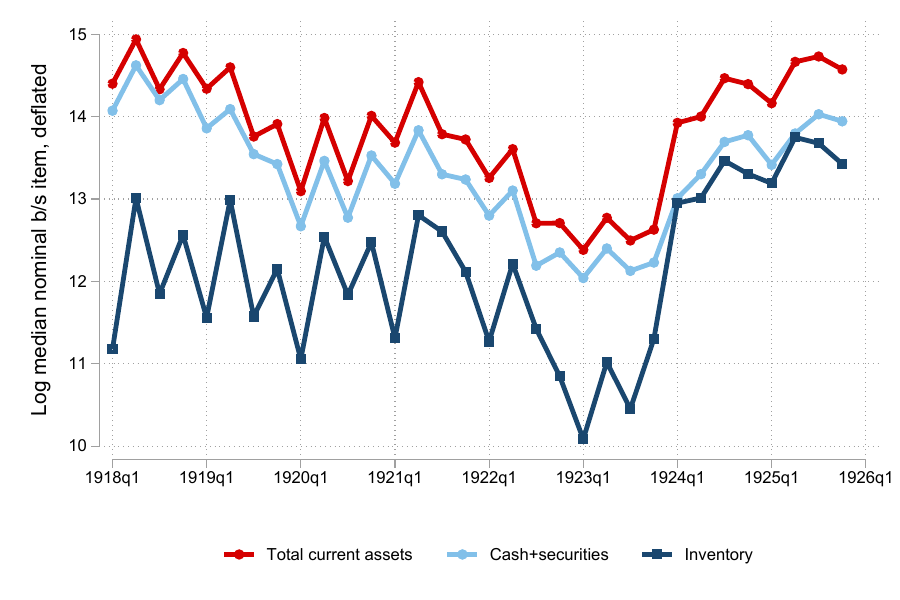}}
      \end{center}

        \footnotesize{\textit{Notes}: This figure plots the evolution of the medians of key balance sheet items in paper marks, deflated by the wholesale price index. The large changes in 1924Q1 occur due to the introduction of revalued Goldmark balance sheets.}
	 \label{fig:saling_med_leverage_deflated}
\end{figure}

\begin{figure}[!ht]
    \caption{Nonfinancial Equity and Bank Equity Returns Based on Indexes from Berliner Börsen Zeitung and Wirtschaft und Statistik.}
        \begin{center}

        \includegraphics[width=0.85\textwidth]{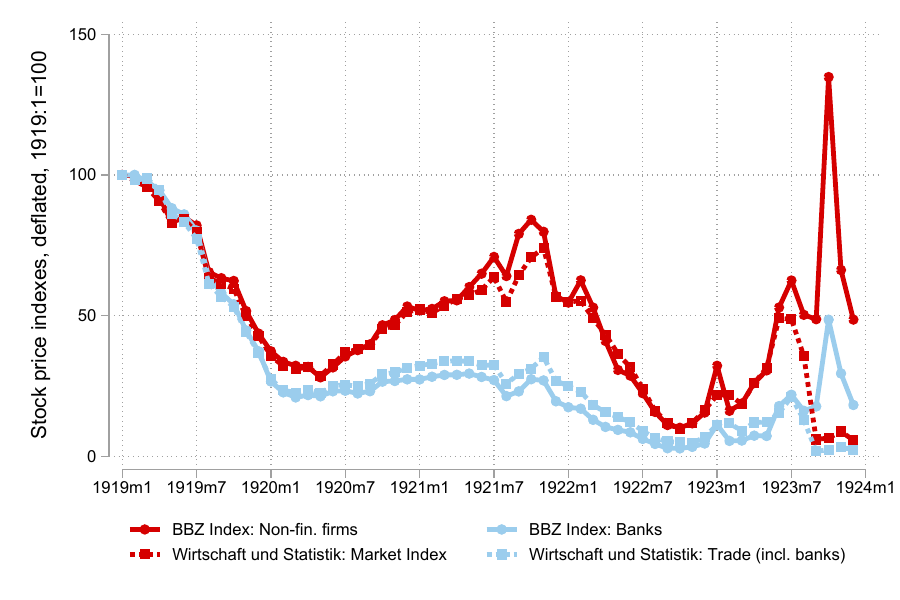}

        \end{center}
        \footnotesize{\textit{Notes}: This figure plots equity indexes for nonfinancial firms and banks. We use two sources. The first is an equal-weighted from our hand-collected stock price data from \textit{Berliner Börsen Zeitung} (BBZ). The second is published stock price indexes from \textit{Wirtschaft und Statistik}. \textit{Wirtschaft und Statistik}'s index for ``Trade,'' includes banks, so we use this series as a comparison for our index of bank stocks from BBZ. }
	 \label{fig:bbz_validation}
\end{figure}

\begin{figure}[htpb]
		\begin{center}
		\caption{Unemployment and Firm Bankruptcies. \label{fig:unemp_def}}
  	     \includegraphics[width=0.9\textwidth]{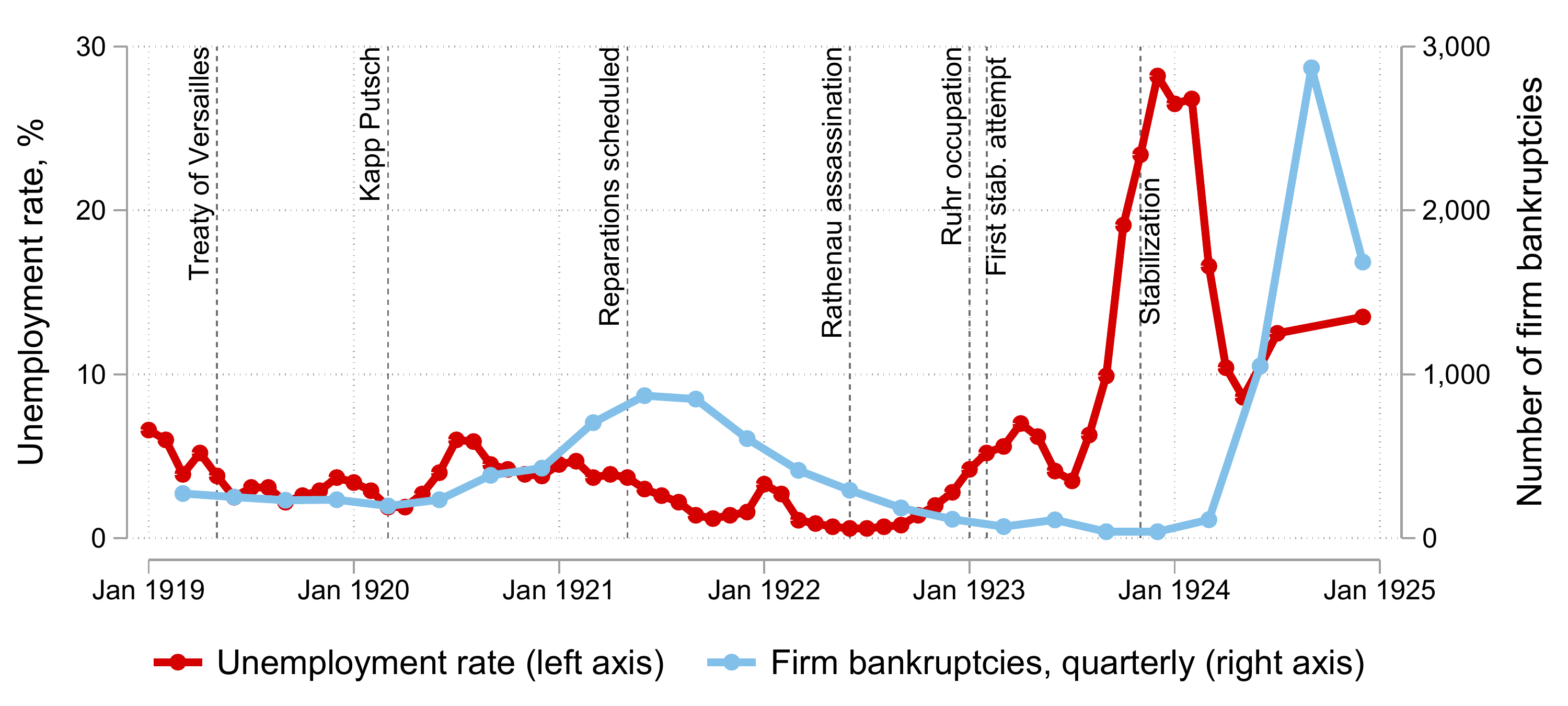}	
  	     \end{center}
        \footnotesize{\textit{Notes}: Quarterly bankruptcies are from the \textit{Vierteljahrshefte zur Statistik des Deutschen Reichs Herausgegeben vom Statistischen Reichsamt} and \textit{Wirtschaft and Statistik}. Unemployment for industries is from \textit{Reichsarbeitsblatt}.}
	
\end{figure}

\begin{figure}[htpb]
		\begin{center}
		\caption{Inflation and Firm Bankruptcies. \label{fig:def_inflation}}
        {\includegraphics[width=0.7\textwidth]{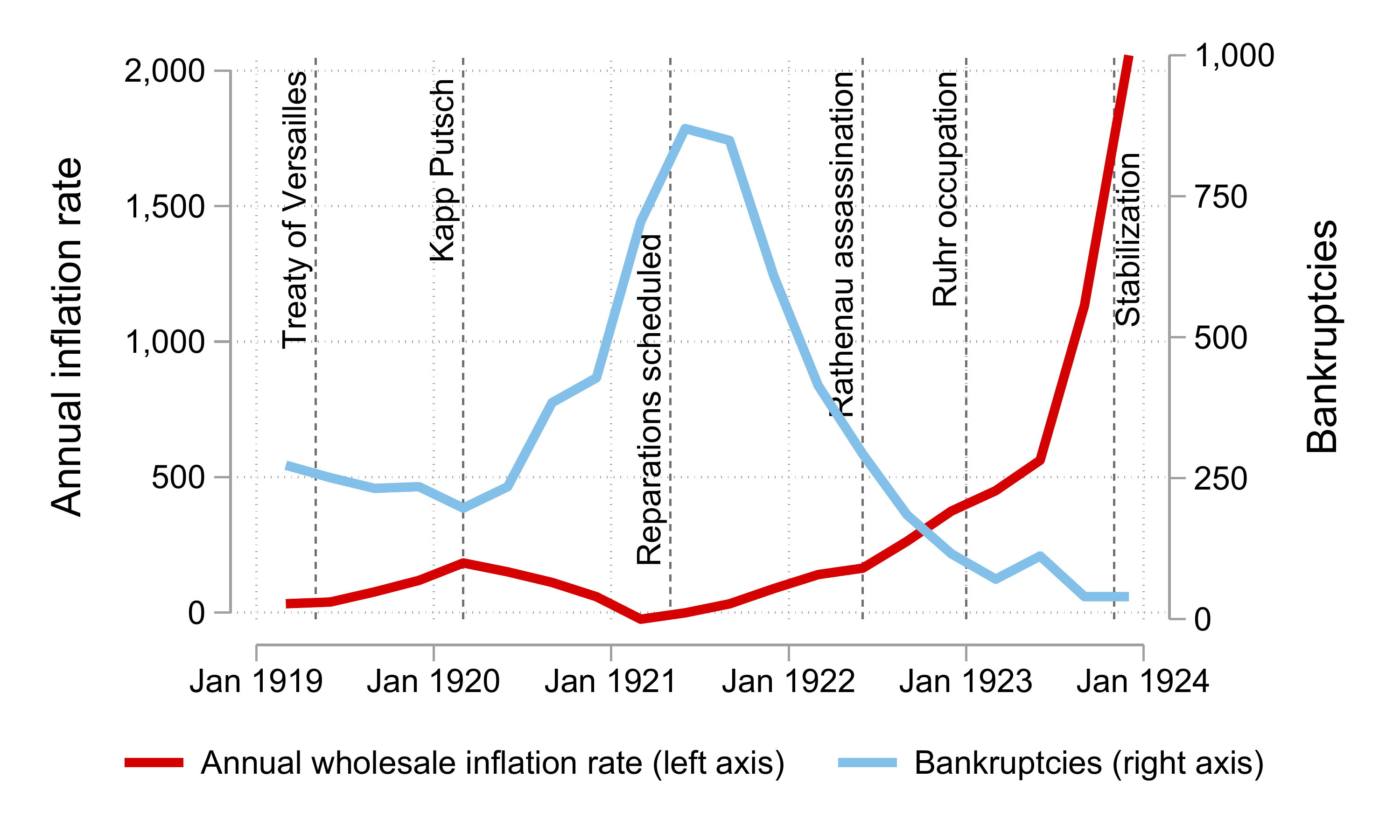}}
  	     
  	     \end{center}
        \vspace{-.5cm}
        \footnotesize{\textit{Notes}: Quarterly counts of firm bankruptcies are obtained from the \textit{Vierteljahrshefte zur Statistik des Deutschen Reichs Herausgegeben vom Statistischen Reichsamt}. Inflation of wholesale prices as reported in \textit{Zahlen zur Geldentwertung}.}
	
\end{figure}

\begin{figure}[htpb]
	\begin{center}
    \caption{Inflation and Firm Bankruptcies: Robustness using the Acceleration in Inflation. \label{fig:fin_PC_accel}}
    \includegraphics[width=0.7\textwidth]{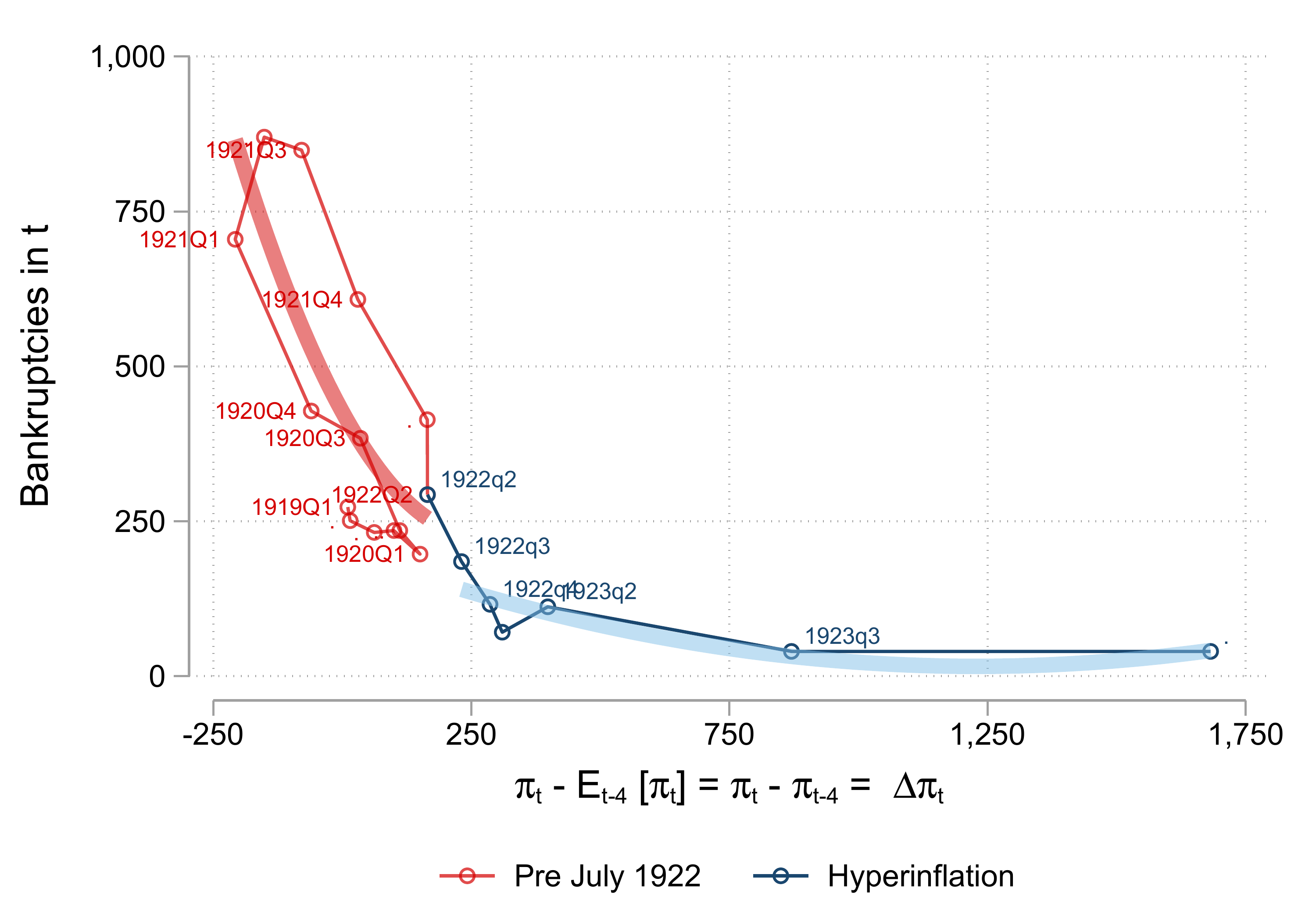}
	\end{center}

    \footnotesize{\textit{Notes}: This figure plots the number of firm bankruptcies in quarter $t$ against inflation over the past four quarters net of expected inflation over the same period. Expected inflation is assumed to be inflation over the past year from quarter $t-8$ to $t-4$. Inflation is calculated as the log change (times 100). Quarterly counts of firm bankruptcies are obtained from the \textit{Vierteljahrshefte zur Statistik des Deutschen Reichs Herausgegeben vom Statistischen Reichsamt}. Inflation of wholesale prices as reported in \textit{Zahlen zur Geldentwertung}.}
	
\end{figure}

\begin{figure}[!htpb]
\caption{Distribution of Firm Leverage in 1917 and 1924.}
        \begin{center}
          \label{fig:lev_hist}
   
\includegraphics[width=0.7\textwidth]{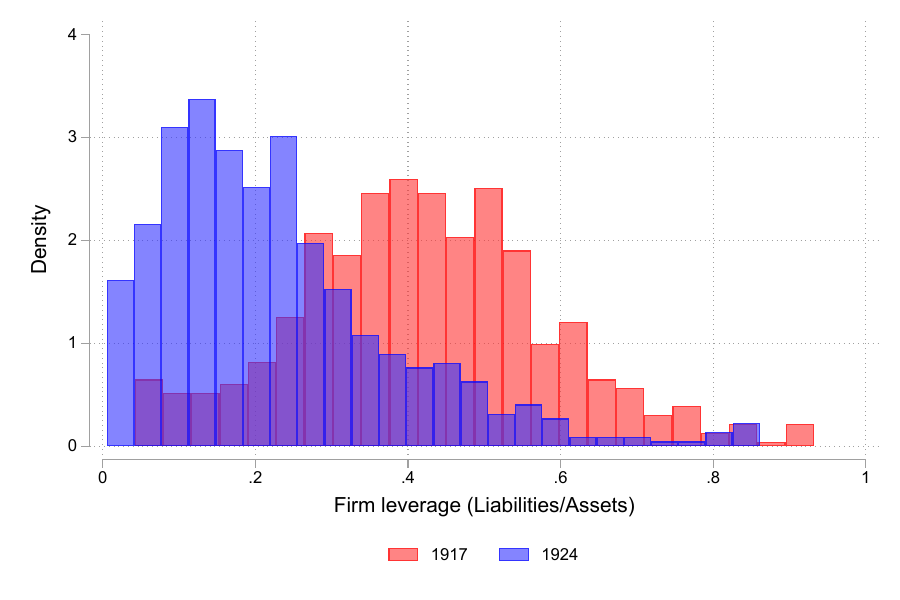}
\end{center}
\vspace{-.7cm}
\footnotesize{\textit{Notes}: This figure shows the distribution of firm book leverage before of the postwar inflation in 1917 and in the aftermath of the hyperinflation in 1924. Leverage is defined as $\sfrac{(Assets-Equity)}{Assets}$. }
\end{figure}

\begin{figure}[htpb]
	\begin{center}
	\caption{Interval between Wage Adjustment Falls during the Inflation: Evidence from Industry-Level Wages. \label{fig:price_interval_additional_wages}}

   \subfloat[Wage adjustments. \label{fig:price_interval_extended}]{\includegraphics[width=0.49\textwidth]{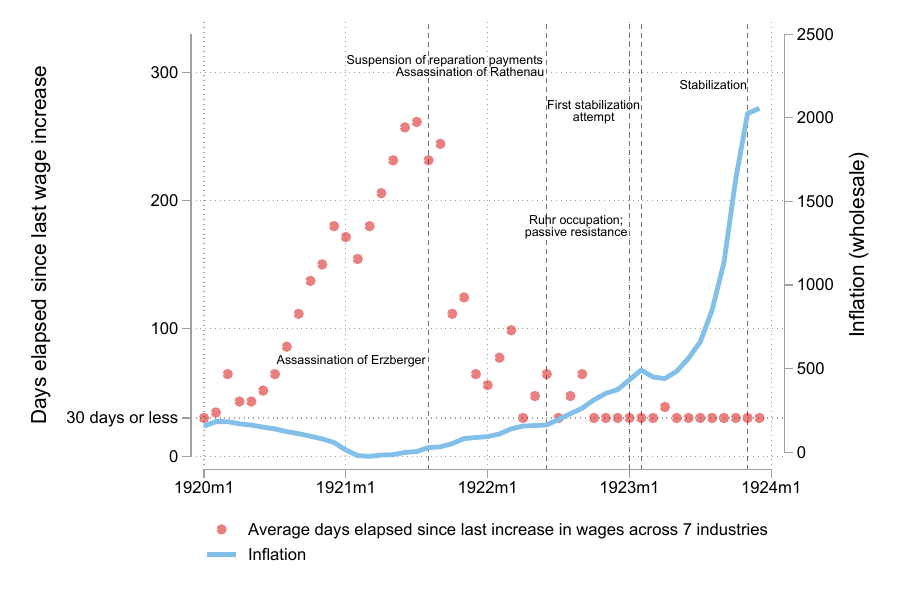}}
   \subfloat[Wage adjustments, by industry.]{\includegraphics[width=0.49\textwidth]{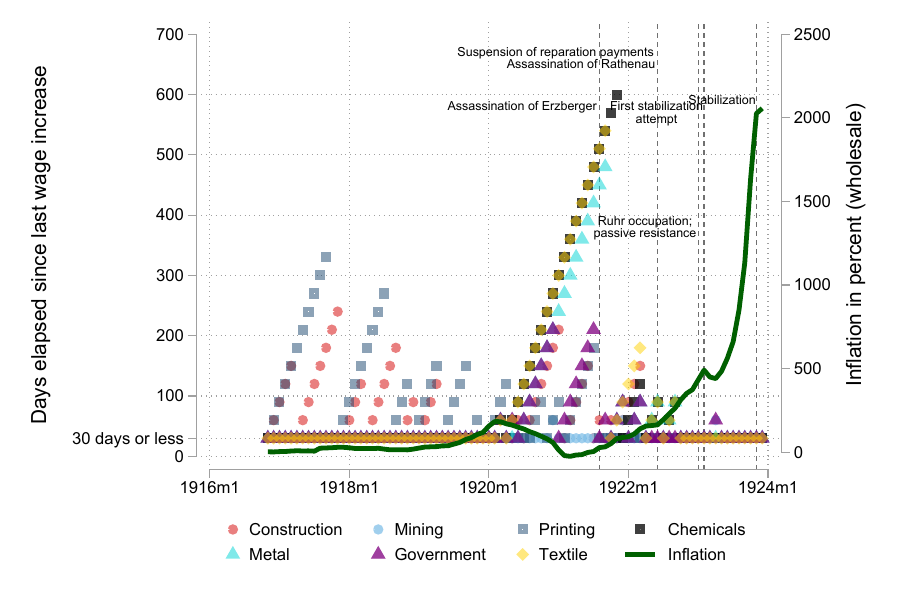}}
    \end{center}
	
    \footnotesize{\textit{Notes}: This figure plots the duration of unchanged wages over time. Wages are as reported in \textit{Zahlen zur Geldentwertung in Deutschland von 1914 bis 1923} and \textit{Wirtschaft und Statistik} (various issues). Inflation is defined as the difference between the log of the wholesale price level in month $t$ and month $t-12$, times 100.}
\end{figure}

\begin{figure}[htpb]
	\begin{center}
	\caption{Interval between Price Adjustment Falls during the Inflation: Evidence from Cost-of-Living Index Prices. \label{fig:price_interval_additional}}

    \subfloat[Frequency of price adjustments by  type of  good.]{\includegraphics[width=0.49\textwidth]{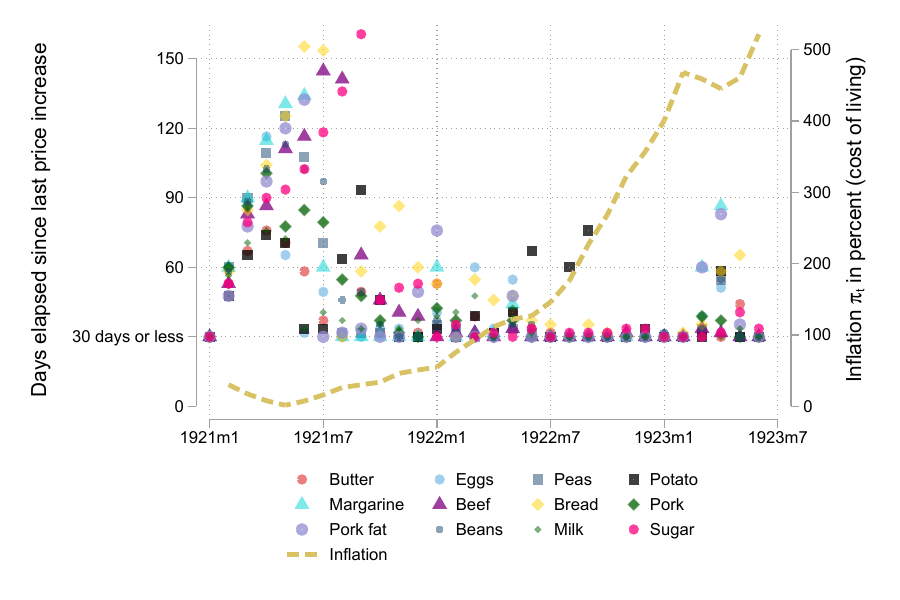}}
   \subfloat[Price adjustment by product for Berlin in 1923.]{\includegraphics[width=0.49\textwidth]{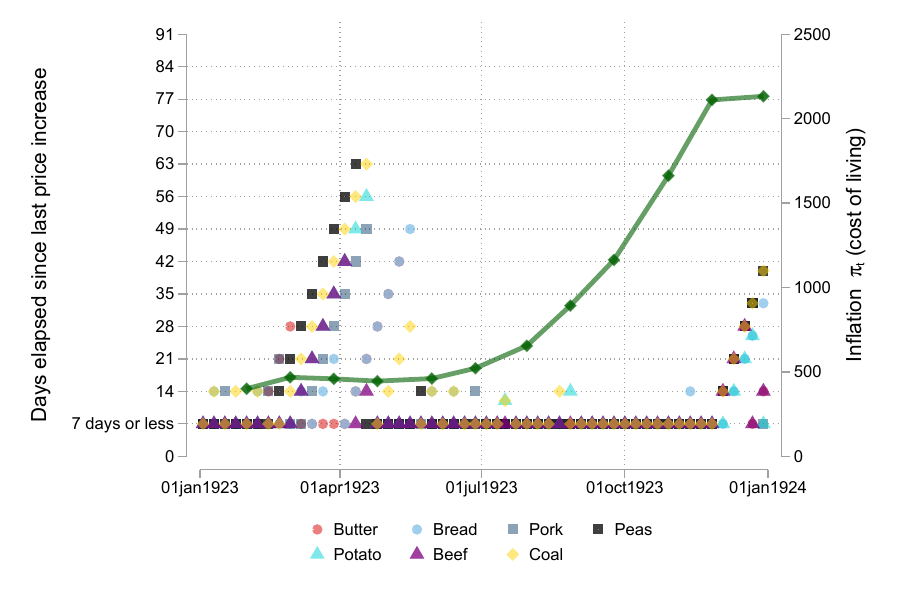}}
     
     \subfloat[Price adjustments of 12 consumption goods by city.]{\includegraphics[width=0.5\textwidth]{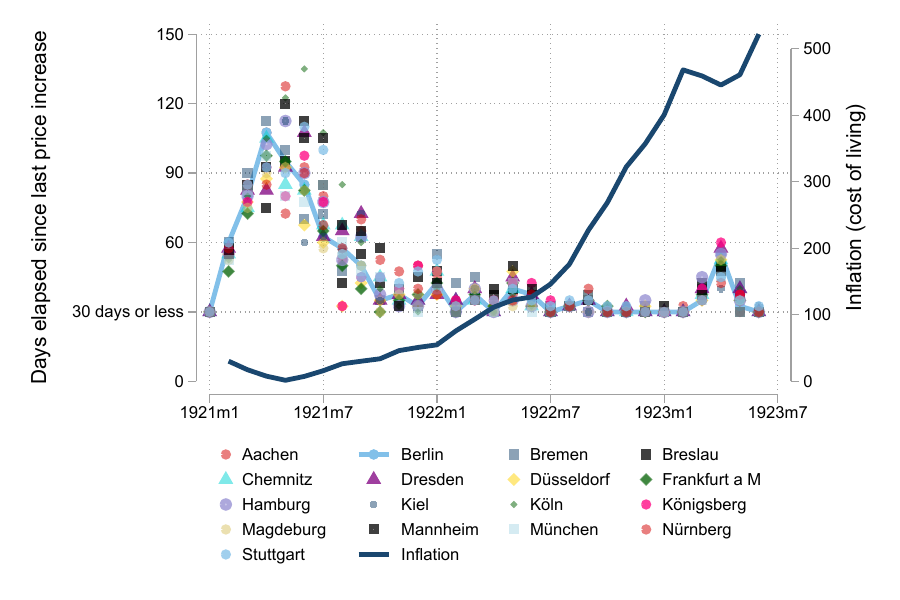}}
     \subfloat[Price adjustments for 95 wholesale-traded products.]{\includegraphics[width=0.5\textwidth]{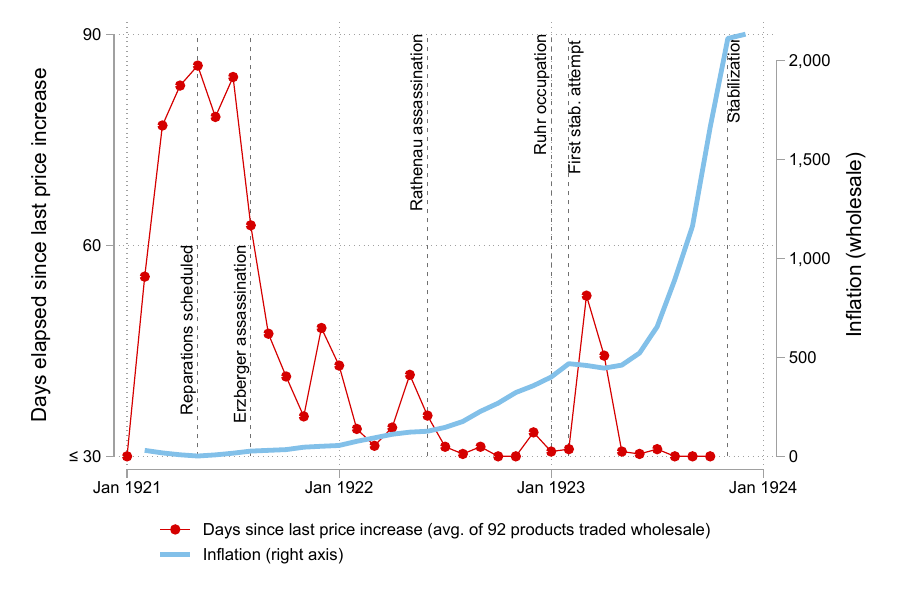}}
    
    \end{center}
	
    \footnotesize{\textit{Notes}: This figure plots the duration of retail product prices, for products underlying the cost-of-living index. Retail prices are as reported in \textit{Zahlen zur Geldentwertung in Deutschland von 1914 bis 1923} and \textit{Wirtschaft und Statistik} (various issues). Inflation is defined as the difference between the log of the cost-of-living index in month $t$ and month $t-12$, times 100.}
\end{figure}

\begin{figure}[htpb]
	\begin{center}
	\caption{Interval between Price Adjustment Falls during the Inflation: Evidence from Newspaper Prices. \label{fig:price_interval_newspapers}}

   \subfloat[Daily newspaper prices, Berliner Börsen Zeitung.]{\includegraphics[width=0.5\textwidth]{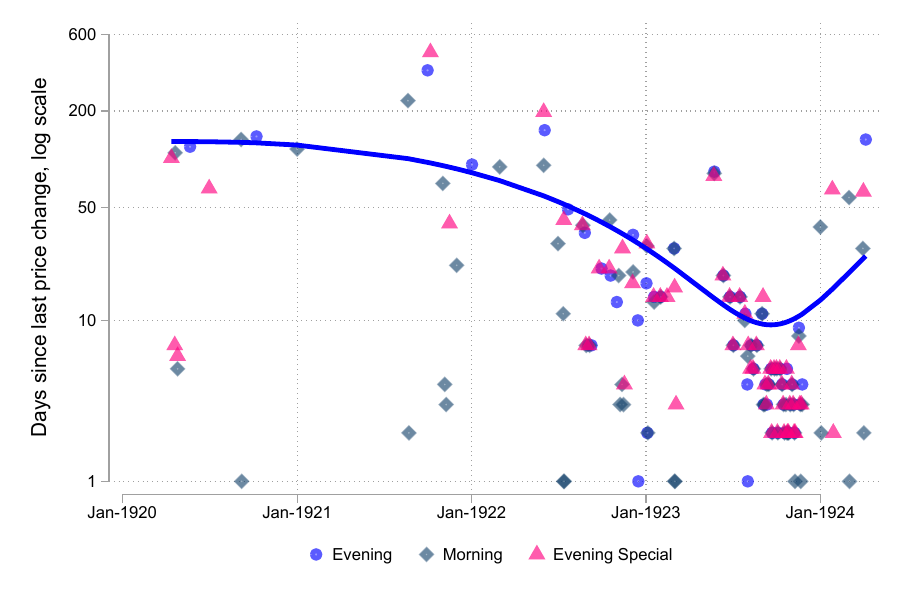}}
   \subfloat[Daily newspaper prices, Berliner Tageblatt und Handels-Zeitung.]{\includegraphics[width=0.5\textwidth]{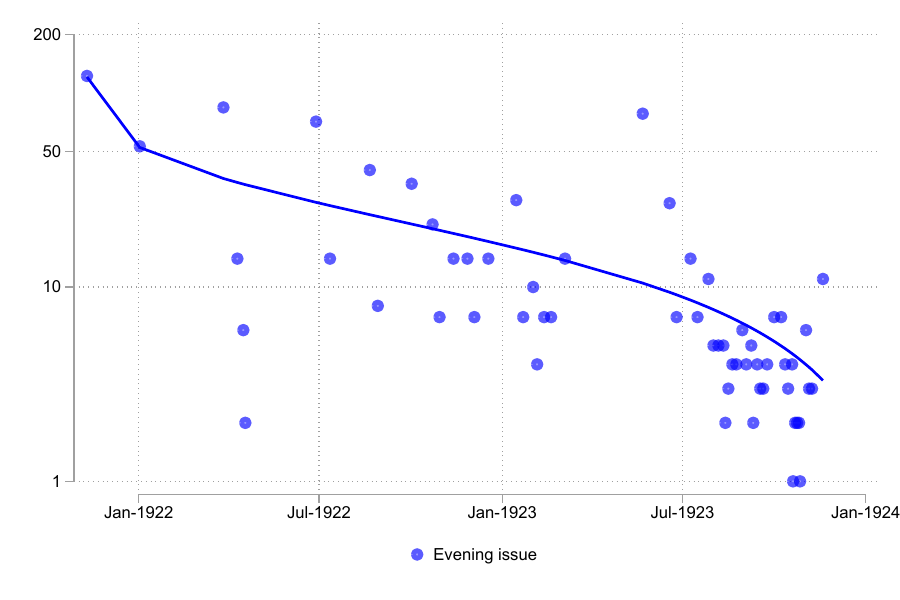}}
   	\end{center}
	
    \footnotesize{\textit{Notes}: This figure plots the duration of unchanged prices for various issues of two German newspapers, the \textit{Berliner Börzen Zeitung} and the \textit{Berliner Tageblatt und Handels-Zeitung}. Daily newspaper prices are hand-collected from scans of the newspapers.}
\end{figure}

\begin{figure}[htpb]
\caption{Real Wages Declined Relative to 1913 during Germany's Inflation, Especially for High-skilled Workers.}
\begin{center}
\subfloat[Real wages of state employees]{\includegraphics[width=0.5\textwidth]{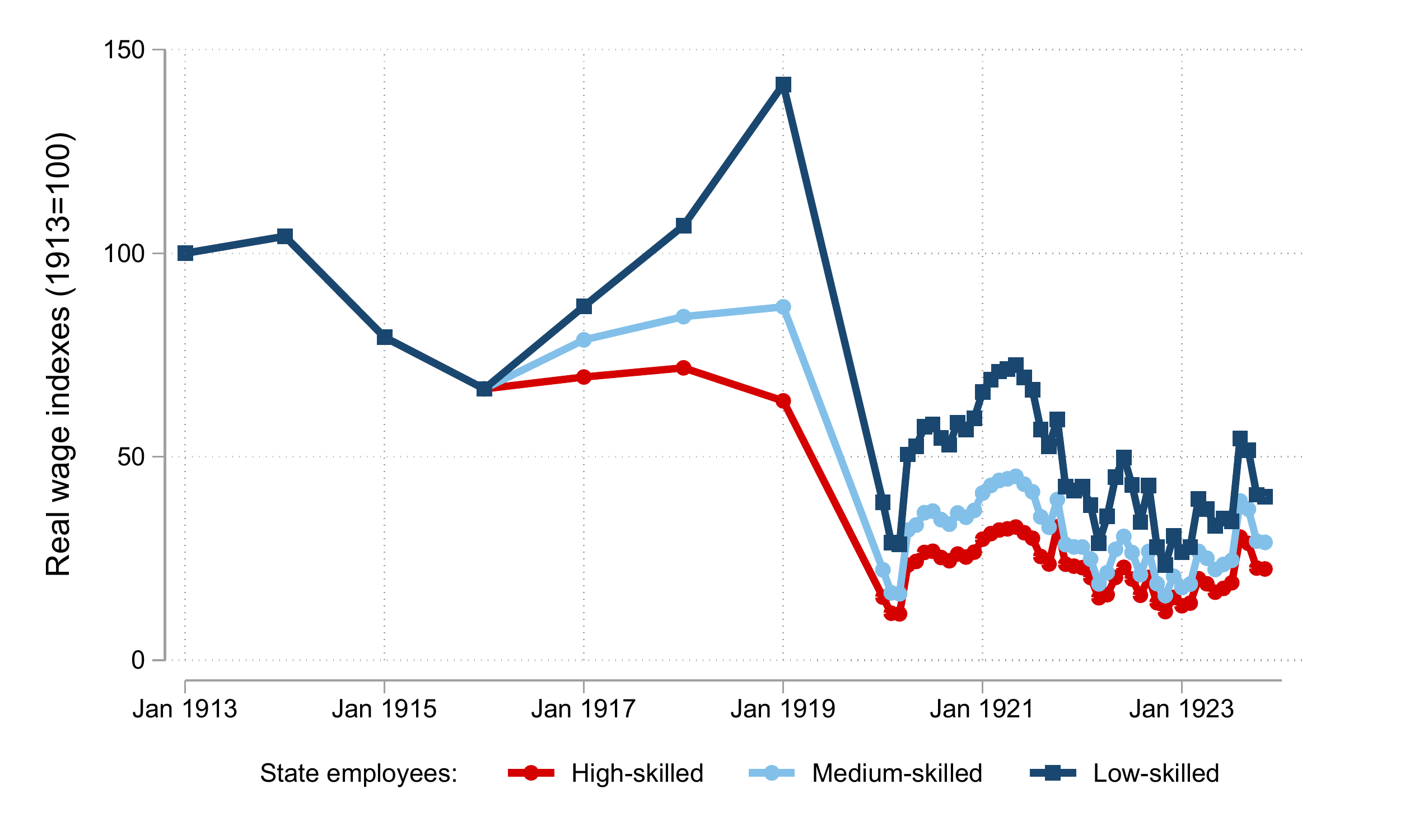}} \hfill
\subfloat[Real wages of high and low skilled workers, six-industry average]{\includegraphics[width=0.5\textwidth]{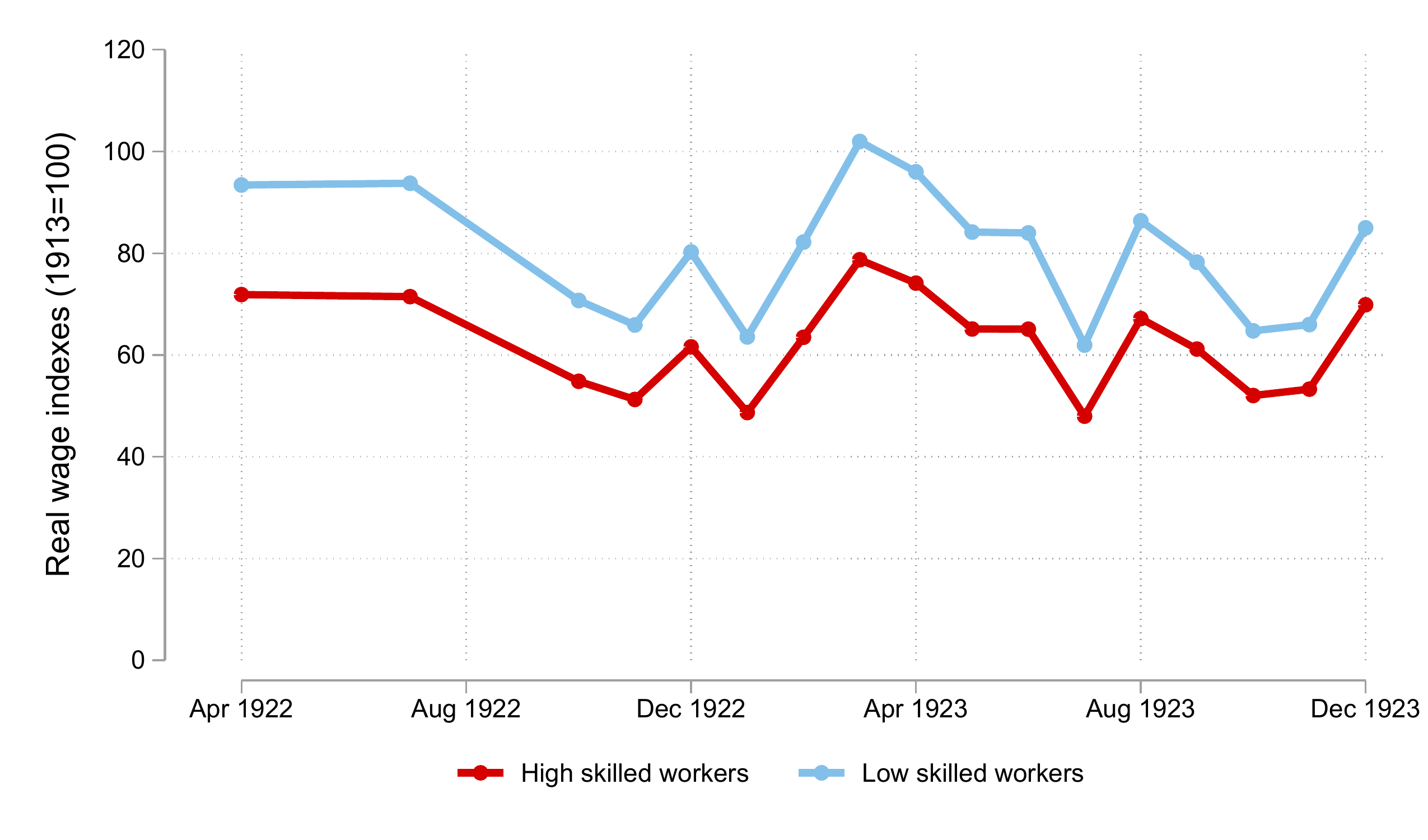}}
\subfloat[Real wages of public railroad workers, Ruhr workers, and book printers]{\includegraphics[width=0.5\textwidth]{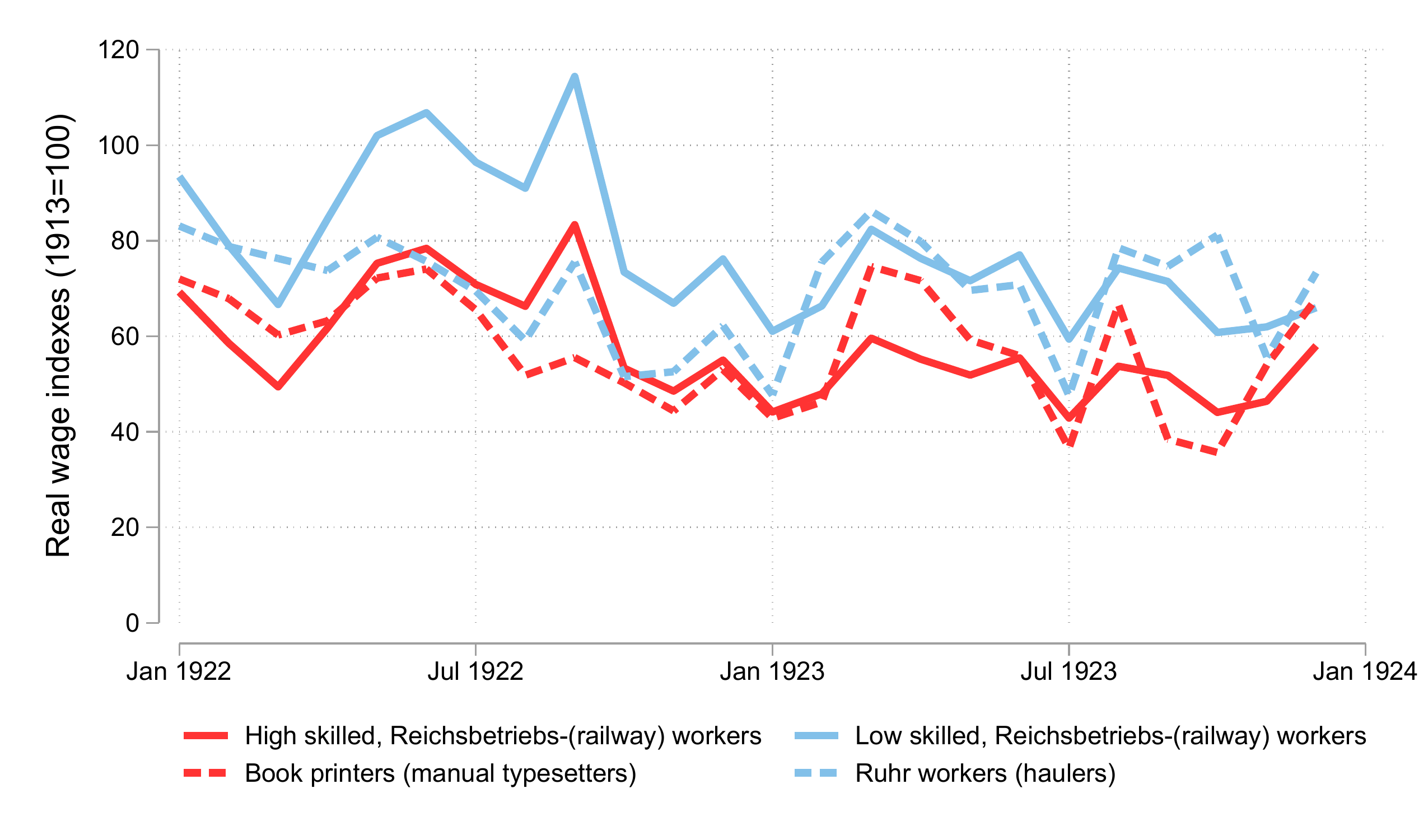}}
\end{center}
\footnotesize{\textit{Notes}: This figure plots the evolution of real wages for various groups of workers and industries. Wage data are from \textit{Wirtschaft und Statistik}. Real wages are deflated by wholesale prices.}
	 \label{fig:real_wages}
\end{figure}

\clearpage

\clearpage

\begin{figure}[!htpb]
    \caption{Employment Dynamics across Low and High Leverage Firms: Robustness.}
    \label{fig:leverage_employment_dynamics_robustness}
        \begin{center}
       \subfloat[Liabilities/Assets in 1917: Log growth.]{
        \includegraphics[width=0.49\textwidth]{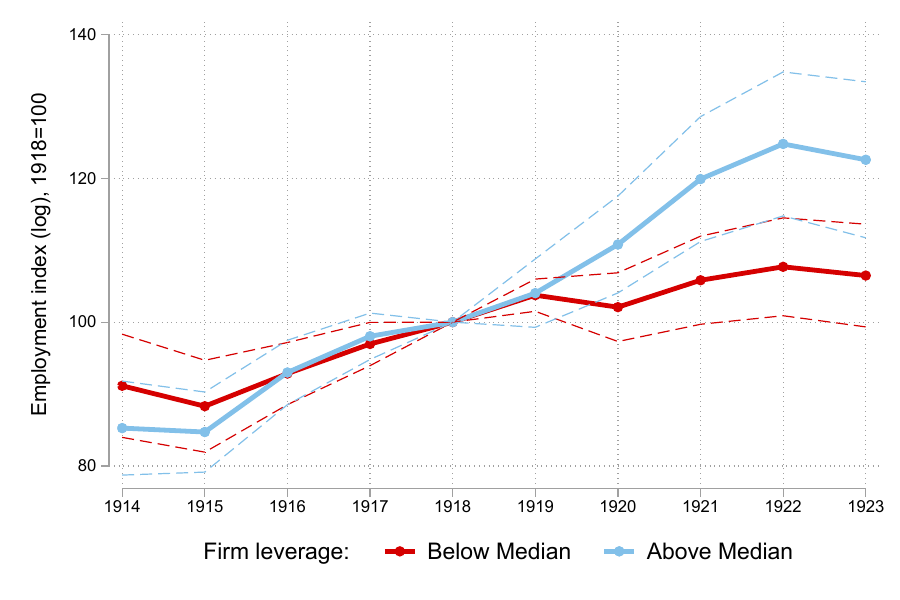}}  
     \subfloat[Grouping by terciles by Liabilities/Assets in 1917.]{
        \includegraphics[width=0.49\textwidth]{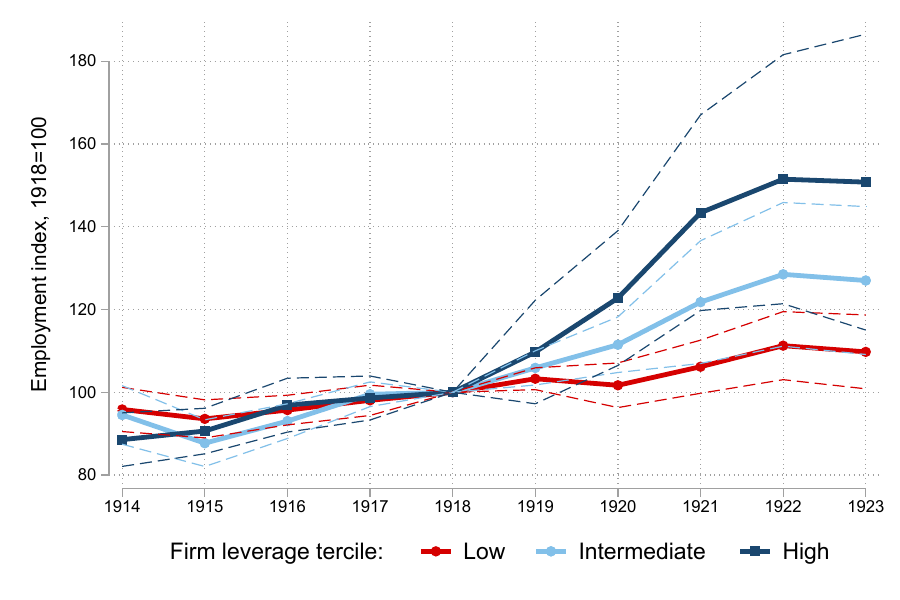}}  

          \subfloat[Above/below median of Debt/Assets in 1917]{
        \includegraphics[width=0.49\textwidth]{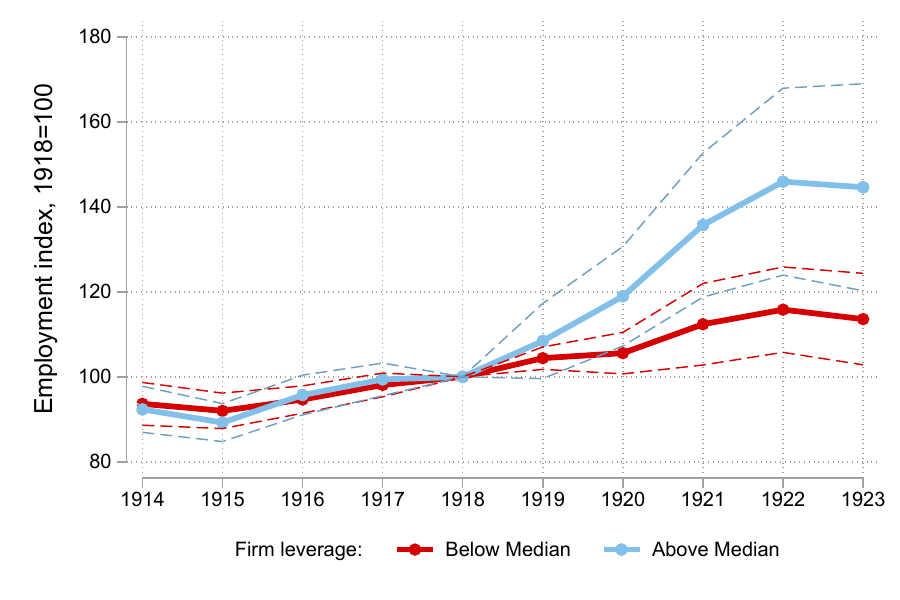}}  \hfill
     \subfloat[Debt/Assets in 1917: Log growth.]{
        \includegraphics[width=0.49\textwidth]{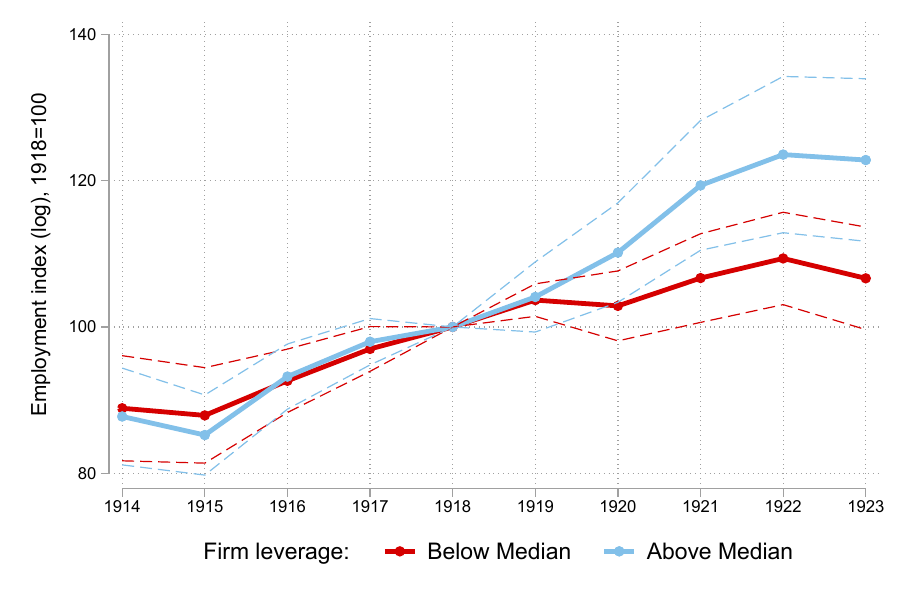}}

     \subfloat[Grouping into terciles by Debt/Assets in 1917.]{
        \includegraphics[width=0.49\textwidth]{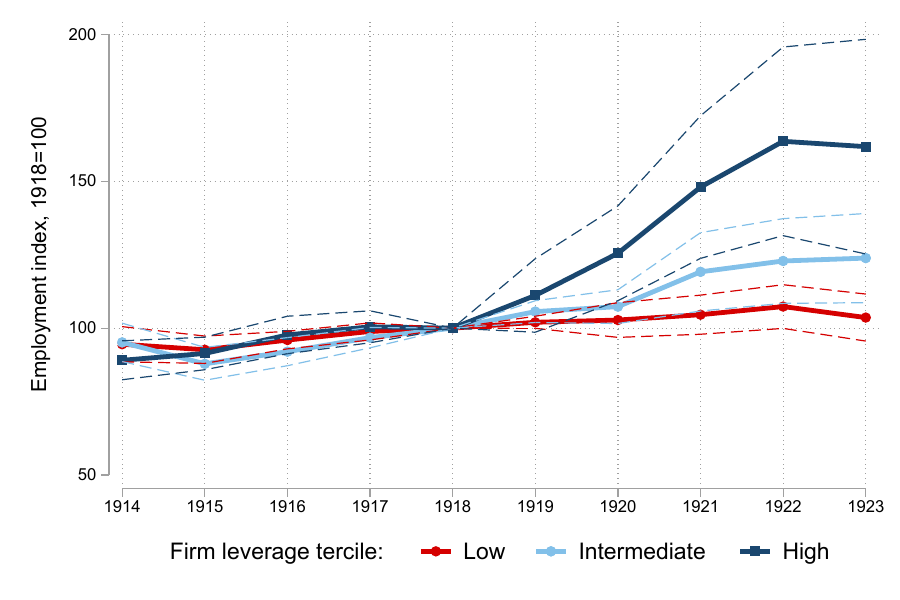}}

\end{center}
\vspace{-.7cm}
\footnotesize{\textit{Notes}: This figure presents robustness tests for \Cref{fig:leverage_employment_dynamics}(a). Panel (a) presents robustness to computing firm growth using log changes (times 100), rather than standard growth rates. We again index all firms to 100 in 1918. Panel (b) presents results when grouping firms by terciles of liabilities-to-assets in 1917. Panel (c) presents results grouping by debt-to-assets in 1917. Panel (d) reports results using log changes (times 100) when grouping by debt-to-assets in 1917. Panel (e) reports results when grouping by terciles of debt-to-assets in 1917. Dashed lines represent 95\% confidence bands. }

\end{figure}

\begin{figure}[!ht]
    \caption{Firm Leverage and Firm Employment: Robustness to Inclusion of Various Sets of Controls.  \label{fig:dd_emp_robust_all}}
        \begin{center}
        \subfloat[Measuring $Leverage_i$ with liabilities-to-assets in 1917.]{	\includegraphics[width=.6\textwidth]{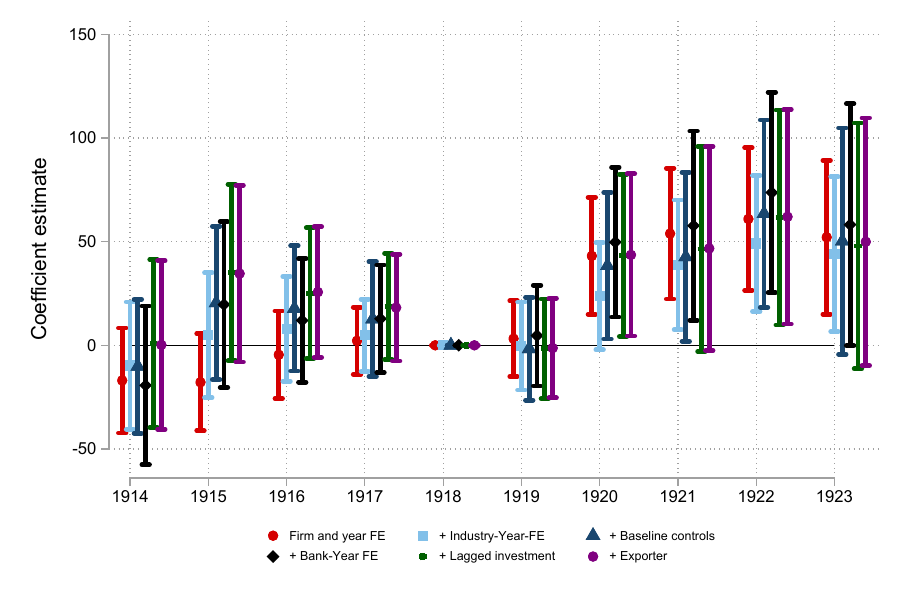}  }   
	
	 \subfloat[Measuring $Leverage_i$ with debt-to-assets in 1917.]{
	\includegraphics[width=.6\textwidth]{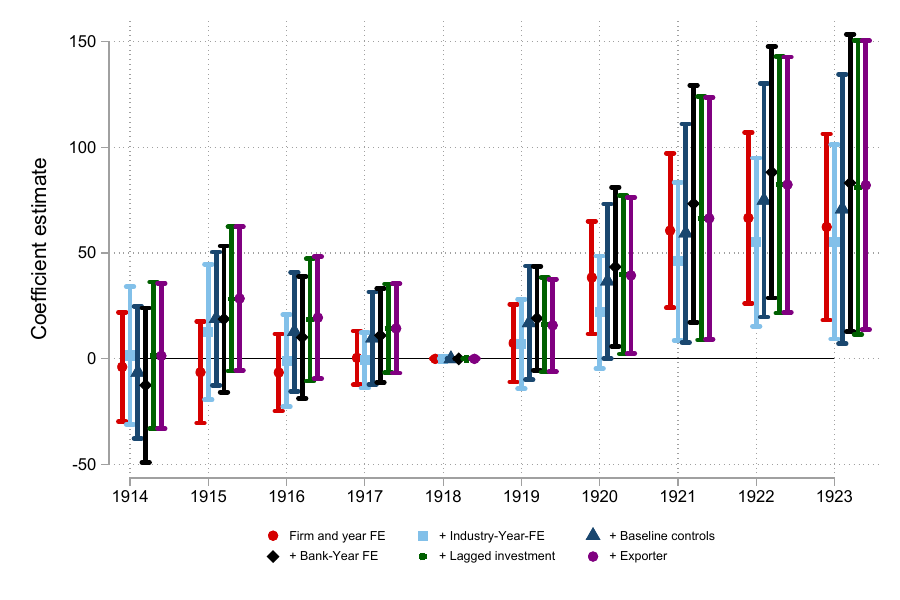}   
	}
	  \end{center}
        \footnotesize{\textit{Notes}: This figure is similar to \Cref{fig:leverage_employment_dynamics}, but it reports estimates for various other control sets. ``Firm and year FE'' are from estimates of \eqref{eq:dynamic} with only firm and year fixed effects. ``+ Industry-Year FE'' are from a specification that adds industry-year fixed effects. ``+ Baseline controls'' further include firm controls interacted with year fixed effects. Firm controls are listed in the note to \Cref{fig:leverage_employment_dynamics}. ``+ Bank-Year FE'' is a specification that adds fixed effects for a connection to one of seven major banks. It also includes a fixed effect for firms with a connection to a bank that is not one of the major banks and a fixed effect for firms without any banking connections. This specification also includes the distance to Berlin, another proxy of exposure to shifts in credit supply. ``+ Lagged investment'' is from a specification that also controls for investment in 1917 and 1918. Investment is defined as $\frac{FixedAssets_{it}- FixedAssets_{it-1}}{TotalAssets_{it-1}}.$ ``+ Export'' also controls for firm export status in 1921. Bank fixed effects and all controls are always interacted with a set of year fixed effects. Error bars represent 95\% confidence intervals from standard errors clustered at the firm level.
        }
\end{figure}

\begin{figure}[!ht]
    \caption{ Employment Dynamics across Low and High Leverage Firms throughout the Post-Inflation Boom.}
    \label{fig:cyclicality}
        \begin{center}
       \subfloat[Unemployment rate.]{
        \includegraphics[width=0.49\textwidth]{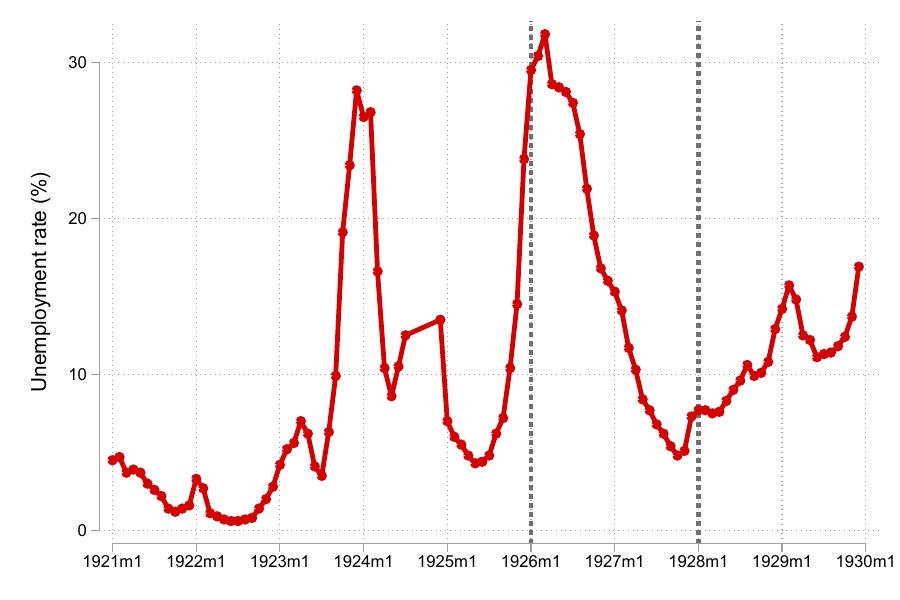}
        }

       \subfloat[Leverage in 1917.]{
        \includegraphics[width=0.49\textwidth]{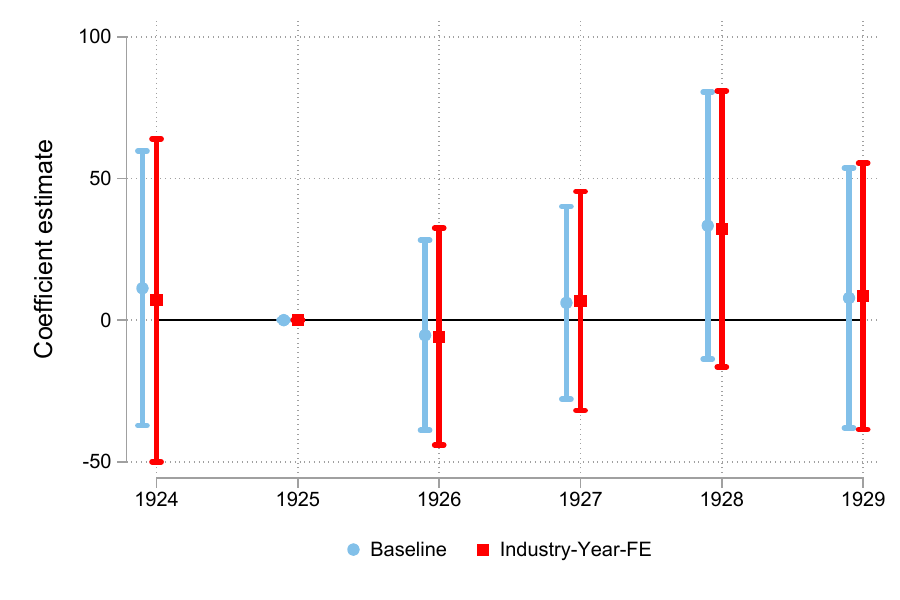}}  
        \hfill
        \subfloat[Leverage in 1925.]{
        \includegraphics[width=0.49\textwidth]{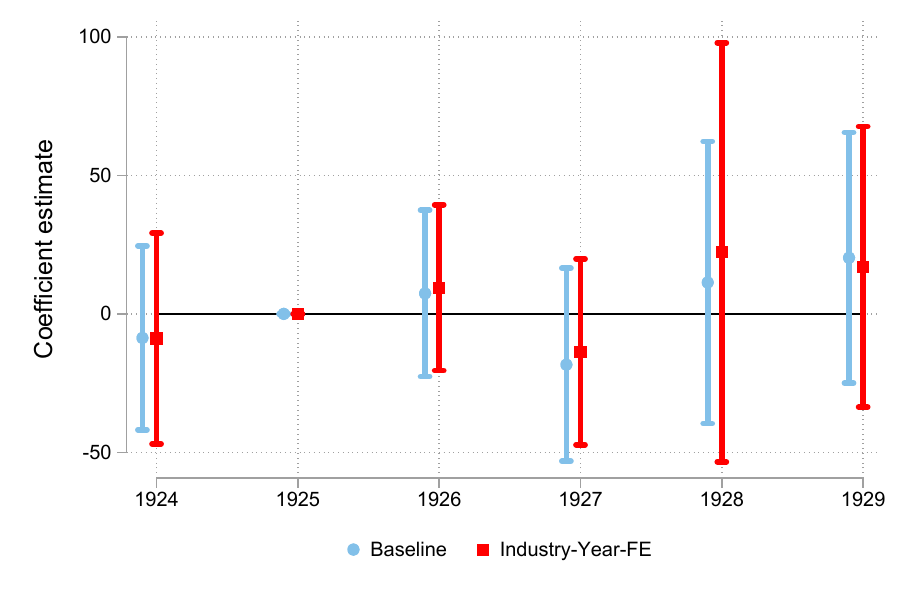}}  
\end{center}
\vspace{-.7cm}
\footnotesize{\textit{Notes}: Panel (a) plots the unemployment rate in Germany from 1921 to 1931. The figure illustrates the post-inflation economic expansion from 1926 to 1928. We mark the beginning and end of this period with vertical dashed lines. Panels (b) and (c) present the sequence of estimates  $\{\beta_k\}$ from estimating
  \[\ln(Employment_{it}) = \alpha_i + \delta_{st} + \sum_{k\neq 1925} \beta_k Leverage_{i} \mathbf{1}_{k=t}  + \sum_{k\neq 1925}  X_{i} \Gamma_k \mathbf{1}_{k=t} + \epsilon_{it}.\] In panel (b), we measure $Leverage_{i}$ as the ratio of liabilities to assets in 1917. In panel (c), we measure $Leverage_{i}$ as the ratio of liabilities to assets in 1925, right before the 1926-1928 expansion.  The baseline specification includes firm and year fixed effects. "Industry-Year FE" refers to a specification that also controls for industry-by-year fixed effects. Errors bars represent 95\% confidence intervals based on standard errors clustered at the firm level.  }

\end{figure}

\begin{figure}[!ht]
    \caption{ Firm Leverage and Firm Employment: Robustness to Balanced Sample Restriction.  \label{fig:dd_emp_robust_samples}}
        \begin{center}
	\includegraphics[width=.8\textwidth]{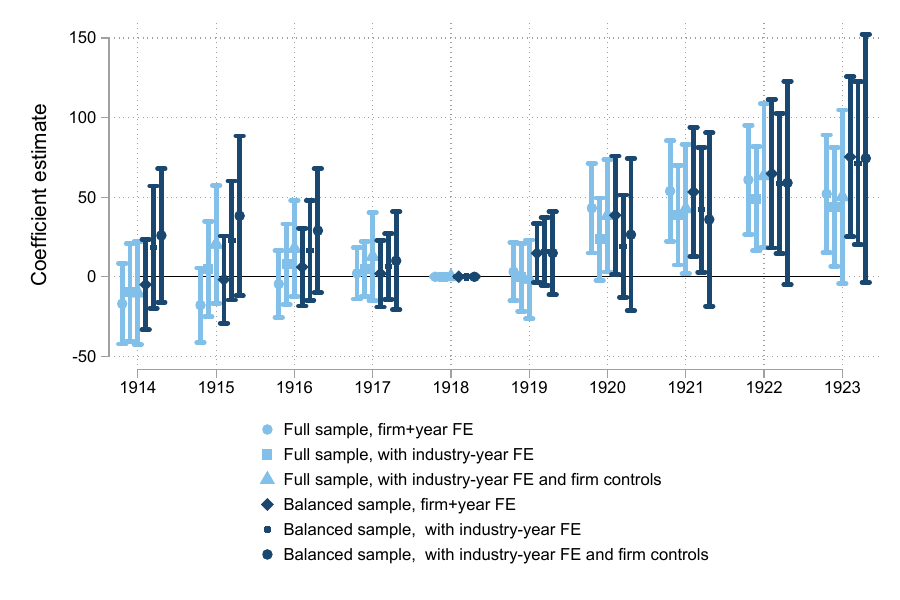}     
	\end{center}
        \footnotesize{\textit{Notes}: This figure is similar to \Cref{fig:leverage_employment_dynamics}, but it reports estimates on a balanced panel of firms. Estimates denoted by ``Full sample" are equivalent to the estimates in \Cref{fig:leverage_employment_dynamics}(b). Estimates denoted by ``Balanced sample'' restrict to the sample of firms that report employment in \textit{Saling's} in every year between 1916 and 1923. Error bars represent 95\% confidence intervals from standard errors clustered at the firm level. 
        
        }
\end{figure}

\begin{figure}[!ht]
	\begin{center}
	\caption{Firm Leverage, Interest Expenses, and Production Expenses. }\label{fig:dd_expenses}
	
\subfloat[Interest Expenses.]{
\includegraphics[width=0.49\textwidth]{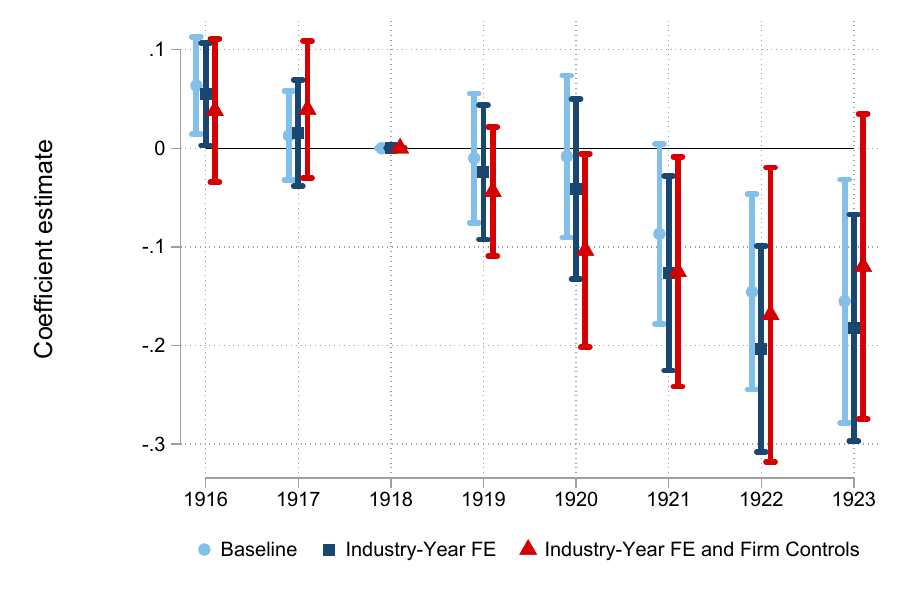}
}
\hfill
\subfloat[Production Expenses.]{\includegraphics[width=0.49\textwidth]{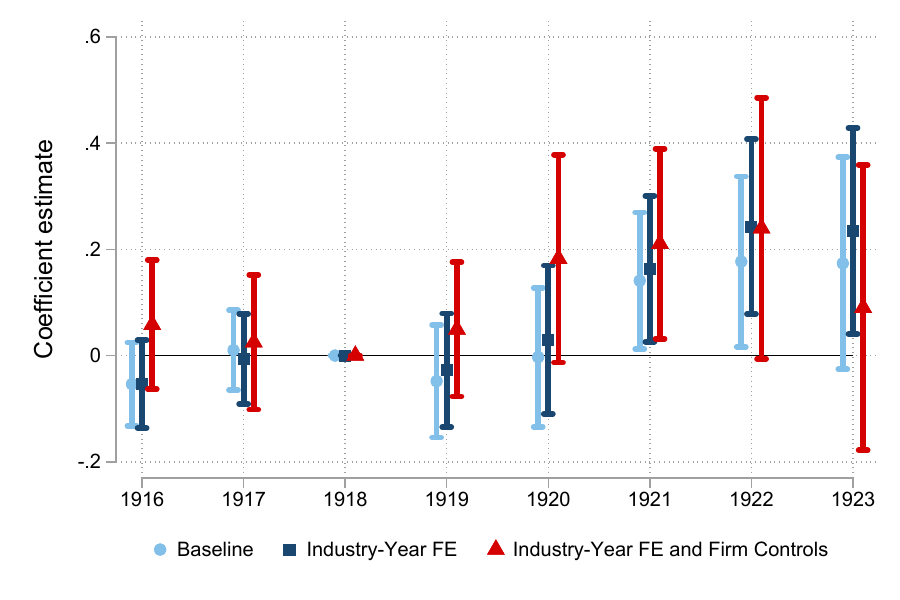}}
\end{center}
\vspace{-.7cm}
\footnotesize{\textit{Notes}:} This figure plots the sequence of estimates  $\{\beta_y\}$ from estimating a version of equation \eqref{eq:dynamic} with the ratio of interest expenses to total expenses (panel a) or production expenses to total expenses (panel b) as the dependent variable. $Leverage_i$ is measured with liabilities-to-assets in 1917. Firm-level control variables are firm size (log of assets), the share of fixed assets to total assets, free cash flow to assets, and profit margin (EBIT-to-revenue), all measured in 1917, and Tobin's Q as of 1918. Error bars represent 95\% confidence intervals based on standard errors clustered at the firm level.
\end{figure}

\begin{figure}[!ht]
    \caption{The Prevalence of Long-Term Bonds: Origination Year, Repayment Start Year, and Final Maturity for Outstanding Bonds of Nonfinancial Firms in 1918 and 1919.}
        \begin{center}

        \includegraphics[width=0.85\textwidth]{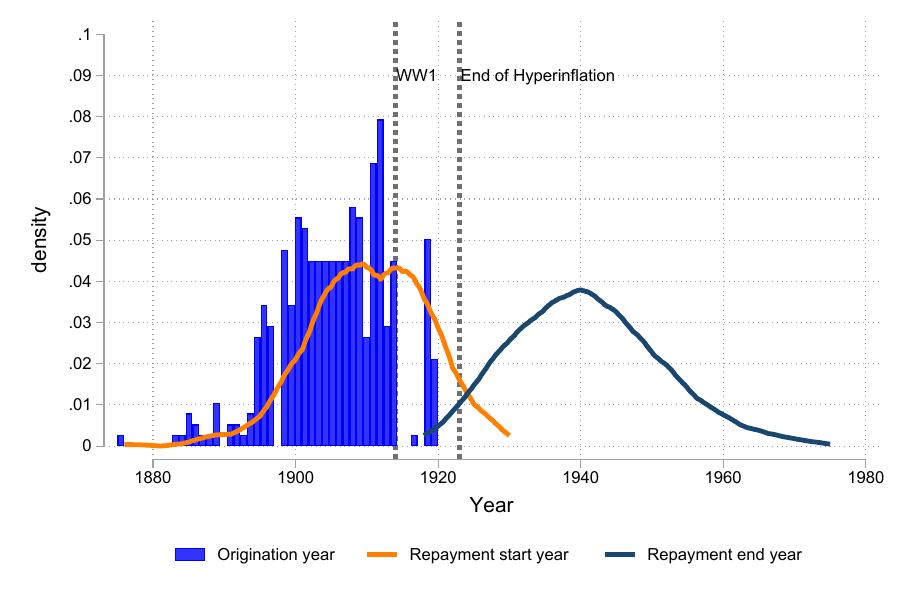}

        \end{center}
        \footnotesize{\textit{Notes}: This figure shows the origination year, repayment start year, and final maturity for outstanding bonds reported by nonfinancial firms in 1918 and 1919. Data obtained from \textit{Saling's B\"{o}rsenjahrbuch} published in 1919 and 1920. $N=417$. 51\% of firms in the sample report information on at least one bond.}
	 \label{fig:bonds_maturity}
\end{figure}

\begin{figure}[!ht]
    \caption{High Leverage Firms' Stock Returns Outperformed Low Leverage Firms during the Inflation.}
    \label{fig:hml_portfolio}
        \begin{center}
	\subfloat[Cumulative Return on High-minus-Low Leverage Porftolio: Monthly Returns Data.]{
	\includegraphics[width=.65\textwidth]{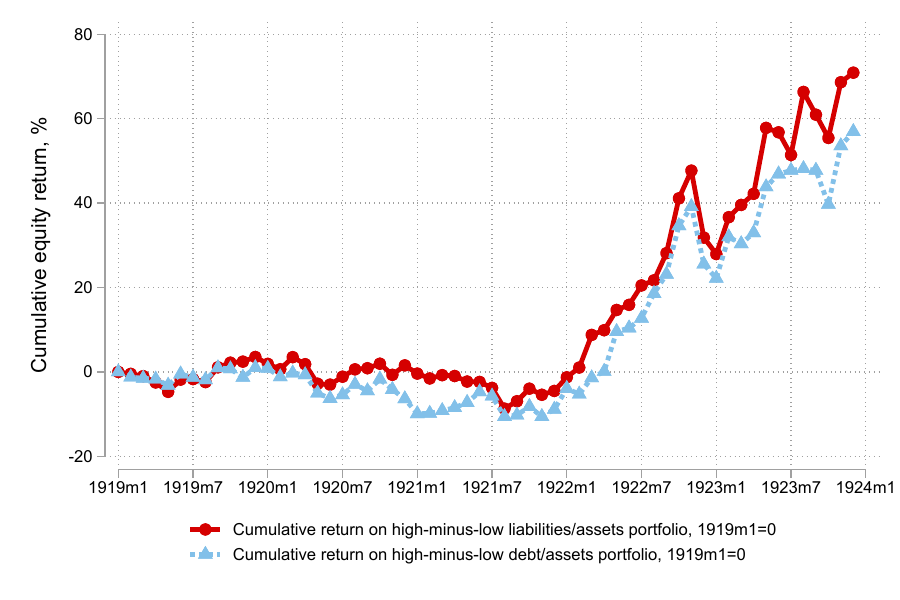}}
	
	\subfloat[Cumulative Return on High-minus-Low Leverage Portfolio vs Cumulative Inflation..]{\includegraphics[width=0.7\textwidth]{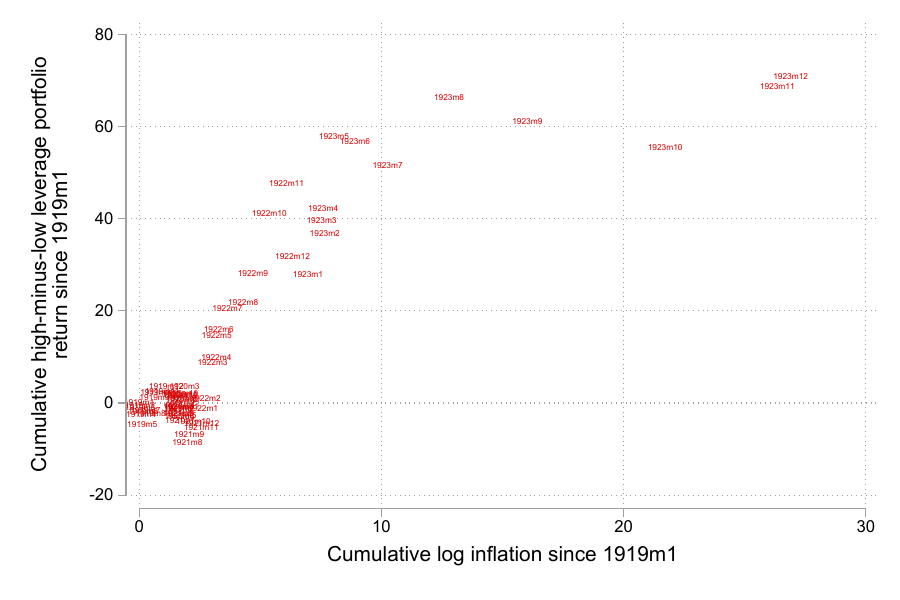} }

	    \end{center}
        \footnotesize{\textit{Notes}: Panel (a) plots the cumulative return on a portfolio that goes long firms in the top quintile of leverage and short firms in the bottom quintile of leverage. Returns are cumulative related to January 1919. Returns are based on portfolios that are resorted at the end of year $t-1$ based on leverage reported in year $t-1$. Panel (b) plots the cumulative return on the high-minus-low leverage portfolio (based on sorting firms by liabilities-to-assets) against cumulative wholesale price inflation. Both variables are measured relative to January 1919. 
        }
\end{figure}

\begin{figure}[!ht]
    \caption{Dividend Yield, 1919-1923.}
        \begin{center}
        \includegraphics[width=0.65\textwidth]{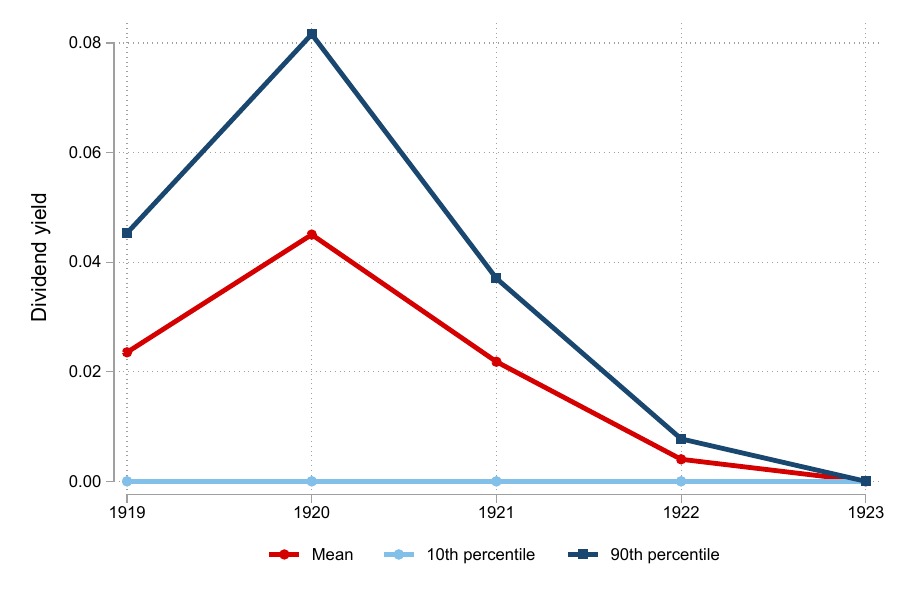}
        \end{center}
        \footnotesize{\textit{Notes}: This figure plots the mean, 10\textsuperscript{th} percentile, and 90\textsuperscript{th} percentile of the dividend yield distribution from 1919 to 1923. The dividend yield for a firm is defined as the total dividend per share paid out in year $t$ relative to the share price at the end of year $t-1$. The analysis is based on firm-level data collected from the \textit{Berliner Börsen Zeitung}.}
	 \label{fig:dividend_yield}
\end{figure}

\begin{figure}[htpb]
    \caption{Reichsbank Balance Sheet in Goldmarks.}
        \begin{center}

	\subfloat[1914-1923]{\includegraphics[width=0.7\textwidth]{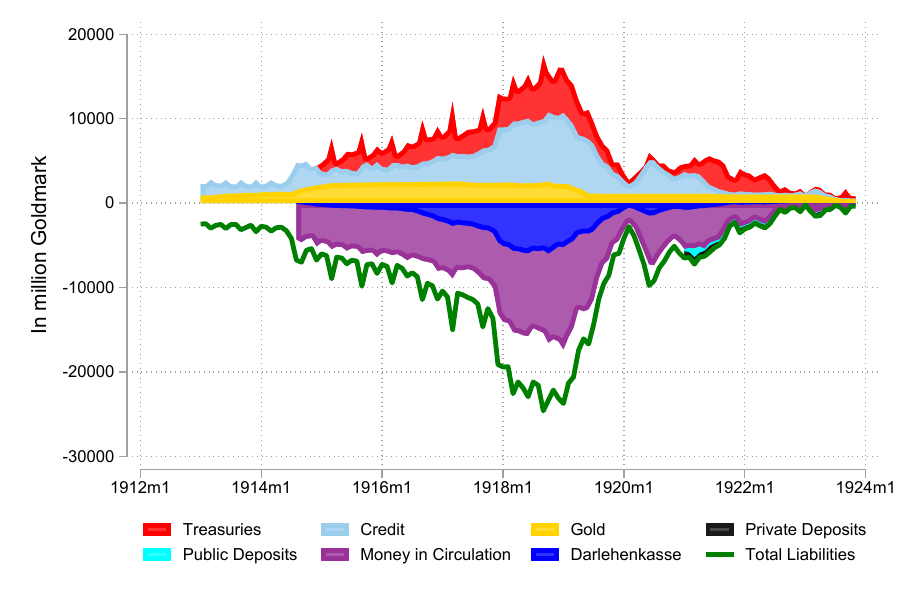}}
	
	\subfloat[1919-1923]{\includegraphics[width=0.7\textwidth]{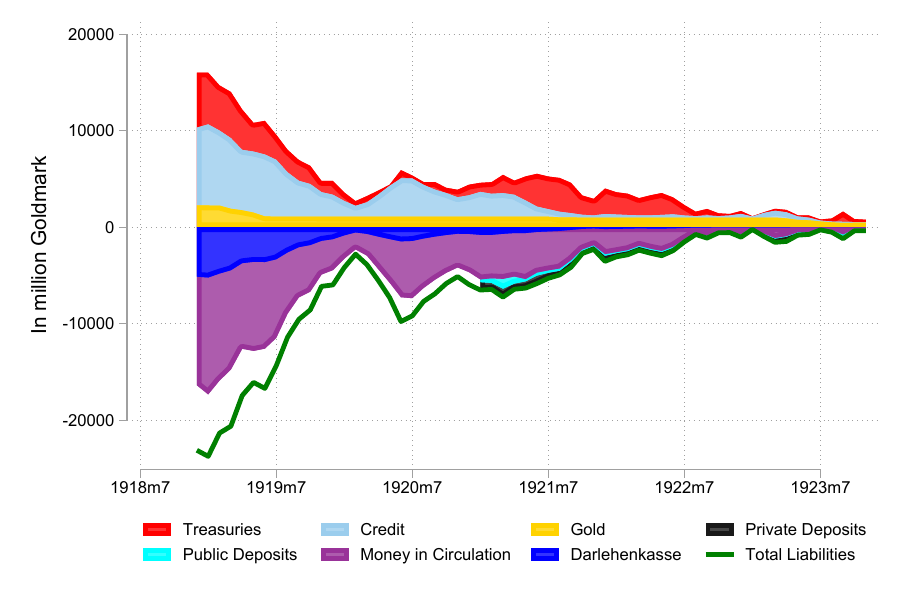}}

        \end{center}
        \footnotesize{\textit{Notes}: Data are from \textit{Zahlen zur Geldentwertung}.}
	 \label{fig:reichsbank_goldmarks}
\end{figure}

\begin{figure}[htpb]
    \caption{Reichsbank Balance Sheet in Papiermarks.}
        \begin{center}

	\subfloat[1914-1921]{\includegraphics[width=0.7\textwidth]{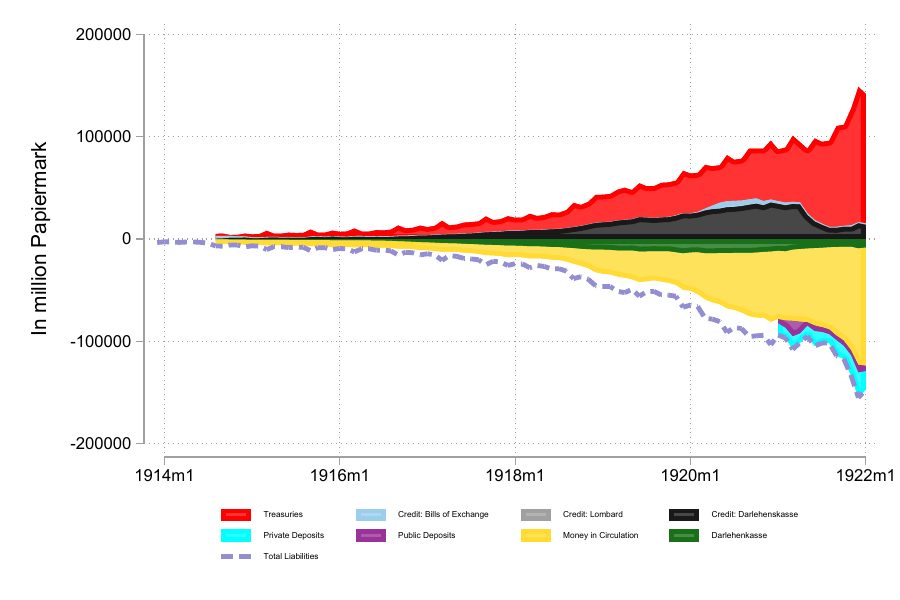}}
	
	\subfloat[1922]{\includegraphics[width=0.49\textwidth]{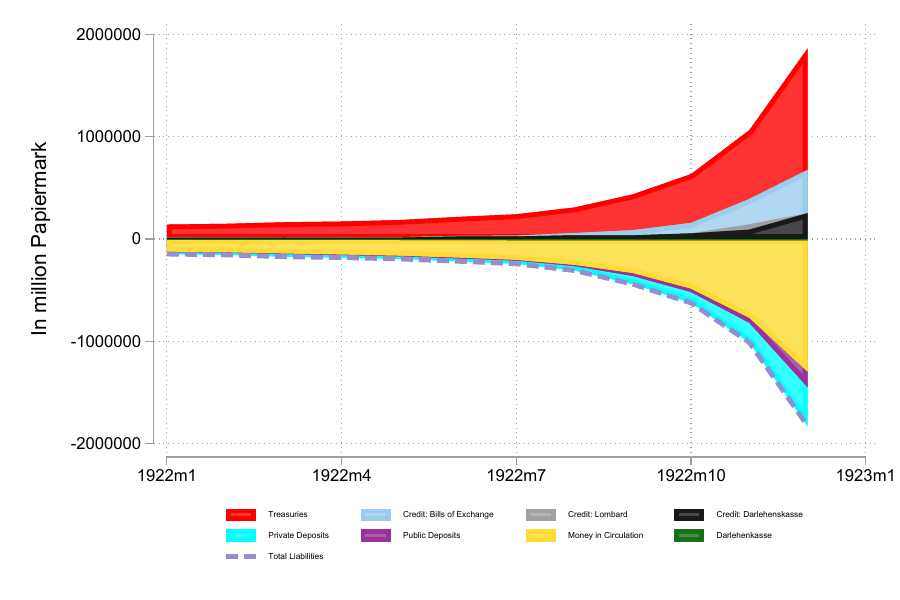}}
        \hfill
    \subfloat[1923]{\includegraphics[width=0.49\textwidth]{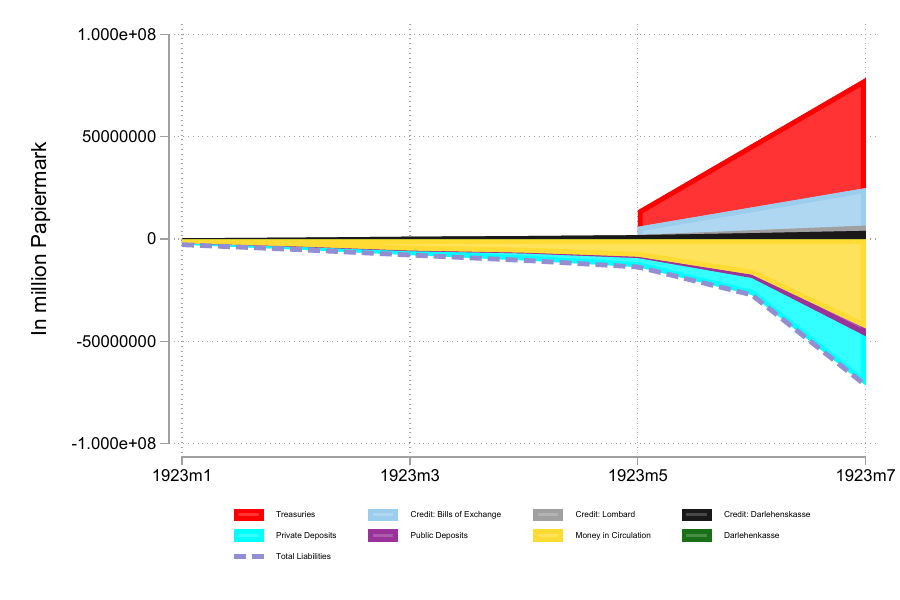}}
        \end{center}
        \footnotesize{\textit{Notes}: Data are as reported in \textit{Zahlen zur Geldentwertung}.}
	 \label{fig:reichsbank_papiermarks}
\end{figure}


\clearpage

\singlespacing

\begin{table}[!ht]
  \centering
  \caption{Time Series Estimates of the Bankruptcy-Inflation Relation.}\label{tab:financial_pc}
        \begin{minipage}{1.0\textwidth}
        \begin{center}
        \scalebox{.8}{
        {
\def\sym#1{\ifmmode^{#1}\else\(^{#1}\)\fi}
\begin{tabular}{l*{6}{c}}
\toprule
                &\multicolumn{1}{c}{(1)}         &\multicolumn{1}{c}{(2)}         &\multicolumn{1}{c}{(3)}         &\multicolumn{1}{c}{(4)}         &\multicolumn{1}{c}{(5)}         &\multicolumn{1}{c}{(6)}         \\
\midrule
Inflation       &    -0.28\sym{**} &    -1.06\sym{***}&    -2.39\sym{**} &    -4.26         &   -0.093\sym{*}  &    -0.43\sym{**} \\
                &   (0.10)         &   (0.29)         &   (0.82)         &   (2.48)         &  (0.044)         &   (0.15)         \\
\addlinespace
\(\text{Inflation}^2\)&                  &  0.00041\sym{**} &                  &    0.011         &                  &  0.00015\sym{*}  \\
                &                  &(0.00015)         &                  &  (0.014)         &                  &(0.000066)         \\
\midrule
Observations    &       20         &       20         &       14         &       14         &        7         &        7         \\
\(R^2\)         &     0.30         &     0.52         &     0.42         &     0.45         &     0.47         &     0.77         \\
Sample          &1919-1923         &1919-1923         &1919-July 1922         &1919-July 1922         &July 1922-1923         &July 1922-1923         \\
Frequency       &Quarterly         &Quarterly         &Quarterly         &Quarterly         &Quarterly         &Quarterly         \\
\bottomrule
\end{tabular}
}
}
        \end{center}
        {\footnotesize \textit{Notes}: This table shows results from estimating the following equation: 
        \[\text{Bankruptcies}_t = \alpha + \beta \times \pi_t+ \epsilon+t,\] were $t$ is quarterly, and $\pi_t$ is the inflation in wholesale prices from $t-4$ to $t$.   Quarterly counts of firm bankruptcies are obtained from the \textit{Vierteljahrshefte zur Statistik des Deutschen Reichs Herausgegeben vom Statistischen Reichsamt}. Inflation of wholesale prices as reported in \textit{Zahlen zur Geldentwertung}. *,**, and *** indicate significance at the 10\%, 5\%, and 1\% level, respectively.
         }
        \end{minipage}
\end{table}%

\begin{table}[!ht]
  \centering
  \caption{ Correlates of Firm Leverage.}\label{tab:lev_correlates}
        \begin{minipage}{1.0\textwidth}
        \begin{center}
        \scalebox{0.9}{
        {
\def\sym#1{\ifmmode^{#1}\else\(^{#1}\)\fi}
\begin{tabular}{l*{7}{c}}
\toprule
                &\multicolumn{7}{c}{\( \text{Liabilities/Assets}_{i,1917} \)}                                                                        \\\cmidrule(lr){2-8}
                &\multicolumn{1}{c}{(1)}         &\multicolumn{1}{c}{(2)}         &\multicolumn{1}{c}{(3)}         &\multicolumn{1}{c}{(4)}         &\multicolumn{1}{c}{(5)}         &\multicolumn{1}{c}{(6)}         &\multicolumn{1}{c}{(7)}         \\
\midrule
\( \ln(\text{Total Assets})_{i,1917} \)&    0.065\sym{***}&                  &                  &                  &                  &                  &    0.063\sym{***}\\
                & (0.0051)         &                  &                  &                  &                  &                  & (0.0063)         \\
\addlinespace
\( \text{Fixed Assets/Total Assets}_{i,1917} \)&                  &    -0.11\sym{***}&                  &                  &                  &                  &   -0.047         \\
                &                  &  (0.032)         &                  &                  &                  &                  &  (0.045)         \\
\addlinespace
\( \text{EBIT margin}_{i,1917} \)&                  &                  &   0.0078\sym{**} &                  &                  &                  &  -0.0013         \\
                &                  &                  & (0.0034)         &                  &                  &                  & (0.0040)         \\
\addlinespace
 \( \text{FCF/Assets}_{i,1917} \)&                  &                  &                  &    -0.15\sym{***}&                  &                  &   -0.097\sym{*}  \\
                &                  &                  &                  &  (0.051)         &                  &                  &  (0.049)         \\
\addlinespace
 \( \text{Tobin's Q}_{i,1918} \)&                  &                  &                  &                  &    0.068\sym{**} &                  &    0.031         \\
                &                  &                  &                  &                  &  (0.034)         &                  &  (0.035)         \\
\midrule
Observations    &      724         &      724         &      670         &      678         &      567         &      724         &      520         \\
\(R^2\)         &     0.20         &    0.024         &    0.011         &    0.016         &   0.0083         &     0.11         &     0.34         \\
Industry Fixed Effects&                  &                  &                  &                  &                  &\checkmark         &\checkmark         \\
\bottomrule
\end{tabular}
}
}
        \end{center}
        {\footnotesize \textit{Notes}: This table shows results from estimating the following cross-sectional regression: 
        \[\text{Leverage}_{i,1917} = \alpha + \beta \times X_{i} + \epsilon_{i},\] where $Leverage_{i,1917}$ is defined as liabilities to assets in 1917 and $X_{i}$ is a firm level variable. Robust standard errors in parentheses. *,**, and *** indicate significance at the 10\%, 5\%, and 1\% level, respectively.}
        \end{minipage}
\end{table}%

\begin{table}[!ht]
  \centering
   \caption{Firm Leverage and Employment---Measuring Leverage as of 1918.}\label{tab:leverage_employment_1918}
        \begin{minipage}{1.0\textwidth}
        \begin{center}
        \scalebox{0.9}{{ \def\sym#1{\ifmmode^{#1}\else\(^{#1}\)\fi} \begin{tabular}{l*{6}{c}}   \toprule  Dependent Variable & \multicolumn{6}{c}{ln(Employment)} \\    \cmidrule(lr){2-7}       
                &\multicolumn{1}{c}{(1)}         &\multicolumn{1}{c}{(2)}         &\multicolumn{1}{c}{(3)}         &\multicolumn{1}{c}{(4)}         &\multicolumn{1}{c}{(5)}         &\multicolumn{1}{c}{(6)}         \\
\midrule
\( \text{Liab/Assets}_{i,1918} \times \mathbf{1}_{t\geq1920} \) &     49.7\sym{***}&     43.9\sym{**} &     42.9\sym{*}  &                  &                  &                  \\
                &   (14.9)         &   (20.4)         &   (22.7)         &                  &                  &                  \\
\( \text{Debt/Assets}_{i,1918} \times \mathbf{1}_{t\geq1920} \) &                  &                  &                  &     52.9\sym{***}&     40.5\sym{*}  &     39.8\sym{*}  \\
                &                  &                  &                  &   (16.1)         &   (20.9)         &   (23.6)         \\
\midrule
Observations    &     2409         &     1947         &     1903         &     2409         &     1947         &     1903         \\
Number of Firms &      361         &      282         &      280         &      361         &      282         &      280         \\
\(R^2_{overall}\)&     0.97         &     0.97         &     0.98         &     0.97         &     0.97         &     0.98         \\
\(R^2_{within}\)&    0.024         &    0.094         &    0.055         &    0.030         &    0.095         &    0.056         \\
Year FE         &\checkmark         &\checkmark         &                  &\checkmark         &\checkmark         &                  \\
Firm FE         &\checkmark         &\checkmark         &\checkmark         &\checkmark         &\checkmark         &\checkmark         \\
Baseline Controls $\times$ $\mathbf{1}_{t\geq1920}$&                  &\checkmark         &\checkmark         &                  &\checkmark         &\checkmark         \\
Industry-Year-FE&                  &                  &\checkmark         &                  &                  &\checkmark         \\
\bottomrule
\end{tabular}
}
}

              \end{center}
        {\footnotesize \textit{Notes}: 
           This table reports results from estimating:
\begin{equation*}
    y_{it} = \alpha_i + \delta_{st} + \beta \times (Leverage_{i,1918} \times  \mathbf{1}_{t\geq 1920} ) + \Gamma \times (X_{i} \times \mathbf{1}_{t\geq 1920})  + \epsilon_{it}.
\end{equation*}
where $y_{it}$ is firm $i$'s number of employees (in logs, multiplied by 100). $Leverage_{i,1918}$ is either the ratio of firm $i$'s financial debt or total liabilities to total assets in year 1918. $\alpha_i$ is a set of firm fixed effects, $\delta_{st}$ is a set of industry-time fixed effects, and $X_i$ is a vector of firm-level controls. Baseline controls are size (log of assets), the ratio of fixed assets to total assets, profit margin (EBIT-to-revenue), free cash flow to assets, and Tobin's Q, all as of 1918.  *,**, and *** indicate significance at the 10\%, 5\%, and 1\% level, respectively. 
         }
        \end{minipage}
\end{table}%

\begin{table}[!ht]

  \centering
   \caption{Firm Leverage and Employment: Controlling for Export Status, Political Connections, and Bank Exposure.}\label{tab:robust1_liab}
        \begin{minipage}{1.0\textwidth}
        \begin{center}
        \scalebox{0.74}{{ \def\sym#1{\ifmmode^{#1}\else\(^{#1}\)\fi} \begin{tabular}{l*{7}{c}}   \toprule  Dependent Variable & \multicolumn{7}{c}{ln(Employment)} \\    \cmidrule(lr){2-8}       
                &\multicolumn{1}{c}{(1)}         &\multicolumn{1}{c}{(2)}         &\multicolumn{1}{c}{(3)}         &\multicolumn{1}{c}{(4)}         &\multicolumn{1}{c}{(5)}         &\multicolumn{1}{c}{(6)}         &\multicolumn{1}{c}{(7)}         \\
\midrule
Liab/Assets$_{i,1917}$ $\times$ $\mathbf{1}_{t\geq1920}$&     41.6\sym{**} &     37.8\sym{*}  &     41.5\sym{*}  &     38.7\sym{*}  &     37.5\sym{*}  &     43.9\sym{**} &     43.6\sym{**} \\
                &   (20.8)         &   (20.4)         &   (21.1)         &   (21.1)         &   (21.6)         &   (22.0)         &   (21.8)         \\
Tobin's Q$_{i,1918}$ $\times$$\mathbf{1}_{t\geq1920} $&     40.9\sym{***}&     41.9\sym{***}&     40.7\sym{***}&     40.4\sym{***}&     42.1\sym{***}&     40.6\sym{***}&     40.4\sym{***}\\
                &   (14.5)         &   (14.4)         &   (14.6)         &   (14.9)         &   (14.6)         &   (14.8)         &   (14.9)         \\
$ \ln(\text{Total Assets})_{i,1917}$ $\times$ $\mathbf{1}_{t\geq1920}$&     0.99         &     1.41         &     0.87         &     0.40         &     1.02         &     1.93         &     2.03         \\
                &   (2.51)         &   (2.54)         &   (2.49)         &   (2.87)         &   (2.69)         &   (2.75)         &   (2.99)         \\
$ \text{FCF/Assets}_{i,1917}$ $\times$ $\mathbf{1}_{t\geq1920}$&    -15.7         &    -15.5         &    -14.9         &    -18.8         &    -17.4         &    -15.3         &    -16.3         \\
                &   (17.9)         &   (17.2)         &   (17.5)         &   (18.0)         &   (18.4)         &   (17.8)         &   (17.1)         \\
Fixed Assets/Total Assets$_{i,1917}$ $\times$$\mathbf{1}_{t\geq1920} $&     13.7         &     15.9         &     13.7         &     9.45         &     11.3         &     15.2         &     16.3         \\
                &   (16.1)         &   (15.9)         &   (15.8)         &   (17.3)         &   (16.6)         &   (16.1)         &   (17.3)         \\
EBIT Margin$_{i,1917}$ $\times$$\mathbf{1}_{t\geq1920} $&    -6.34\sym{***}&    -6.40\sym{***}&    -6.12\sym{***}&    -6.19\sym{***}&    -6.19\sym{***}&    -6.13\sym{***}&    -5.89\sym{***}\\
                &   (1.26)         &   (1.22)         &   (1.26)         &   (1.16)         &   (1.29)         &   (1.16)         &   (1.21)         \\
Exporter$_{i,1921}$ $\times$$\mathbf{1}_{t\geq1920} $&                  &     41.6\sym{*}  &                  &                  &                  &                  &     46.0\sym{*}  \\
                &                  &   (23.2)         &                  &                  &                  &                  &   (26.8)         \\
KM to Berlin (in log)$_{i}$ $\times$$\mathbf{1}_{t\geq1920} $&                  &                  &    -0.74         &                  &                  &                  &     0.26         \\
                &                  &                  &   (1.49)         &                  &                  &                  &   (1.55)         \\
Network dist. to Government$_{i,1918}$ $\times$$\mathbf{1}_{t\geq1920} $&                  &                  &                  &     4.31         &                  &                  &     7.31         \\
                &                  &                  &                  &   (5.21)         &                  &                  &   (4.75)         \\
Network dist. to Reichsbank$_{i,1918}$ $\times$$\mathbf{1}_{t\geq1920} $ &                  &                  &                  &    -8.71         &                  &                  &    -7.01         \\
                &                  &                  &                  &   (5.73)         &                  &                  &   (5.92)         \\
Network dist. to Hugo Stinnes$_{i,1918}$ $\times$$\mathbf{1}_{t\geq1920} $&                  &                  &                  &     3.00         &                  &                  &     3.79         \\
                &                  &                  &                  &   (5.05)         &                  &                  &   (5.49)         \\
Network dist. to Banks$_{i,1918}$ $\times$$\mathbf{1}_{t\geq1920} $&                  &                  &                  &                  &    -2.17         &                  &    -4.31         \\
                &                  &                  &                  &                  &   (6.08)         &                  &   (7.86)         \\
\midrule
Observations    &     1866         &     1866         &     1866         &     1818         &     1818         &     1866         &     1818         \\
Number of Firms &      278         &      278         &      278         &      270         &      270         &      278         &      270         \\
\(R^2_{overall}\)&     0.97         &     0.98         &     0.97         &     0.97         &     0.97         &     0.98         &     0.98         \\
\(R^2_{within}\)&    0.049         &    0.067         &    0.050         &    0.057         &    0.049         &    0.060         &    0.088         \\
Firm FE         &\checkmark         &\checkmark         &\checkmark         &\checkmark         &\checkmark         &\checkmark         &\checkmark         \\
Baseline Controls $\times$ $\mathbf{1}_{t\geq1920}$&\checkmark         &\checkmark         &\checkmark         &\checkmark         &\checkmark         &\checkmark         &\checkmark         \\
Industry-Year-FE&\checkmark         &\checkmark         &\checkmark         &\checkmark         &\checkmark         &\checkmark         &\checkmark         \\
Exporter $\times$ $\mathbf{1}_{t\geq1920}$&                  &\checkmark         &                  &                  &                  &                  &\checkmark         \\
KM to Berlin $\times$ $\mathbf{1}_{t\geq1920}$&                  &                  &\checkmark         &                  &                  &                  &\checkmark         \\
Network dist. to Gov. $\times$ $\mathbf{1}_{t\geq1920}$&                  &                  &                  &\checkmark         &                  &                  &\checkmark         \\
Network dist. to Banks $\times$ $\mathbf{1}_{t\geq1920}$&                  &                  &                  &                  &\checkmark         &                  &\checkmark         \\
Large-Bank-Time FE&                  &                  &                  &                  &                  &\checkmark         &\checkmark         \\
\bottomrule
\end{tabular}
}
}

              \end{center}
        {\footnotesize \textit{Notes}: 
        This table reports results from estimating equation \eqref{eq:diff-in-diff} with a range of additional control variables.  The dependent variable is log firm employment (times 100).  Column 1 reports the estimate corresponding to column 3 in \Cref{tab:leverage_employment}, here with coefficients reported for all control variables. Column 2 adds a control for export status measured based on export activity in 1921 (reported in the 1922 edition of \textit{Saling's}). Column 3 controls for a firm location's distance to Berlin. \Cref{fig:saling_map} presents a map with the geocoded location of firm headquarters. Columns 4-5 control for a firm's board network distance to government officials, the Reichsbank Central Committee of Shareholders (Zentralausschuss), politically connected industrialist Hugo Stinnes, and German banks. See \Cref{app:data} for details on the construction of the board member network distances.  Column 6  adds fixed effects for a connection to one of seven major banks based on the banks that paid out a firm's dividends (\textit{Zahlstellen}). It also includes a fixed effect for firms with a connection to a bank that is not one of the major banks and a fixed effect for firms without any banking connections. Column 7 includes all controls and fixed effects jointly. All these additional controls and fixed effects are interacted with the post-1919 indicator.
      Standard errors in parentheses are clustered at the firm level.    *,**, and *** indicate significance at the 10\%, 5\%, and 1\% level, respectively.
         }
        \end{minipage}
\end{table}%

\begin{table}[!ht]

  \centering
   \caption{Firm Leverage and Employment: Controlling for Export Status, Political Connections, and Bank Exposure.}\label{tab:robust1_debt}
        \begin{minipage}{1.0\textwidth}
        \begin{center}
        \scalebox{0.74}{{ \def\sym#1{\ifmmode^{#1}\else\(^{#1}\)\fi} \begin{tabular}{l*{7}{c}}   \toprule  Dependent Variable & \multicolumn{7}{c}{ln(Employment)} \\    \cmidrule(lr){2-8}       
                &\multicolumn{1}{c}{(1)}         &\multicolumn{1}{c}{(2)}         &\multicolumn{1}{c}{(3)}         &\multicolumn{1}{c}{(4)}         &\multicolumn{1}{c}{(5)}         &\multicolumn{1}{c}{(6)}         &\multicolumn{1}{c}{(7)}         \\
\midrule
Debt/Assets$_{i,1917}$ $\times$$\mathbf{1}_{t\geq1920}$&     49.4\sym{**} &     45.8\sym{**} &     48.9\sym{**} &     45.1\sym{**} &     46.5\sym{**} &     50.8\sym{**} &     50.6\sym{**} \\
                &   (22.6)         &   (21.9)         &   (23.4)         &   (22.8)         &   (23.4)         &   (23.2)         &   (23.9)         \\
Tobin's Q$_{i,1918}$ $\times$$\mathbf{1}_{t\geq1920} $&     40.7\sym{***}&     41.8\sym{***}&     40.7\sym{***}&     40.0\sym{***}&     41.7\sym{***}&     40.9\sym{***}&     40.4\sym{***}\\
                &   (14.5)         &   (14.4)         &   (14.6)         &   (15.0)         &   (14.5)         &   (15.0)         &   (15.0)         \\
$ \ln(\text{Total Assets})_{i,1917}$ $\times$ $\mathbf{1}_{t\geq1920}$&     0.18         &     0.62         &     0.18         &    -0.23         &    0.081         &     0.89         &     1.25         \\
                &   (2.50)         &   (2.50)         &   (2.50)         &   (2.87)         &   (2.71)         &   (2.80)         &   (2.96)         \\
$ \text{FCF/Assets}_{i,1917}$ $\times$ $\mathbf{1}_{t\geq1920}$&    -12.8         &    -12.9         &    -12.7         &    -15.7         &    -13.9         &    -12.8         &    -13.2         \\
                &   (17.9)         &   (17.4)         &   (17.7)         &   (18.2)         &   (18.5)         &   (17.8)         &   (17.6)         \\
Fixed Assets/Total Assets$_{i,1917}$ $\times$$\mathbf{1}_{t\geq1920} $&     7.15         &     9.90         &     7.15         &     3.37         &     5.14         &     7.64         &     8.64         \\
                &   (15.2)         &   (15.2)         &   (15.1)         &   (16.3)         &   (15.7)         &   (15.2)         &   (16.7)         \\
EBIT Margin$_{i,1917}$ $\times$$\mathbf{1}_{t\geq1920} $&    -6.03\sym{***}&    -6.10\sym{***}&    -5.97\sym{***}&    -5.93\sym{***}&    -5.88\sym{***}&    -5.83\sym{***}&    -5.71\sym{***}\\
                &   (1.05)         &   (0.97)         &   (1.08)         &   (0.98)         &   (1.11)         &   (0.98)         &   (1.06)         \\
Exporter$_{i,1921}$ $\times$$\mathbf{1}_{t\geq1920} $&                  &     40.6\sym{*}  &                  &                  &                  &                  &     46.5\sym{*}  \\
                &                  &   (21.5)         &                  &                  &                  &                  &   (25.2)         \\
KM to Berlin (in log)$_{i}$ $\times$$\mathbf{1}_{t\geq1920} $&                  &                  &    -0.23         &                  &                  &                  &     0.77         \\
                &                  &                  &   (1.47)         &                  &                  &                  &   (1.56)         \\
Network dist. to Government$_{i,1918}$ $\times$$\mathbf{1}_{t\geq1920} $&                  &                  &                  &     3.95         &                  &                  &     6.90         \\
                &                  &                  &                  &   (5.25)         &                  &                  &   (4.95)         \\
Network dist. to Reichsbank$_{i,1918}$ $\times$$\mathbf{1}_{t\geq1920} $ &                  &                  &                  &    -7.88         &                  &                  &    -6.37         \\
                &                  &                  &                  &   (5.47)         &                  &                  &   (5.91)         \\
Network dist. to Hugo Stinnes$_{i,1918}$ $\times$$\mathbf{1}_{t\geq1920} $&                  &                  &                  &     2.87         &                  &                  &     4.65         \\
                &                  &                  &                  &   (4.89)         &                  &                  &   (5.46)         \\
Network dist. to Banks$_{i,1918}$ $\times$$\mathbf{1}_{t\geq1920} $&                  &                  &                  &                  &    -2.86         &                  &    -5.77         \\
                &                  &                  &                  &                  &   (6.22)         &                  &   (8.30)         \\
\midrule
Observations    &     1866         &     1866         &     1866         &     1818         &     1818         &     1866         &     1818         \\
Number of Firms &      278         &      278         &      278         &      270         &      270         &      278         &      270         \\
\(R^2_{overall}\)&     0.98         &     0.98         &     0.98         &     0.98         &     0.97         &     0.98         &     0.98         \\
\(R^2_{within}\)&    0.059         &    0.076         &    0.059         &    0.064         &    0.058         &    0.069         &    0.096         \\
Firm FE         &\checkmark         &\checkmark         &\checkmark         &\checkmark         &\checkmark         &\checkmark         &\checkmark         \\
Baseline Controls $\times$ $\mathbf{1}_{t\geq1920}$&\checkmark         &\checkmark         &\checkmark         &\checkmark         &\checkmark         &\checkmark         &\checkmark         \\
Industry-Year-FE&\checkmark         &\checkmark         &\checkmark         &\checkmark         &\checkmark         &\checkmark         &\checkmark         \\
Exporter $\times$ $\mathbf{1}_{t\geq1920}$&                  &\checkmark         &                  &                  &                  &                  &\checkmark         \\
KM to Berlin $\times$ $\mathbf{1}_{t\geq1920}$&                  &                  &\checkmark         &                  &                  &                  &\checkmark         \\
Network dist. to Gov. $\times$ $\mathbf{1}_{t\geq1920}$&                  &                  &                  &\checkmark         &                  &                  &\checkmark         \\
Network dist. to Banks $\times$ $\mathbf{1}_{t\geq1920}$&                  &                  &                  &                  &\checkmark         &                  &\checkmark         \\
Large-Bank-Time FE&                  &                  &                  &                  &                  &\checkmark         &\checkmark         \\
\bottomrule
\end{tabular}
}
}

              \end{center}
        {\footnotesize \textit{Notes}: 
           { This table is analogous to \Cref{tab:robust1_liab}, but it uses firm debt-to-assets in 1917 as the measure of firm leverage (rather than liabilities-to-assets in 1917). See \Cref{tab:robust1_liab} for additional details on the empirical specification and control variables. Standard errors in parentheses are clustered at the firm level.    *,**, and *** indicate significance at the 10\%, 5\%, and 1\% level, respectively. }
         }
        \end{minipage}
\end{table}%

\begin{table}[!ht]
  \centering
   \caption{ Firm Leverage and Employment: Sample Restrictions Based on War-Relevant Industries and Firm Size.}
        \begin{minipage}{1.0\textwidth}
        \begin{center}
        \scalebox{0.7}{
        { \def\sym#1{\ifmmode^{#1}\else\(^{#1}\)\fi}  \begin{tabular}{l*{7}{c}}   \toprule  Dependent Variable & \multicolumn{7}{c}{ln(Employment)} \\    \cmidrule(lr){2-8}       
                &\multicolumn{1}{c}{(1)}         &\multicolumn{1}{c}{(2)}         &\multicolumn{1}{c}{(3)}         &\multicolumn{1}{c}{(4)}         &\multicolumn{1}{c}{(5)}         &\multicolumn{1}{c}{(6)}         &\multicolumn{1}{c}{(7)}         \\
\midrule
Liab/Assets$_{i,1917}$ $\times$$\mathbf{1}_{t\geq1920}$&     52.6\sym{***}&     33.0\sym{*}  &     79.8\sym{***}&     57.4\sym{***}&     58.1\sym{***}&     78.7\sym{***}&     48.4\sym{***}\\
                &   (15.2)         &   (16.9)         &   (27.5)         &   (17.3)         &   (20.1)         &   (28.2)         &   (18.0)         \\
War-relevant industry $\times \mathbf{1}_{t\geq1920} $&     7.82         &                  &                  &     6.73         &     4.01         &     1.66         &     7.19         \\
                &   (5.01)         &                  &                  &   (5.41)         &   (5.64)         &   (6.41)         &   (5.09)         \\
\midrule
Observations    &     2355         &     1463         &      892         &     1857         &     1660         &     1104         &     2052         \\
Number of Firms &      351         &      213         &      138         &      231         &      205         &      132         &      256         \\
\(R^2_{overall}\)&     0.97         &     0.97         &     0.97         &     0.95         &     0.93         &     0.88         &     0.97         \\
\(R^2_{within}\)&    0.039         &   0.0097         &    0.074         &    0.041         &    0.035         &    0.051         &    0.067         \\
Year FE         &\checkmark         &\checkmark         &\checkmark         &\checkmark         &\checkmark         &\checkmark         &\checkmark         \\
Firm FE         &\checkmark         &\checkmark         &\checkmark         &\checkmark         &\checkmark         &\checkmark         &\checkmark         \\
Sample          &All firms         &w/ war-rel.         &w/o war-rel.         &<p90(size)         &<p80(size)         &<p50(size)         &All firms         \\
Non-para. size control&                  &                  &                  &                  &                  &                  &\checkmark         \\
\bottomrule
\end{tabular}
}
}
        \end{center}
        {\footnotesize \textit{Notes}: This table reports results from estimating equation \eqref{eq:diff-in-diff} for various sub-samples and additional controls. ``/w war-rel." and ``w/o war-rel." refer to estimations only on the sample of war-relevant industries and only the sample excluding war-relevant industries, respectively. War-relevant industries are mines and smelters, chemical factories, inking plants, salt pans, railway supplies and machines, metals industry, shipping and port companies, and other transport companies. This classification follows the definition of \citet{Kocka1984}. ``p<90(size)'' refers to a specification that is estimated on the sample of firms below the 90\% percentile in employment (measured as the average before the inflation over 1914-1918). ``Non-para. size control'' refers to a specification that controls for ten deciles of size interacted with the post-1919 indicator.  Standard errors in parentheses are clustered at the firm level.    *,**, and *** indicate significance at the 10\%, 5\%, and 1\% level, respectively.
        
         }
        \end{minipage}
        \label{tab:war_exposure}
\end{table}%

\begin{table}[!ht]

  \centering
   \caption{Firm Leverage and Employment---Instrumenting with Pre-WWI Leverage.}\label{tab:preWWI_IV}
        \begin{minipage}{1.0\textwidth}
        \begin{center}
        \scalebox{0.9}{{ \def\sym#1{\ifmmode^{#1}\else\(^{#1}\)\fi} \begin{tabular}{l*{4}{c}}   \toprule  Dependent Variable & \multicolumn{4}{c}{ln(Employment)} \\    \cmidrule(lr){2-5}       
                &\multicolumn{1}{c}{(1)}         &\multicolumn{1}{c}{(2)}         &\multicolumn{1}{c}{(3)}         &\multicolumn{1}{c}{(4)}         \\
\midrule
\( \text{Liab./Assets}_{i,1917} \times \mathbf{1}_{t\geq1920} \)&     75.1\sym{***}&                  &                  &                  \\
                &   (27.6)         &                  &                  &                  \\
\( \text{Liab./Assets}_{i,1918} \times \mathbf{1}_{t\geq1920} \) &                  &     63.5\sym{**} &                  &                  \\
                &                  &   (32.1)         &                  &                  \\
\( \text{Debt/Assets}_{i,1917} \times \mathbf{1}_{t\geq1920} \)&                  &                  &     64.3\sym{***}&                  \\
                &                  &                  &   (24.4)         &                  \\
\( \text{Debt/Assets}_{i,1918} \times \mathbf{1}_{t\geq1920} \)&                  &                  &                  &     59.6\sym{**} \\
                &                  &                  &                  &   (28.4)         \\
\midrule
Observations    &     2254         &     2292         &     2254         &     2292         \\
Number of Firms &      337         &      343         &      337         &      343         \\
\(R^2_{overall}\)&    0.031         &    0.025         &    0.044         &    0.032         \\
Year FE         &\checkmark         &\checkmark         &\checkmark         &\checkmark         \\
Firm FE         &\checkmark         &\checkmark         &\checkmark         &\checkmark         \\
IV              &\( L/A_{i,preWWI} \)         &\( L/A_{i,preWWI} \)         &\(D/A_{i,preWWI} \)         &\(D/A_{i,preWWI} \)         \\
First-stage F-stat&    128.5         &     76.8         &    118.7         &     89.0         \\
\bottomrule
\end{tabular}
}
}

              \end{center}
        {\footnotesize \textit{Notes}:  This table presents results from estimating \eqref{eq:diff-in-diff} and instrumenting $Leverage_{i}$ with leverage before WWI. We measure leverage before WWI as the average over 1910-1913. The estimation period is 1914-1923. *,**, and *** indicate significance at the 10\%, 5\%, and 1\% level, respectively.
         }
        \end{minipage}
\end{table}%

\begin{table}[!ht]
  \centering
  
   \caption{Firm Leverage and Employment: Excluding Industries with High Export Activity.}\label{tab:exporter_status}
        \begin{minipage}{1.0\textwidth}
        \begin{center}
        \scalebox{0.7}{
        { \def\sym#1{\ifmmode^{#1}\else\(^{#1}\)\fi} \begin{tabular}{l*{6}{c}}   \toprule  Dependent Variable & \multicolumn{6}{c}{ln(Employment)} \\    \cmidrule(lr){2-7}       
                &\multicolumn{1}{c}{(1)}         &\multicolumn{1}{c}{(2)}         &\multicolumn{1}{c}{(3)}         &\multicolumn{1}{c}{(4)}         &\multicolumn{1}{c}{(5)}         &\multicolumn{1}{c}{(6)}         \\
\midrule
 \( \text{Liab/Assets}_{i,1917} \times  \mathbf{1}_{t\geq1920} \) &     66.4\sym{***}&     71.8\sym{***}&     91.4\sym{*}  &    170.7\sym{***}&     51.3\sym{***}&     29.0         \\
                &   (18.3)         &   (25.0)         &   (51.6)         &   (60.2)         &   (13.8)         &   (20.0)         \\
\midrule
Observations    &     1460         &     1194         &      288         &      220         &     2067         &     1690         \\
Number of Firms &      226         &      183         &       40         &       29         &      311         &      251         \\
\(R^2_{overall}\)&     0.97         &     0.97         &     0.98         &     0.98         &     0.97         &     0.97         \\
\(R^2_{within}\)&    0.050         &    0.080         &    0.092         &     0.26         &    0.026         &    0.097         \\
Year FE         &\checkmark         &\checkmark         &\checkmark         &\checkmark         &\checkmark         &\checkmark         \\
Firm FE         &\checkmark         &\checkmark         &\checkmark         &\checkmark         &\checkmark         &\checkmark         \\
Baseline Controls $\times$ $\mathbf{1}_{t\geq1920}$&                  &\checkmark         &                  &\checkmark         &                  &\checkmark         \\
Sample          &Ex. mining+metal         &Ex. mining+metal         &Non-tradable         &Non-tradable         & Tradable         & Tradable         \\
\bottomrule
\end{tabular}
}
}
        \end{center}
        {\footnotesize \textit{Notes}:  This table is similar to \Cref{tab:leverage_employment} but excludes industries with high export activity. Columns 1-2 exclude mining industries (mines and smelters) and metals industries. These are industries that were very active in exports. Columns 3-4 present estimates for the sample of non-tradable industries, defined as construction, services (hotel companies), transport, utilities, and other industries. 
        Control variables are defined as in \Cref{tab:leverage_employment}. We do not include industry fixed effects, as this exercise excludes the majority of industries in \textit{Saling's}. *,**, and *** indicate significance at the 10\%, 5\%, and 1\% level, respectively.
         }
        \end{minipage}
\end{table}%

\begin{table}[!ht]
  \centering
   \caption{Firm Leverage and Production Expenses.}\label{tab:s_prod_cost}
        \begin{minipage}{1.0\textwidth}
        \begin{center}
        \scalebox{0.8}{{   \def\sym#1{\ifmmode^{#1}\else\(^{#1}\)\fi}  \begin{tabular}{l*{4}{c}}   \toprule  Dependent Variable & \multicolumn{4}{c}{Production Expenses/Tot. Expenses} \\    \cmidrule(lr){2-5}       
                &\multicolumn{1}{c}{(1)}         &\multicolumn{1}{c}{(2)}         &\multicolumn{1}{c}{(3)}         &\multicolumn{1}{c}{(4)}         \\
\midrule
Liab/Assets$_{i,1917}$ $\times$ $\mathbf{1}_{t\geq1920}$&     0.13\sym{***}&     0.16\sym{**} &                  &                  \\
                &  (0.051)         &  (0.079)         &                  &                  \\
Debt/Assets$_{i,1917}$ $\times$$\mathbf{1}_{t\geq1920}$&                  &                  &     0.17\sym{***}&    0.097         \\
                &                  &                  &  (0.049)         &  (0.077)         \\
\midrule
Observations    &     3226         &     2291         &     3226         &     2291         \\
Number of Firms &      520         &      335         &      520         &      335         \\
\(R^2_{overall}\)&     0.63         &     0.66         &     0.64         &     0.66         \\
\(R^2_{within}\)&   0.0067         &    0.024         &    0.011         &    0.020         \\
Year FE         &\checkmark         &                  &\checkmark         &                  \\
Firm FE         &\checkmark         &\checkmark         &\checkmark         &\checkmark         \\
Baseline Controls $\times$ $\mathbf{1}_{t\geq1920}$&                  &\checkmark         &                  &\checkmark         \\
Industry-Year-FE&                  &\checkmark         &                  &\checkmark         \\
\bottomrule
\end{tabular}
}
}

              \end{center}
        {\footnotesize \textit{Notes}: 
           {  This table reports results from estimating \eqref{eq:diff-in-diff}, with firm $i$'s share of production expenses to total expenses as the dependent variable. $Leverage_{i}$ is either firm $i$'s ratio of financial debt or total liabilities to total assets in 1917.  Baseline controls are firm-level controls, as defined in \Cref{tab:leverage_employment}. The estimation period is from 1916 through 1923. Standard errors in parentheses are clustered at the firm level. *,**, and *** indicate significance at the 10\%, 5\%, and 1\% level, respectively.}
         }
        \end{minipage}
\end{table}%

\begin{table}[htpb]
  \begin{center}
  \begin{threeparttable}
\caption{Interest Rates, Volume, and Maturity Structure of Bonds Outstanding in 1918 and 1919.}
\centering
\small
   \begin{tabular}{lcccccc}
\toprule
 & Mean & p5 & p25 & Median & p75 & p95 \\
 \cmidrule(lr){2-2} 
 \cmidrule(lr){3-3}
 \cmidrule(lr){4-4}
 \cmidrule(lr){5-5}
 \cmidrule(lr){6-6}
  \cmidrule(lr){7-7} 
  \midrule
  \multicolumn{7}{c}{Panel A: Interest rates and volume} \\ \midrule
Interest rate (ppt) & 4&4&4.25&4.5&4.5&5\\
 
Volume (in thousand RM) &  3961&500&1000&2000&5000&13000\\
  
\midrule
  \multicolumn{7}{c}{Panel B: Origination and maturity}  \\ \midrule
Origination year &  1906&1895&1901&1906&1911&1919\\
  
Repayment start year & 1910&1897&1905&1910&1916&1922\\
  
Repayment end year & 1941&1926&1933&1940&1947&1961\\
  
\bottomrule
\end{tabular}
\label{tab:bond_terms}
\end{threeparttable}\end{center}
{\footnotesize
Notes: This table provides information on all outstanding bonds issued by firms in our sample. Data are obtained from \textit{Saling's B\"{o}rsen-Jahrbuch} published in 1919 and 1920 (covering outstanding bonds in 1918 and 1919). The table is based on a sample of $N=417$ firms. 51\% of firms in the sample report information on at least one bond.}
\end{table}

\begin{table}[!ht]
  \centering
   \caption{Firm Leverage, Long-term Debt, Interest Expenses, and Employment.}\label{tab:leverage_financials-LTDebt}
        \begin{minipage}{1.0\textwidth}
        \begin{center}
        \scalebox{0.9}{
        {   \def\sym#1{\ifmmode^{#1}\else\(^{#1}\)\fi}   \begin{tabular}{l*{6}{c}}   \toprule  Dependent Variable  & \multicolumn{3}{c}{ \( \frac{\text{Int. Expenses}}{ \text{Tot. Expenses} } \) }  & \multicolumn{3}{c}{ln(Employment)}   \\    \cmidrule(lr){2-4} \cmidrule(lr){5-7}        
                &\multicolumn{1}{c}{(1)}         &\multicolumn{1}{c}{(2)}         &\multicolumn{1}{c}{(3)}         &\multicolumn{1}{c}{(4)}         &\multicolumn{1}{c}{(5)}         &\multicolumn{1}{c}{(6)}         \\
\midrule
Liab/Assets$_{i,1917}$$\times$$\mathbf{1}_{t\geq1920}$&  -0.0055         &   -0.014         &    0.017         &     45.9\sym{***}&     41.7\sym{***}&     30.5         \\
                &  (0.041)         &  (0.040)         &  (0.048)         &   (17.6)         &   (15.0)         &   (19.1)         \\
Liab/Assets$_{i,1917}$ $\times$ LTD/Debt $\times$$\mathbf{1}_{t\geq1920}$&    -0.31\sym{***}&                  &                  &     49.3         &                  &                  \\
                &   (0.11)         &                  &                  &   (37.9)         &                  &                  \\
Liab/Assets$_{i,1917}$ LTD Tercile$_2$ $\times$ $\times$$\mathbf{1}_{t\geq1920}$&                  &   -0.088         &   -0.025         &                  &     16.7         &     21.5         \\
                &                  &  (0.078)         &  (0.076)         &                  &   (34.7)         &   (38.1)         \\
Liab/Assets$_{i,1917}$ LTD Tercile$_3$ $\times$ $\times$$\mathbf{1}_{t\geq1920}$&                  &    -0.21\sym{**} &    -0.28\sym{***}&                  &     34.7         &     69.8         \\
                &                  &  (0.083)         &   (0.11)         &                  &   (26.9)         &   (45.2)         \\
\midrule
Observations    &     3227         &     3253         &     2341         &     2340         &     2355         &     1910         \\
Number of Firms &      515         &      519         &      337         &      348         &      351         &      280         \\
\(R^2_{overall}\)&     0.68         &     0.68         &     0.65         &     0.97         &     0.97         &     0.97         \\
\(R^2_{within}\)&     0.11         &    0.097         &     0.15         &    0.035         &    0.058         &     0.10         \\
Firm FE         &\checkmark         &\checkmark         &\checkmark         &\checkmark         &\checkmark         &\checkmark         \\
Year FE         &\checkmark         &\checkmark         &\checkmark         &\checkmark         &\checkmark         &\checkmark         \\
Baseline Controls $\times$ $\mathbf{1}_{t\geq1920}$&                  &                  &\checkmark         &                  &                  &\checkmark         \\
\bottomrule
\end{tabular}
}
}
        \end{center}
        {\footnotesize \textit{Notes}: 
        This table reports results from estimating either
                \begin{align*}
	y_{it} & = \alpha_i + \delta_{st} + \beta_1  \left(\frac{Liabilities}{Assets}\right)_{i,1917} \times \;  \left(\frac{LTD}{Liabilities}\right)_{i,1917}  \;  \mathbf{1}_{t\geq 1920}  \\ & + \beta_2  \left(\frac{Liabilities}{Assets}\right)_{i,1917}  \mathbf{1}_{t\geq 1920} + \beta_3   \;  \left(\frac{LTD}{Liabilities}\right)_{i,1917}  \;  \mathbf{1}_{t\geq 1920}+ \Gamma \; X_{i} \; \mathbf{1}_{t\geq 1920}  + \epsilon_{it},
	\end{align*}
	(in columns 1 and 4) or
        \begin{align*}
	y_{it} = \alpha_i + \delta_{st} + \sum_{k} \beta_{k} \; Tercile_{i,k}  \; \left(\frac{Liabilities}{Assets}\right)_{i,1917} \;  \mathbf{1}_{t\geq 1920}  + \Gamma \; X_{i} \; \mathbf{1}_{t\geq 1920}  + \epsilon_{it},
	\end{align*}
in the remaining columns. $Tercile_{i,k}$ is an indicator for being in the $k$\textsuperscript{th} tercile of the long-term debt to total debt distribution in 1917 $(\frac{LTD}{Debt}_{i,1917})$. The dependent variable is either interest expenses as a share of total expenses (columns 1-3) or log employment times 100 (columns 4-6). Baseline controls are as in \Cref{tab:leverage_employment}. Standard errors in parentheses are clustered at the firm level. *,**, and *** indicate significance at the 10\%, 5\%, and 1\% level, respectively.
         }
        \end{minipage}
\end{table}

\begin{table}[!ht]
  \centering
  \caption{Leverage and Stock Returns: Fama-Macbeth Regressions, 1919m1-1923m12.}\label{tab:fama_macbeth}
        \begin{minipage}{1.0\textwidth}
        \begin{center}
        \scalebox{0.9}{
        {
\def\sym#1{\ifmmode^{#1}\else\(^{#1}\)\fi}
\begin{tabular}{l*{6}{c}}
\toprule
                &\multicolumn{1}{c}{(1)}         &\multicolumn{1}{c}{(2)}         &\multicolumn{1}{c}{(3)}         &\multicolumn{1}{c}{(4)}         &\multicolumn{1}{c}{(5)}         &\multicolumn{1}{c}{(6)}         \\
\midrule
cons            &    -3.85         &    -1.78         &    -1.11         &    -3.47         &    -0.42         &     0.18         \\
                &   (3.51)         &   (2.46)         &   (2.10)         &   (3.53)         &   (2.38)         &   (2.02)         \\
\addlinespace
\(\text{Liab/Assets}_{i,t-1} \)&     2.00\sym{*}  &     4.56\sym{***}&     4.38\sym{***}&                  &                  &                  \\
                &   (1.15)         &   (1.24)         &   (0.99)         &                  &                  &                  \\
\addlinespace
\(\text{Debt/Assets}_{i,t-1} \)&                  &                  &                  &     1.66         &     2.32\sym{**} &     2.14\sym{**} \\
                &                  &                  &                  &   (1.01)         &   (1.06)         &   (0.82)         \\
\addlinespace
\( \text{Size}_{i,t-1} \)&                  &    0.088         &     0.15         &                  &     0.11         &     0.18         \\
                &                  &   (0.35)         &   (0.18)         &                  &   (0.35)         &   (0.18)         \\
\addlinespace
\( \text{MtB}_{i,t-1} \)&                  &  -0.0047\sym{***}&  -0.0051\sym{***}&                  &  -0.0045\sym{***}&  -0.0049\sym{***}\\
                &                  & (0.0012)         & (0.0011)         &                  & (0.0011)         & (0.0011)         \\
\addlinespace
\( \text{Market beta}_i \)&                  &                  &    -0.75         &                  &                  &    -0.75         \\
                &                  &                  &   (3.14)         &                  &                  &   (3.15)         \\
\midrule
Observations    &    37367         &    37367         &    37365         &    37367         &    37367         &    37365         \\
\bottomrule
\end{tabular}
}
}
        \end{center}
        {\footnotesize \textit{Notes}: This table presents the average coefficients from the  \cite{FamaMacbeth1973}  regression procedure. Specifically, we first run cross-sectional regressions of the form:
$$R_{i,t+1} = \alpha^t + \beta^t_1 Leverage_{i,t} + \beta^t_2 Size_{i,t} + \beta^t_3 MtB_{i,t} + \beta^t_4 MarketBeta_{i} + \epsilon_{i,t+1},  $$ for each month $t$ using data from 1919m1 to 1923m12. We then report the time series average of the estimated cross-sectional coefficients, e.g.,  $ \hat \beta_i = \frac{1}{T} \sum_{t=1919m2}^{1923m23} \hat \beta^t_i$ and compute standard errors following \cite{FamaMacbeth1973}. Market-to-book is measured as the market value of equity relative to book value, as reported in \textit{Berliner Boersen Zeitung}. Following the empirical asset pricing literature, size is measured with the log of market cap. Market cap is estimated as the price-to-book ratio reported in \textit{Berliner Boersen Zeitung} times book equity values reported in \textit{Saling's}. Following \citet[ch. 12.3]{cochrane2009asset}, market beta is estimated in the full sample over 1919m1-1923m12 and is constant within firms over time.  *,**, and *** indicate significance at the 10\%, 5\%, and 1\% level, respectively. 
         }
        \end{minipage}
\end{table}%

   	\clearpage
\renewcommand\thefigure{\thesection.\arabic{figure}}
\renewcommand\thetable{\thesection.\arabic{table}}

\setcounter{figure}{0}
\setcounter{table}{0}

\section{Data Appendix}\label{app:data}

\subsection{Firm-level Financial Data}\label{firm-level-financial-data}

We obtain firm financial information from \emph{Saling's
Börsen-Jahrbuch}, a German investor manual published annually throughout our sample and made available in digital format by the \href{https://digi.bib.uni-mannheim.de}{University of Mannheim}.\footnote{In particular, we use the second volume of this series, known as \emph{Saling's Börsen-Papiere - Zweiter (finanzieller) Teil (Berliner Börse)}. Note also that its editions are listed as if spanning two consecutive years, such as ``1921/22'', but their publication date is actually about the middle of the first of the two years (e.g., approximately ``June 1921'').} For each firm, \emph{Saling's} lists a header followed by sections on various topics such as the composition of
the board, dividend payments, firm history, existing bonds, etc. We
focus on five key sections. First, we compile general information about the firm, including its name, any name variants, the location of its headquarters, and its industry. These are represented in the two blue boxes in Figure \ref{fig:salings-snippet}. Second, we obtain balance sheet information, reported in \emph{Saling's} in paragraph form, as shown in the first dashed green box of \cref{fig:salings-snippet}. Third, we obtain firms' income statements, as shown in the second green box. Fourth, we obtain employment counts, as shown in the red box. Finally, we also obtain information on outstanding bonds, executives and supervisory board members, and export status, as described in more detail below in this Appendix.

{
\setstretch{1.0}
  \begin{figure}[!ht]
    
    \begin{center}
    \caption{Annotated Extract of the 1927 Edition of \emph{Saling's Börsen-Jahrbuch}. \label{fig:salings-snippet}}
    \includegraphics[width=0.9\textwidth]{"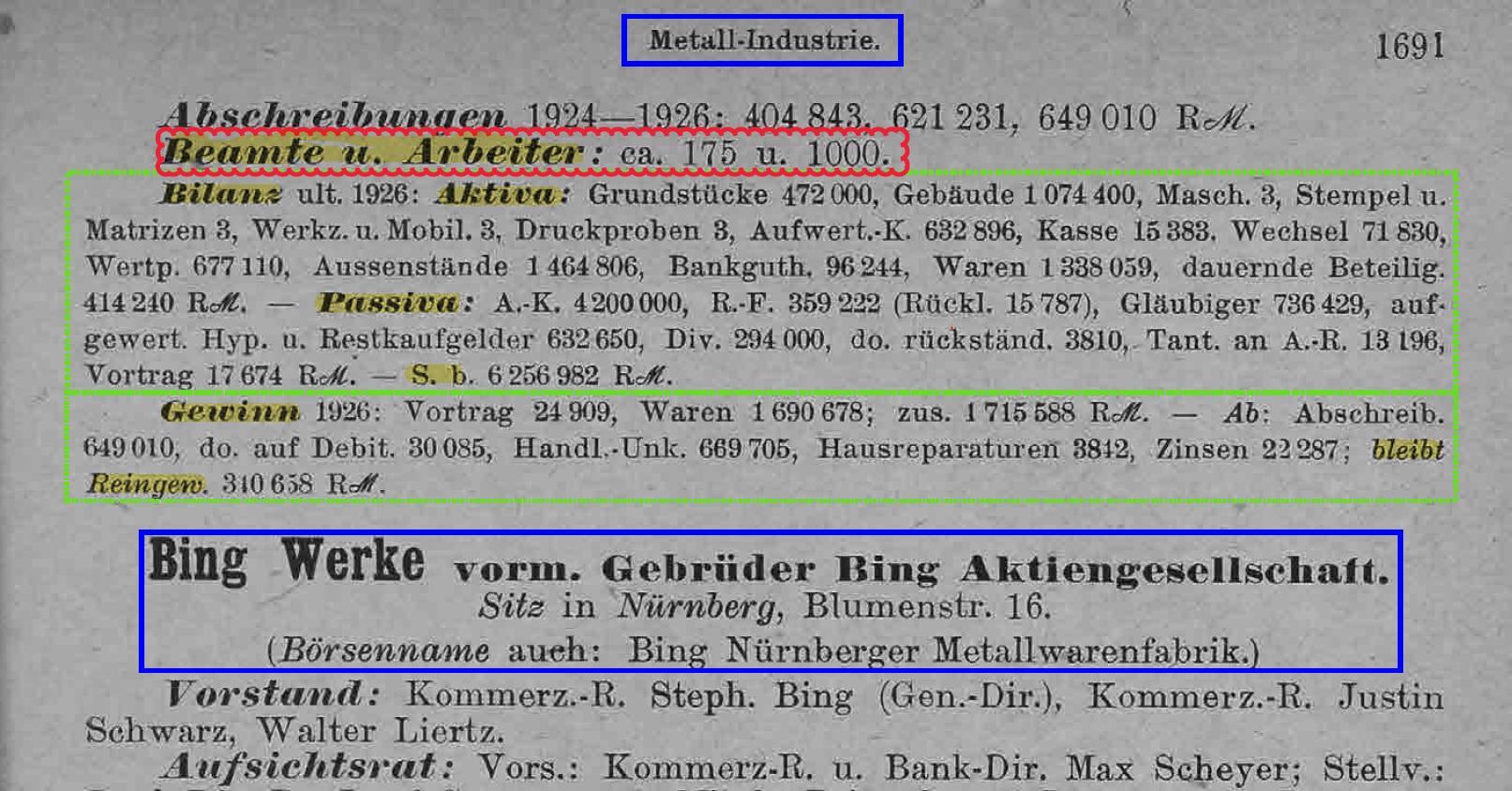"}
    \end{center}
        \footnotesize{\textit{Notes}: This scanned image contains information for the firms H. Berthold AG (green boxes)  and Bing Werke (second blue box). It illustrates some of the key sections of \emph{Saling's}
that we identify and process to construct our balance sheet,
income statement, and employment data. These are, from top to bottom: i) the page header indicating the industry of the firms in the current book chapter (blue solid box), ii) employment count (red curved box), iii) a balance sheet (first solid green box), iv) an income statement (second solid green box), and v) the firm header containing its name and the location of its headquarters. Some of the keywords used to detect blocks are also highlighted in yellow.}
    
  \end{figure}
}

\paragraph*{Data extraction}\label{data-extraction}

Properly extracting the data for our analysis requires understanding and addressing three key challenges. First, and most straightforward, is the digitization and optical character recognition (OCR) process itself. The challenges here are similar to those of other historical documents. Second, balance sheet (and income statement) information was not standardized in \emph{Saling's} at the time, so the information was not
available in tabular form but in a rather unwieldy textual form. Moreover, balance sheet items were not standardized to belong to only certain categories, but instead reflected open-ended categories mostly defined by the firms themselves. Lastly, the reports had substantial but subtle changes through time, including multiple currencies as well as post-inflation balance sheet revaluation.

To address these three challenges, we applied a series of data extraction and validation methods, which can be summarized in the following six steps:

\begin{enumerate}
\item
  Download all scanned pages from \emph{Saling's} from the
  \href{https://digi.bib.uni-mannheim.de}{University of Mannheim}
  digital repository, for the years 1916-1929, using the images with the highest available quality.
\item These pages often had scanning artifacts that made optical character recognition (OCR) challenging. We address these problems through the methods discussed in \citet{Correia2022}. Then, we feed the cleaned-up pages into multiple OCR engines (Amazon Textract and Google Cloud Vision).
\item We apply several ad-hoc algorithms to detect the different blocks of text that contain our relevant information. To identify industries, we look for the top-most centered block of text in each page. To identify firm headers, we search for text with certain characteristics (large font, centered, with large margin above) as well as for certain  keywords (``Sitz in'', ``Börsenname'', etc.). Similarly, we identify  balance sheets and income statements through keywords (``Bilanz'', ``Aktiva'', etc.) with text possessing certain characteristics (i.e.,~indented and at the beginning of a paragraph).
\item The previous step might have not detected all relevant blocks due to, e.g., ~OCR typos, so we apply several additional strategies to select any other remaining blocks. For instance, for firm headers we:
  \begin{enumerate}
  \item Digitize the firm index located at the beginning of the book, and correlate it with the pages where firm headers were detected.
  \item Search for paragraphs that only appear once per firm (such as
    ``Vorstand'' in \cref{fig:salings-snippet}) and flag any instances of these paragraphs that appear consecutively without being interspersed by firm headers.
  \item Exploit the panel dimension of the dataset to detect firms that are missing in certain years but not in others.
  \end{enumerate}

\begin{table}[!ht]
\caption{Standardized balance sheet schema}\label{table:salings-balancesheet}
\small\centering
\begin{tabular}{ll}
\toprule
\textbf{Assets} & \textbf{Liabilities} \\ 
\midrule
    Current assets                      & Current liabilities \\
    \hspace{3mm} Cash and receivables   & Bank debt           \\
    \hspace{3mm} Securities             & Long-term debt      \\
    \hspace{3mm} Inventory              & Other debt        \\
    \hspace{3mm} Other                  & Other liabilities          \\
    Long-term assets                    & Equity   \\
    Other assets                        & \hspace{3mm} Paid-in equity      \\
                                        & \hspace{3mm} Reserves            \\
                                        & \hspace{3mm} Accumulated profits \\
\bottomrule
\end{tabular}
\end{table}

\item Once properly identified, the information in each block is converted to key-value pairs and standardized. For this, we apply certain automated steps (remove abbreviations, apply spell-checkers, etc.) and then manually construct a crosswalk from the mostly ad-hoc labels into standardized labels.  

\Cref{table:salings-balancesheet} shows the schema of our standardized balance sheets based on a coarsened version of Germany's General Commercial Code (Handelsgesetzbuch). Coarsening the balance sheet into less detailed items is necessary because the original items are often not sufficiently informative to break balance sheet items down further. For instance, \Cref{table:salings-crosswalk} lists a snippet of this crosswalk, showing ten of the more than 3,000 unique labels that map into the ``long-term assets'' category. \Cref{table:saling-incomestatement} shows the schema of our standardized income statement. Profit, calculated as total revenue minus expenses, is reported before dividends, and maps into the profit item in the standardized balance sheet.

\item Lastly, we manually review all extracted data against their scanned image, with the main goal of finding any typos in the digitized balance sheet values. We pay particular attention to cases where balance sheet identities do not hold.
\end{enumerate}

\begin{table}[!ht]
    \caption{Sample of labels that map into ``long-term assets'', with their English translation.}
    \label{table:salings-crosswalk}
    \small
    \centering
    \begin{tabular}{ll}
    \toprule
    \multicolumn{1}{c}{\textbf{German label}} & \multicolumn{1}{c}{\textbf{English translation}} \\
    \midrule
    Gebäude & Building \\
    Grundstücke & Plots of land \\
    Grundstück und Gebäude & Land and building \\
    Fabrikgebäude & Factory building \\
    Betriebsgebäude & Operations building \\
    Grubenfelder & Mining fields \\
    Verwaltungsgebäude & Administrative building \\
    Bahnhofsanlage & Train station facility \\
    Grundstück in Hannover & Property in Hanover \\
    Grundstück einschließlich Gleisanschluss & Land including railway connection \\
    \bottomrule
    \end{tabular}
\end{table}%

\begin{table}[htbp]
  \centering
  \caption{Schema for standardized income statement items}
    \begin{tabular}{l}
    \toprule
    \textbf{Revenue} \\
\cmidrule(l{.5em}){1-1}
    \textbf{Expenses} \\
\hspace{3mm}   Operating Expenses \\
\hspace{3mm}  \hspace{3mm} Materials \\
\hspace{3mm}   \hspace{3mm} Salaries \\
\hspace{3mm}   \hspace{3mm} Taxes \\
\hspace{3mm}   \hspace{3mm} Other \\
\cmidrule(l{1.2em}){1-1}
\hspace{3mm}   Interest Expenses \\
\hspace{3mm}   Depreciation \\
\hspace{3mm}   Debt Appreciation \\
\hspace{3mm}   Other Expenses \\
\cmidrule(l{1.2em}){1-1}
\hspace{3mm}   Reserves \\
\hspace{3mm}   \hspace{3mm} Capital \\
\hspace{3mm}   \hspace{3mm} Pensions \\
\hspace{3mm}   Balance Carryforward \\
\cmidrule(l{.5em}){1-1}
    \textbf{Profit = Revenue - Expenses} \\
  \bottomrule
    \end{tabular}%
  \label{table:saling-incomestatement}%
\end{table}%

\paragraph*{Identification of fiscal years}\label{identification-of-fiscal-years}

Whenever a firm's fiscal year does not coincide with the calendar year, there might be ambiguities in the exact as-of dates listed in the income statements. For instance, an income statement for the period ``1925-26'' might correspond to any range of dates throughout this period. Whenever such ambiguities arrive, we clear them through the \emph{Geschäftsjahr} (``fiscal year'') section of each firm's
report.

\paragraph*{Data quality through the hyperinflation}\label{data-quality-through-the-hyperinflation}

Balance sheet numbers during the hyperinflation years are particularly challenging to digitize. Because of inflation, the numbers became extremely long. \textit{Saling's} often switched to reporting data in millions, billions, or trillions with decimal values as subindices. For instance, Figure \ref{fig:salings-inflation} illustrates some of the difficulties in extracting the information for that period, as the numbers were extremely long and complex (``\(1 407 368._{85}\) Bill. \emph{M}'').

{
\setstretch{1.0}
  \begin{figure}[htpb]
    \begin{center}
    \caption{Extract of the 1924 Edition of \emph{Saling's} for the Firm
\emph{Bayerische Spiegelglasfabriken AG}. \label{fig:salings-inflation}}
    \includegraphics[width=0.8\textwidth]{"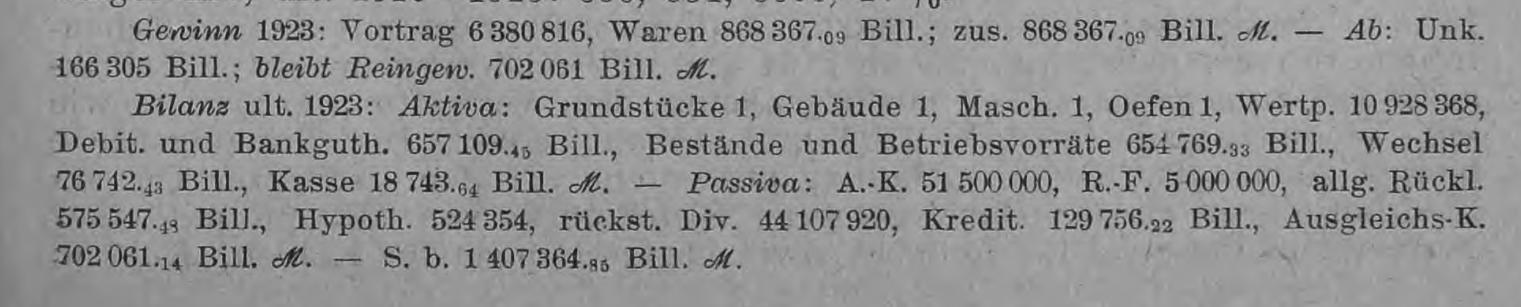"}
    \end{center}
    {\footnotesize \textit{Notes:} Notice how numbers are prefixed by "Bill." to indicate they are in trillions of Marks (English trillion is equivalent to a German \textit{billionen}). Further, decimal parts are reported as subindices following the integers.}
  \end{figure}
}

A more fundamental measurement challenge arises due to the impact of inflation on accounting. Several historical sources indicate that balance sheet statements were not adjusted for inflation, leading to measurement error problem during the hyperinflation, especially in 1923. For example, the 1923 financial report of Darmst\"{a}dter und Nationalbank stated that ``the figures in our balance sheet and profit- and-loss statement are, as in those of all German companies, unfit for any serious scrutiny, and to examine them in detail is folly'' \citep{Sweeney1934}. Similarly, \cite{Hoffmann2020} provide examples of firms in 1923, noting that the calculation of balance sheets and income in paper marks ``lost its economic meaning'' and that firms only reported financial statements out of legal obligation. As we discuss in section \ref{sec:data}, we are therefore extremely cautious in drawing inferences based on financial statements from 1923. Note that our main analysis (e.g., Figure \ref{fig:leverage_employment_dynamics} and Table \ref{tab:leverage_employment}) is not affected by measurement error induced by inflation, as we sort firms by leverage from balance sheets constructed before the inflation in 1918-1919.

The impact of hyperinflation on the reliability of financial accounts was understood by contemporaries. Some firms began voluntarily reporting balance sheets in Goldmarks by the middle of 1923. In December 1923, the government required all firms to prepare new opening balance sheets by January 1, 1924 in Goldmarks through the regulation on Goldmark accounts (\textit{Verordnung über Goldbilanzen})  in December 1923.

\begin{figure}[!ht]
    \caption{Balance Sheet Dynamics in Saling: Ratios.}
	\begin{center}
        \includegraphics[width=0.6\textwidth]{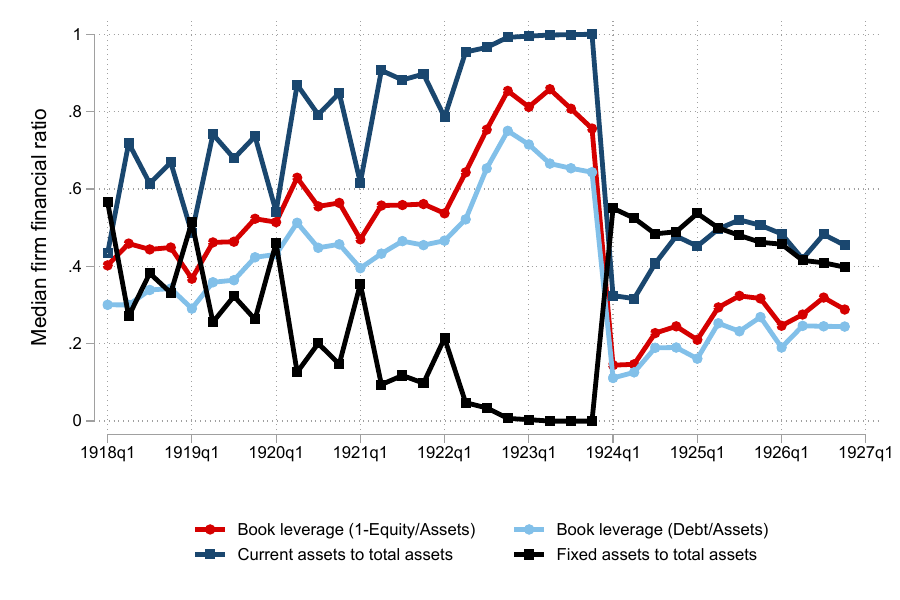} 
	\end{center}
        \footnotesize{\textit{Notes}: This figure plots the evolution of the median of key balance sheet ratios in the \textit{Saling's} data over time. The large changes in 1924Q1 occur due to the introduction of revalued Goldmark balance sheets.}

	 \label{fig:saling_med_leverage_ratio}
\end{figure}

Treating the Goldmark balance sheets as correct, we can compare the changes in balance sheet items before and after the introduction of Goldmark balance sheets to understand how hyperinflation distorted accounting. This reveals that accounting distortions caused by inflation are not symmetric across the balance sheet items. Instead, the revalued Goldmark balance sheets reveal that ``real'' positions such as fixed assets, inventories, and book equity are systematically more likely to be undervalued than ``nominal'' items such as cash, other short-term assets, and debt.\footnote{\cite{Sweeney1934} also uses the Goldmark balance sheets as the correct balance sheets to illustrate the misleading nature of the paper mark balance sheets during the hyperinflation. Based on a case study of one firm, \cite{Sweeney1934} finds that current assets and current liabilities are more likely to be correct in the 1923 paper mark balance sheets, while measurement error is most severe for less liquid, long-term assets, such as fixed assets and book equity. However, it should be noted that even the revalued Goldmark balance sheets may have undervalued real assets, as uncertainty about the costs of stabilization and whether it would succeed led to conservative valuations \citep[][p. 242]{Graham1931}.} Appendix \Cref{fig:saling_med_leverage_ratio} illustrates that the introduction of Goldmark balance sheets leads to large positive revaluations of fixed assets, inventories, and book equity. An implication is that the level of leverage (liabilities-to-assets) is significantly overstated by balance sheets during the hyperinflation.

\begin{figure}[!ht]
    \caption{Saling Firm-Level Balance Sheet Data: Data Quality.}
        \begin{center}
        \includegraphics[width=0.7\textwidth]{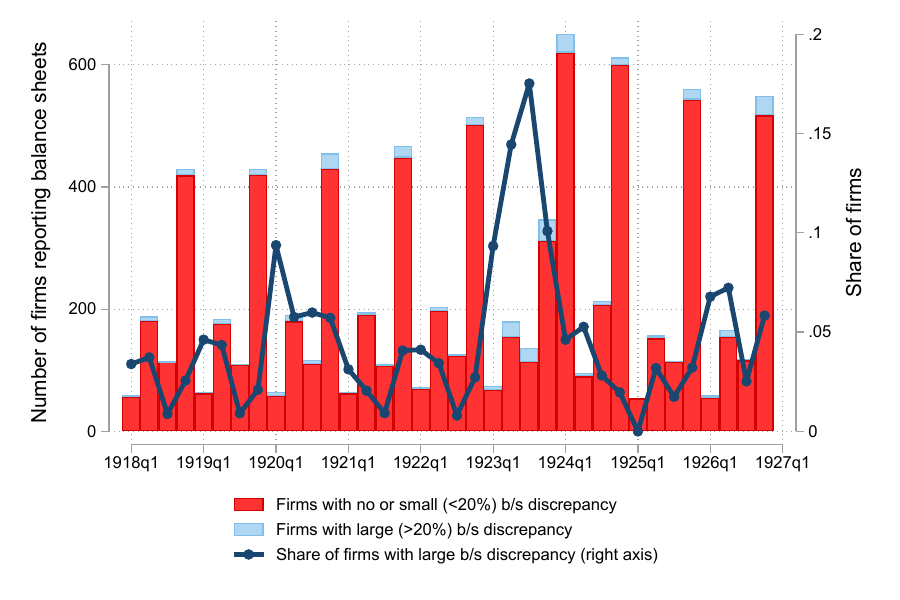}
        \end{center}
	 \label{fig:saling_bs_discrepancy}
        \footnotesize{\textit{Notes}: This figure plots the number and share of firms with no or small balance sheet discrepancies and large balance sheet discrepancies. Balance sheet discrepancies are defined as instances of a larger than 20\% pairwise difference between either the sum of assets, the sum of liabilities and equity, or reported total assets.}
\end{figure}

Clerical errors in balance sheets reported in Saling's also became much more common in 1923. For example, \Cref{fig:saling_bs_discrepancy} shows that violations of the balance sheet accounting identity spikes to about 40\% in 1923, from around 5\% in other years. Specifically, we test whether the sum of all assets equals the sum of all liabilities and equity, as well as whether these sums equal total assets (reported separately). Failure of accounting identities to hold for 5\% of balance sheets outside of the hyperinflation is likely due to clerical errors or cases where firms do not report small balance sheet items.

\hypertarget{firm-level-employment-data}{%
\subsection{Firm-level Employment
Data}\label{firm-level-employment-data}}
\addcontentsline{toc}{subsection}{Firm-level Employment Data}
Roughly one-third of all firms report information on the number of employees. Some firms report the total employment while other firm distinguish between blue-collar and white-collar workers. For our analysis, we always use the total number of employees. The red box in \Cref{fig:salings-snippet} shows an example of how a firm reports the number of employees (both blue collar ``Arbeiter'' and white collar ``Beamte'').

In contrast with balance sheet and income statement data, reporting of employment data is less standardized. Table \ref{table:salings-employment} lists some of the different section headers and keywords used throughout the text.   We therefore hand-collect the employment data in \textit{Saling's}. We collect employment information from 1916 through 1933 in order to perform placebo tests on years after the hyperinflation.

\begin{table}[!ht]
    \caption{Sample of keywords used to detect employment figures, with their English translation.}
    \label{table:salings-employment}
    \small
    \centering
    \begin{tabular}{ll}
    \toprule
    \multicolumn{1}{c}{\textbf{German keyword}} & \multicolumn{1}{c}{\textbf{English translation}} \\
    \midrule
Beamte und Arbeiter & Civil Servants and Workers \\
Gesamtbelegschaft & Total Workforce \\
Arbeiterzahl & Number of Workers \\
Zahl der Arbeiter durchschnittlich & Average Number of Workers \\
Zahl d.~Beamten u. Arbeiter & Number of Civil Servants and Workers \\
Arbeiter und Angestellte & Blue-collar and White-collar Workers \\
Arb.  & Workers \\
    \bottomrule
    \end{tabular}
\end{table}%

\begin{figure}[htpb]
    \caption{Employment Growth in \textit{Saling's} Compared with the Aggregate Unemployment Rate.}

        \begin{center}

        \subfloat[Time series comparison.]{\includegraphics[width=0.5\textwidth]{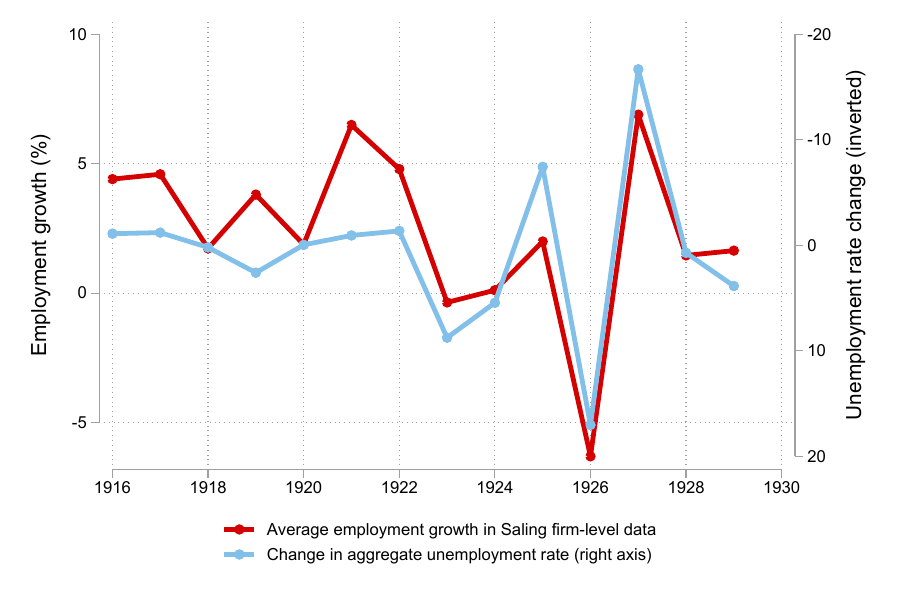}}
        \subfloat[Scatterplot.]{\includegraphics[width=0.5\textwidth]{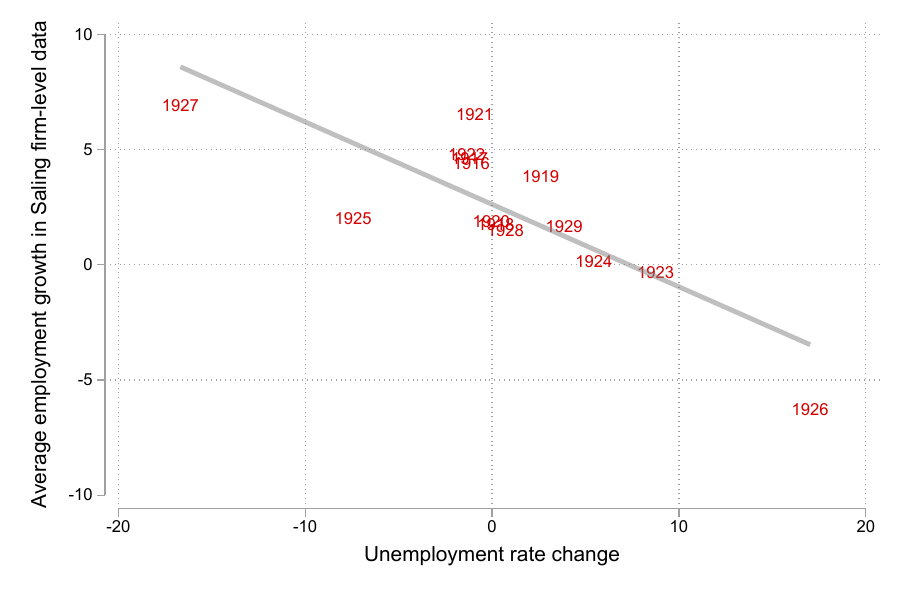}}

        \end{center}

        \footnotesize{\textit{Notes}: This figure validates self-reported employment in the \textit{Saling's} data by comparing it with the change in the aggregate unemployment rate. Aggregate employment growth in \textit{Saling's} is computed as the average of the change in firm log employment (multiplied by 100). The annual unemployment rate is the annual average of the monthly rate and is from the \textit{Reichsarbeitsblatt}.}
	 \label{fig:saling_emp_vs_unemp}
\end{figure}

As a check on the quality of the employment data reported in \textit{Saling's}, \Cref{fig:saling_emp_vs_unemp} compares average employment growth in \textit{Saling's} with the change in the aggregate unemployment rate (on an inverted scale). While these variables need not exactly coincide, it is reassuring that the two variables co-move reasonably closely, suggesting that the employment information in \textit{Saling's}  captures the aggregate fluctuations in employment reasonably well. Employment growth among firms in \textit{Saling's} captures the boom in the 1920-1922 period, the slowdown in 1923 and after the stabilization in 1924, the boom in the latter part of the 1920s, and the slowdown in 1928-1929. Consistent with this visual impression, the $R^2$ in a regression of average firm employment growth in \textit{Saling's} on the change in the unemployment rate is 66\%.

The main analysis in the paper uses the employment data as it is reported in \textit{Saling's}. The panel on firm employment over 1914-1923 is unbalanced and contains some gaps, as sometimes firms do not report in a given year. To address this concern, for robustness, we consider a balanced sample over 1916-1923. Figure \ref{fig:dd_emp_robust_samples} shows the main result in the paper is robust to this restriction.

\subsection{Long Term Bond Data}\label{long-term-bond-data}

We  collect granular data on outstanding bonds reported in \textit{Saling's}. Next to balance sheets and income statements, the manual reports details on the terms and history of outstanding long-term bonds. \Cref{fig:salings-bond-example} shows an example of a bond issued by the firm ``Friedr. Bayer \& Co.'' (a predecessor of the still existing Bayer AG) reported in the 1920 edition of \textit{Saling's}. Data on long-term bonds are reported in a subsection of the description of a firm's financials. The header of each subsection indicates whether a bond is a regular bond (``\textit{Anleihe}'') or a mortgage (``\textit{Hypotheken-Anleihe}''). Further details on the contractual terms of each bond are then reported in a non-standardized text format. The description typically contains the original volume of the bond, the origination date, the interest payment, whether the bond has a prepayment option, and further details on the amortization schedule. Bonds can  differ in their amortization schedules. Most bonds have specified dates when amortization starts and ends (and the bonds hence matures). Alternatively, some bonds only report an origination date but no final maturity date. However, these bonds report the scheduled amortization rate, thus allowing us to calculate the implicit final maturity date.  For a subset of publicly traded bonds, the manual also reports the end-of-year bond price (if available).  

Given the non-standardized reporting format, we collect all bond information by hand. We go through each page of the manual and search for reported bonds. For each bond we find, we then collect all available terms on the bond via the editor displayed in \Cref{fig:salings-bond-editor}. that provides a standardized sheet of available variables.

{
\setstretch{1.0}
  \begin{figure}[htpb]
    \centering
    \caption{Sample bond information from the 1920 edition of \emph{Saling's}, for the firm Bayer AG.}
    \includegraphics[width=0.95\textwidth]{"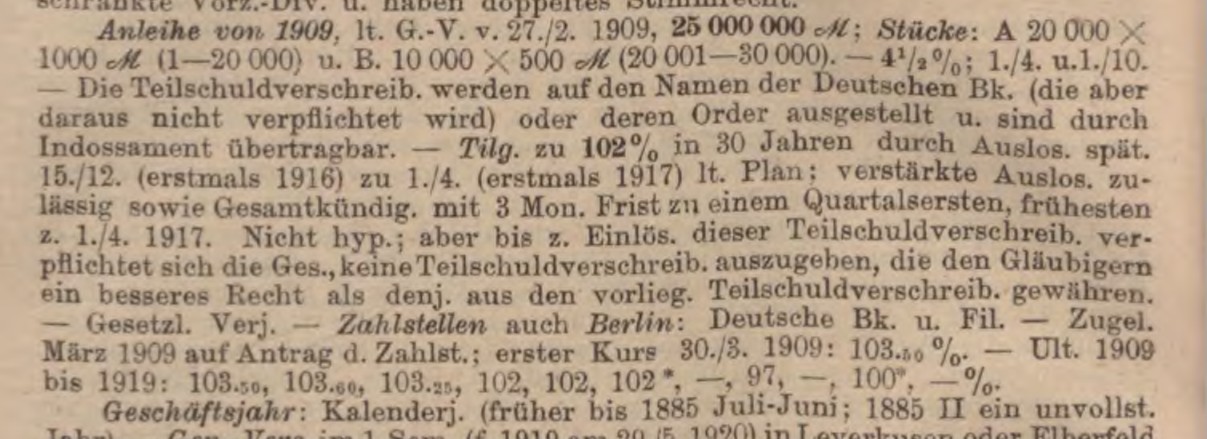"}
    \label{fig:salings-bond-example}
  \end{figure}
}

{
\setstretch{1.0}
  \begin{figure}[htpb]
  \begin{center}
    \caption{Screenshot of bond editor app including snippet of scanned bond information.}
    \includegraphics[width=0.95\textwidth]{"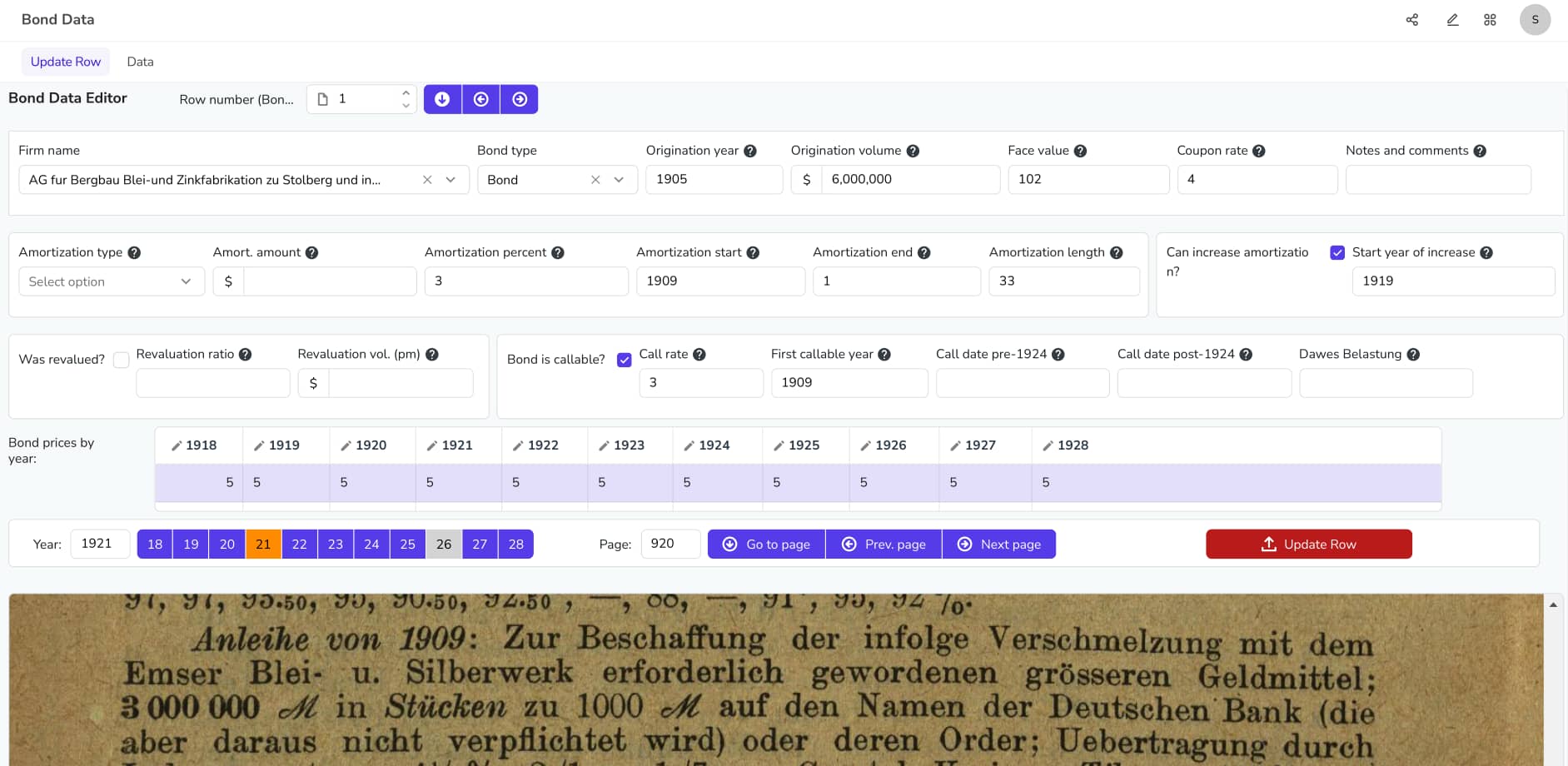"}
    \end{center}
    \footnotesize{\textit{Notes}: This image shows the digitization process for bond data. We track a given bond issuance across multiple years in order to record its different characteristics and history.}
    \label{fig:salings-bond-editor}
  \end{figure}
}

\subsection{Stock Price Data}\label{stock-price-data}

\begin{figure}[!ht]
    \begin{center}
    \caption{Example of Firm-Level Stock Price Data from \textit{Berliner Börsen-Zeitung.} }
    \includegraphics[width=0.65\textwidth]{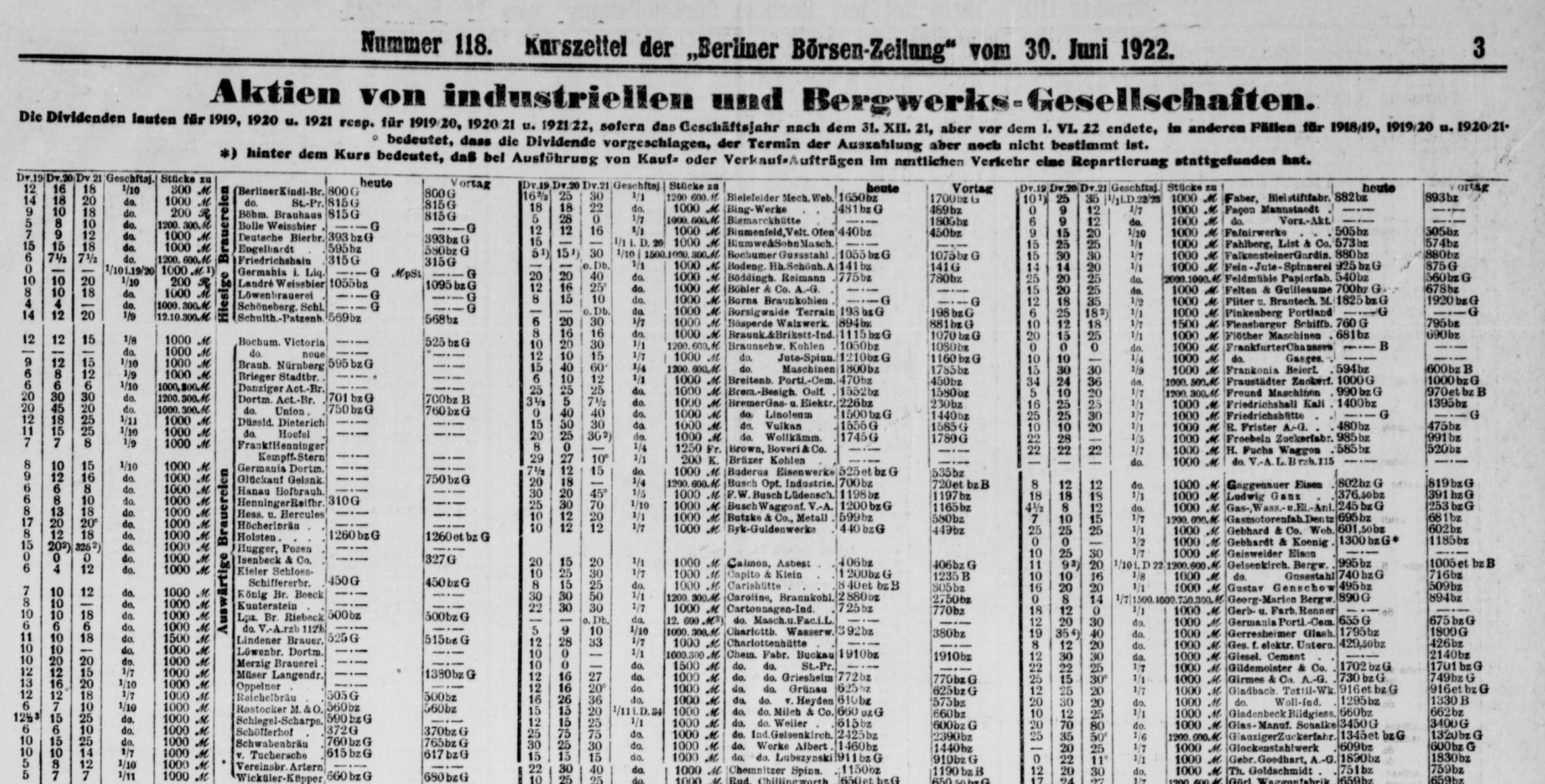}\label{fig:bbz_example}
    \end{center}
    \footnotesize{\textit{Notes}: Stock prices of industrial firms from \textit{Berliner Börsen-Zeitung}, 1922.}
\end{figure}

We obtain stock price data from \textit{Berliner Börsen- Zeitung}, a contemporary newspaper, published twice each workday (morning and evening edition) and once on Saturdays and Sundays. We collect the end-of-month prices for each stock between 1919 and 1924 and map firm names between stock prices and Saling's data based on string matching with a manual overlay. We also collect information on dividends. \Cref{fig:saling_bs_discrepancy} compares the equal-weighted price index we construct from these data with price indexes reported in \textit{Wirtschaft und Statistik}. The series track each other closely, except for a divergence in the final months of the inflation when inflation was extremely high. A reason for this divergence can be due to comparing prices for different days, which can make a significant difference during hyperinflation.

We construct annual real total log returns as $\ln(1+R_t) = \ln\left( \frac{P_t/W_t+D_t/W_t}{P_{t-1}/W_{t-1}} \right),$ where $P_t$ is the price at the end of year $t$, $D_t$ is the dividend in year $t$, and $W_t$ is the wholesale price index at the end of year $t$. 

\subsection{Details on Construction of Specific Variables}\label{variable-construction}

To construct an estimate of Tobin's Q as of the end of 1918, we combine stock prices from \textit{Berliner Börsen- Zeitung} with book equity values from \textit{Saling's}. We rely on the observation that stock prices at the time in both the \textit{Berliner Börsen Zeitung} and \textit{Salings} are reported as a percentage of book value of paid in equity capital. We follow the convention applied in contemporary studies and proxy the value of debt with its book value. We thus construct Tobin's Q as:
\begin{align*}
TobinsQ_{i,1918} = \frac{P^{BBZ}_{i,Jan1919} BookEquity_{i,1918} + Debt_{i,1918}  }{ TotalBookAssets_{i,1918} } 
\end{align*} 
To calculate Tobin's Q as of the end of 1918, we use the January 1919 stock price report in \textit{Berliner Börsen Zeitung}. We similarly construct firm market capitalization as $ P^{BBZ}_{i,t} BookEquity_{i,t}.$

We estimate free cash flow to the firm using the standard formula that adjusts EBIT for depreciation (a non-cash expense) and incorporates investment in fixed assets and net working capital:
\begin{align*} 
FCF_{it} &= (1-\tau)EBIT_t + Depreciation_t \\ &- (FixedAssets_{i,t} -FixedAssets_{i,t-1} + Depreciation)  \\ &- (NWC_{i,t} - NWC_{i,t-1}). 
\end{align*}
Since we do not always know the tax rate, we set it to zero. Since free cash flow is only used as a control variable, this does not materially affect the findings and results are robust to other plausible values of $\tau$. We the construct free cash flow to assets as $FCF_{i,t}/TotalAssets_{i,t}.$

\subsection{Data Quality Control and Data Restrictions}\label{data-restrictions}
Throughout the data digitization process, we apply several quality checks. A key quality check stems from exploiting simple accounting identities that must hold within balance sheets and income statements. For instance, we flag balance sheets if assets and liabilities are not equal. We also flag income statements if the reported net profit is not equal to the operating revenue net of the expenses. Further, we flag financial ratios that by definition cannot exceed or fall short of a certain value. We then manually check all flagged observations to identify and correct mistakes whenever possible.

To further ensure that results are not driven by outliers that result from human mistakes made during the digitization process or errors in the underlying historical data source, we winsorize firm-level data. Specifically, we winsorize variables constructed from balance sheets and income statements at the 1\textsuperscript{st} and the 99\textsuperscript{th} percentiles in each year.  
Moreover, for a subset of variables constructed based on income statements that turn out to be especially noisy, we apply more rigid winsorization and trimming. In particular, we trim the share of interest expenses and production costs as a share of total expenses at the 95\textsuperscript{th} percentile, as extremely high shares of interest expenses or production costs are implausible and are thus likely driven by data errors. 

\begin{figure}[!ht]
    \caption{Number of Firms in Reporting in Saling After Sample Restriction.}
        \begin{center}

        \subfloat[Number of firms reporting B/S by currency of reporting.]{
        \includegraphics[width=0.55\textwidth]{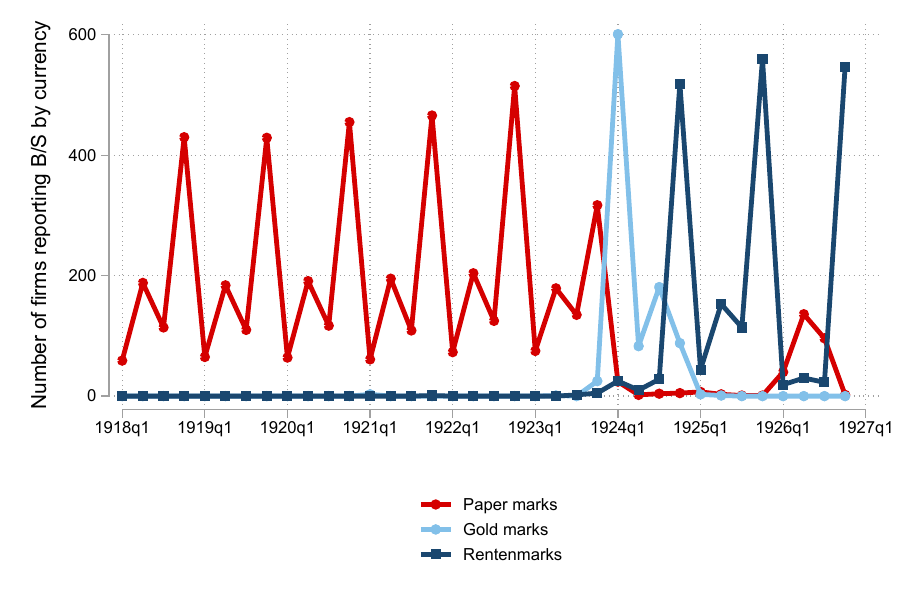}}

        \subfloat[Number of firms reporting I/S by currency of reporting.]{
        \includegraphics[width=0.55\textwidth]{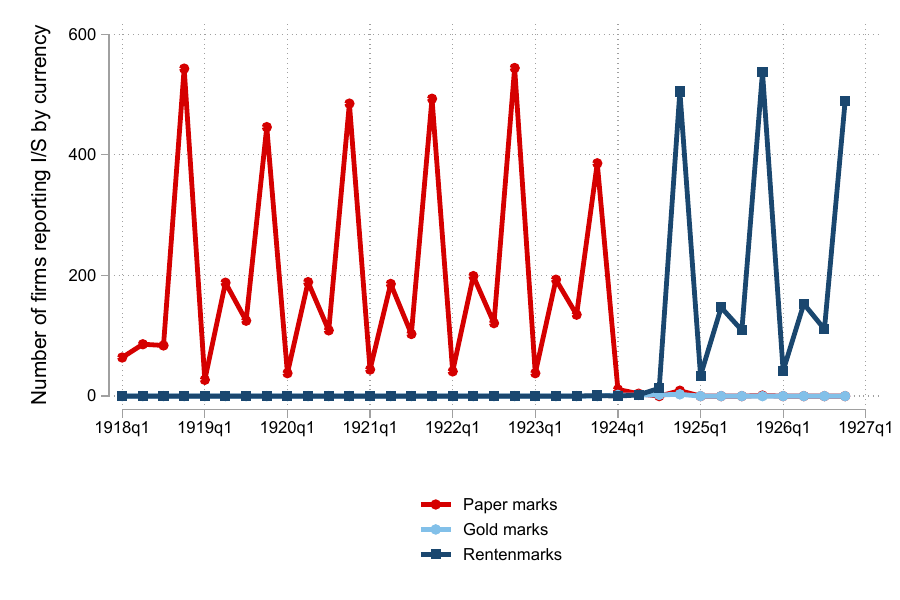}}
        \end{center}
        \footnotesize{\textit{Notes}: This figure plots the number of firms reporting balace sheets and income statements by quarter and currency of reporting in the \textit{Saling's} data. The sample period is 1918Q1-1926Q4. The majority of firms report balance sheets at the end of the year (fourth quarter). The spike in balance sheets in 1924Q1 is the new revalued Goldmark balance sheets. Rentenmark balance sheets refers to Rentenmarks or Reichsmarks, which have the same value.}
	 \label{fig:N_saling}
\end{figure}

Finally, we drop firms with a headquarter located outside of Germany's post-WWI borders or firms with balance sheets and income statements that are not denominated in paper marks, Reichsmarks, or Goldmarks. We also exclude insurance companies, credit banks, and mortgage banks from our analysis. \Cref{fig:N_saling} reports the number of firms reporting balance sheets and income statements, by currency, after imposing these sample restrictions.

\subsection{Executives and Supervisory Board Members}\label{app:data-board}

We collect data on executives and supervisory board memberships. These data allow us to control for firm proximity to sources of political or economic power. In particular, we use the \textit{Register der Vorstands- und Aufsichtsratsmitglieder} on pages XXV-LXXIV of the 1920 edition of \emph{Saling's Börsen Papiere - Zweiter (finanzieller) Teil (Berliner Börse)}. This section lists individuals who belonged to at least one board of directors or supervisory board of the firms listed in the volume, and maps these individuals to their respective firms. Further, this list also includes the main profession of each individual as well as academic and honorary titles. For more institutional background on executives and supervisory board structures in interwar Germany, see, e.g., \citet{Ferguson2008,Huber2021}.

We identify 9,640 firm-board member pairs, corresponding to 1,002 firms and 5,924 individuals. 4,520 individuals belonged to only a single board and thus are not useful for establishing connections across firms. This leaves us with 5,120 ``useful pairs'' that have individuals belonging to multiple boards.

\begin{figure}[htpb]
\caption{Firm interconnectedness through the board members networks.}\label{fig:network}
\begin{center}

\subfloat[Interconnection to government officials.]{\includegraphics[width=0.49\textwidth]{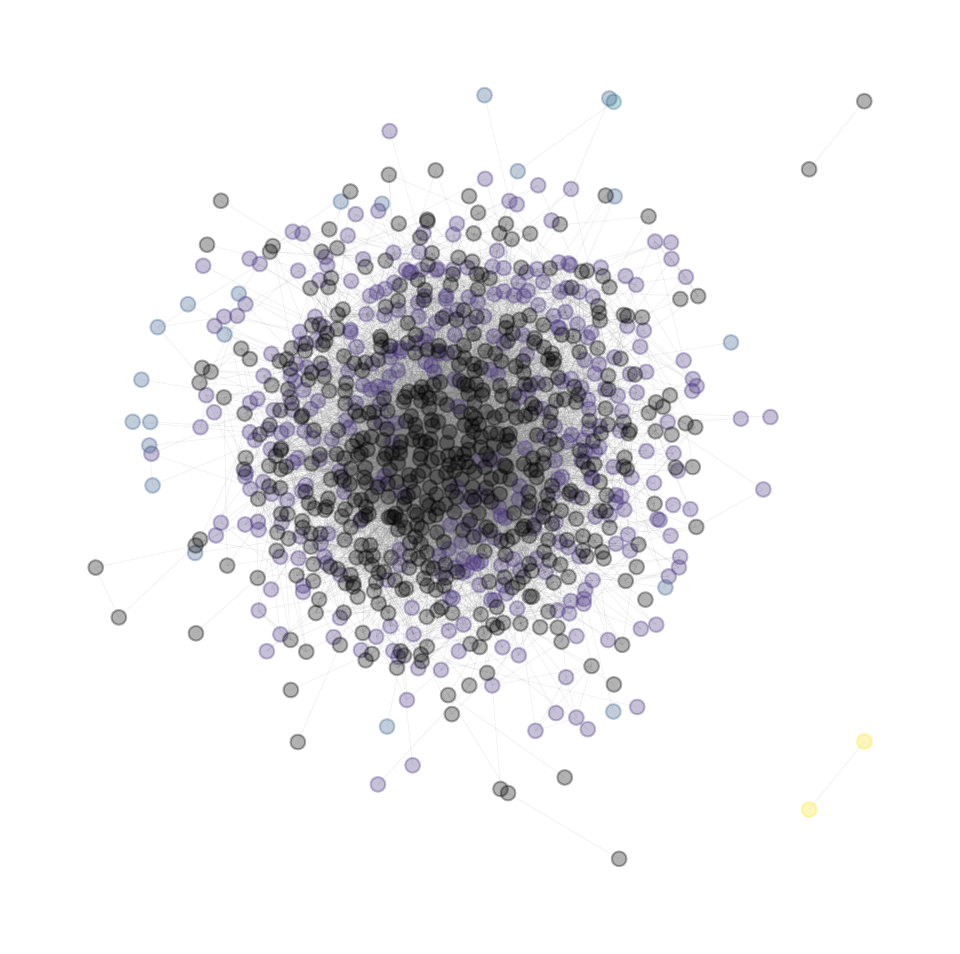}}\label{fig:network_gov}
\hfill
\subfloat[Interconnection to the Reichsbank central committee.]{\includegraphics[width=0.49\textwidth]{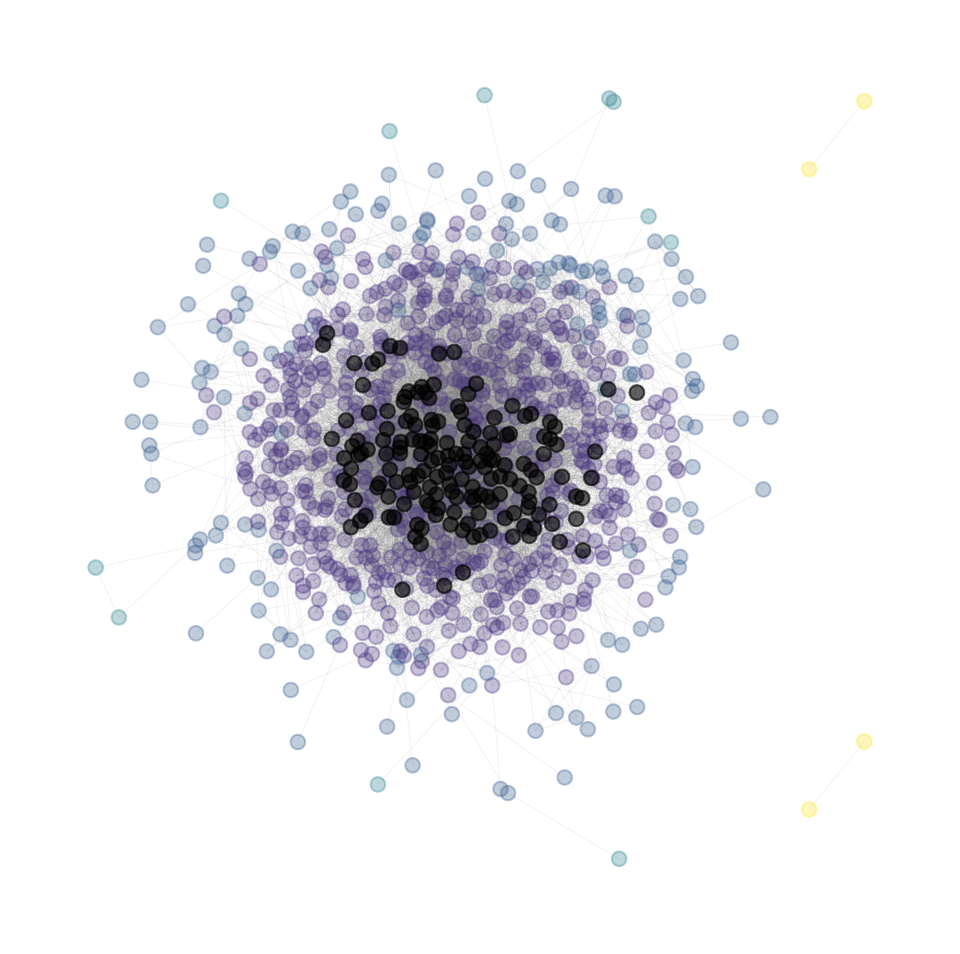}}\label{fig:network_reichsbank}

\subfloat[Interconnection to Hugo Stinnes.]{\includegraphics[width=0.49\textwidth]{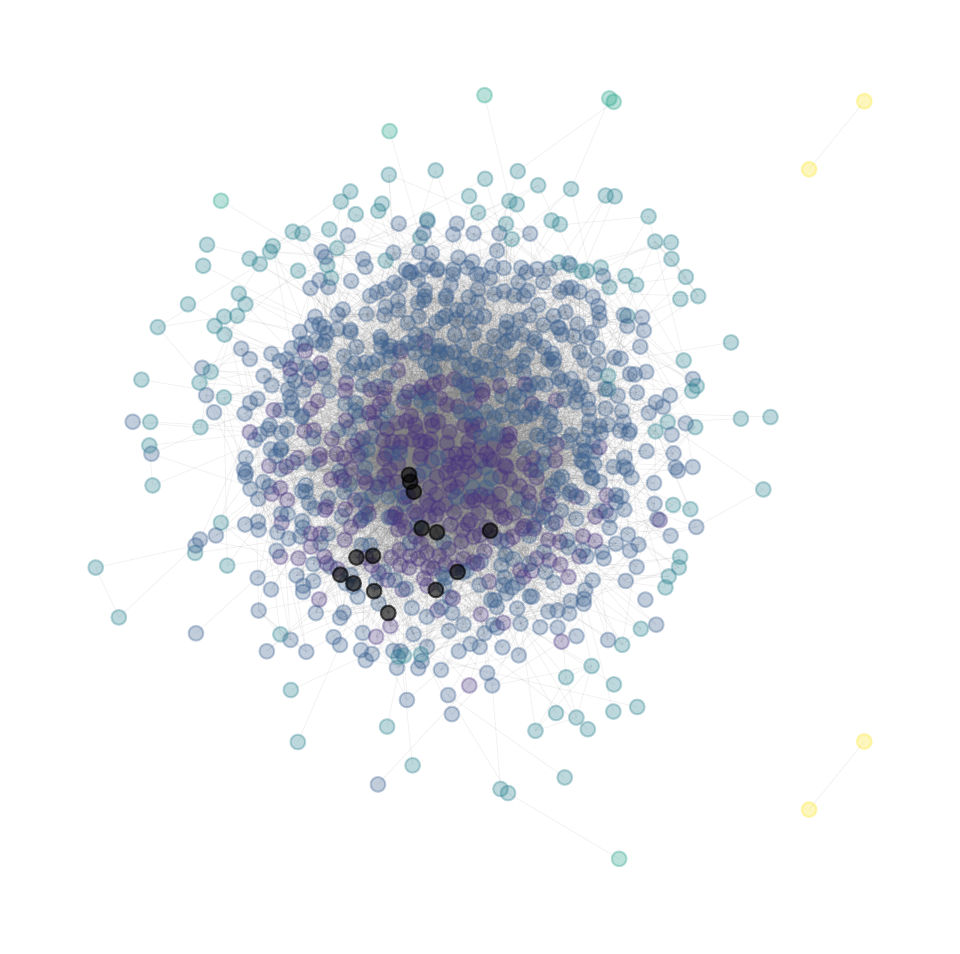}}\label{fig:network_stinnes}
\hfill
\subfloat[Interconnection to German Banks.]{\includegraphics[width=0.49\textwidth]{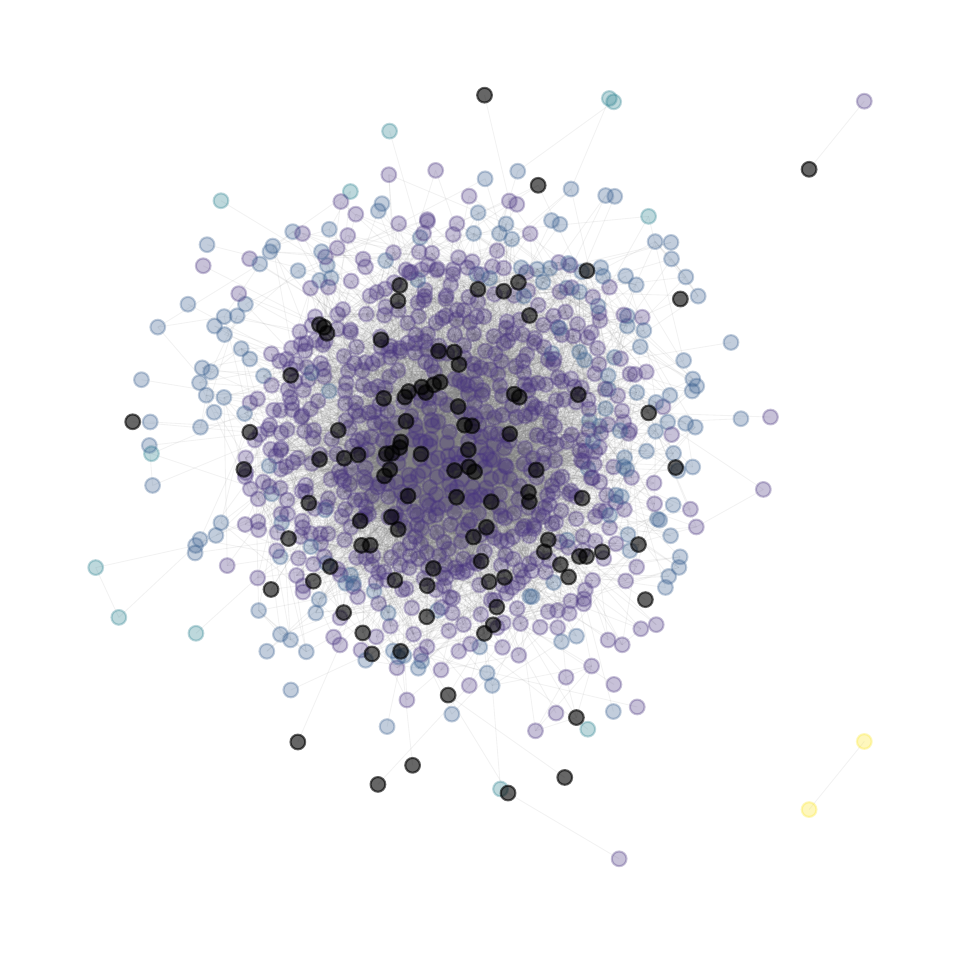}}\label{fig:network_bank}
        \end{center}
        \footnotesize{\textit{Notes}: Data are from \textit{Register der Vorstands- und Aufsichtsratsmitglieder}, on pages XXV-LXXIV of the 1920 edition of \emph{Saling's Börsen Papiere - Zweiter (finanzieller) Teil (Berliner Börse)}. Black nodes correspond to direct connections; blue, dark green, and light green nodes correspond to firms with 2--4 degrees-of-separation; and yellow nodes correspond to firms not connected to a given source. Isolated nodes---firms not connected to any other firm---have been removed in order to simplify the visualization.
        
        Panel A identifies connection to government officials by the stated profession of the board members. Panel B identifies Reichsbank connection through board members that also belonged to the Reichsbank's central committee. Panel C identifies connections to Hugo Stinnes through his own role as board member. Lastly, Panel D identifies connections to German Banks through board members that also belonged to boards of German Banks as enumerated in Saling's.}
\end{figure}

\Cref{fig:network} visualizes the network formed by these firms (all four panels have the same layout). Each node corresponds to a firm, and each edge corresponds to a connection between firms due to a common board member. The figure reveals that most firms belong to the largest sub-graph and are thus connected to each other.

Each panel of this figure is colored differently to represent the degrees of separation of each firm to a potential source of political or economic power. Nodes in black correspond to firms that have a direct connection to a given source. Nodes with two degrees-of-separation are in blue; with three are in dark green; with four in light green; etc. Nodes not connected to a given source are colored in yellow.

For Panel A, we identify connections to the government through the board member job titles listed by Saling's. For instance, some board members are clearly identified as government employees such as ``Staatsminister'' (state minister), ``Regierungsrat'' (senior civil servant member),  ``Oberbürgermeister'' (mayor) as well as through military titles such as ``Generalmajor'' and ``Vizeadmiral''.

In Panel B, we identify connections to the Reichsbank through board members that also belong to the Reichsbank's Zentralausschuss (Central Committee of Shareholders), as listed on page 497 of the same edition of Saling's.

In Panel C, we identify connections to the politically connected industrialist Hugo Stinnes (dubbed the ``inflation'' king for his ability to use credit access to purchase real assets) through his direct participation in a given firm's board.

Lastly, in Panel D, we identify connections to German banks through individuals who were also board members in any of the banks listed in the ``Banken'' section of the same edition of Saling's (pages 505--743).

\subsection{Exporter Status}\label{sec:exporter}
We collect additional information on export status at the firm-level. Here, we exploit that \textit{Saling's B\"{o}rsen-Jahrbuch}, contains a detailed description of each firm's activities, purpose, and developments. For many firms, these descriptions include information on a firm's exporting activities. This information allows us to proxy whether a firm is an exporter. In effect, we manually create firm-level dummies that indicate whether the investors manual mentions exports as part of the overall profile of the firm

As an example, the 1920 edition of  \textit{Saling's} contains the firm ``Rheinische Gerbstoff-und Farbholz-Extract-Fabrik Gebr. Muller AG'' located in Benrath. This firm specializes in producing paint. The firm description notes that: ``Currently the factory mainly produces quebracho extract, both in liquid and solid form, the latter especially for export, which currently accounts for about a third of total production.''\footnote{``Zurzeit stellt die Fabrik in der Hauptsache Quebracho-Extrakt her, sowohl in flüssiger als in fester Form, letztere bes. für den Export, der zurzeit ungefähr ein Drittel der Gesamtprod. beträgt.'' See page 1269 of the the 1920 edition of \textit{Saling's B\"{o}rsen-Jahrbuch}.
} Thus, we assign a dummy that indicates that this firm is an active exporter.
Another example is the firm ``Th. Flother Maschinenbau AG'' located in Gassen and Breslau, which was active in producing manufacturing goods for railways and transportation. This firm states in its description: ``The company is heavily involved in exports to Romania and Russia.''\footnote{The original reads as ``Die Gesellschaft ist am Export nach Rumänien
 u. Russland stark beteiligt.'' on page 1354 of the 1922 edition of \textit{Saling's B\"{o}rsen-Jahrbuch}.} Again, we classify this firm as an exporter.

We note that this measure is not free of caveats. For instance, we cannot typically quantify how important exports are for a given firm in a given year. Moreover, we may capture historical export activities that are not relevant in a given year (although they may be informative nonetheless). In principle, some exporters may not disclose this in \textit{Saling's}, though we believe that firms would generally have an incentive to advertise their export activities as a signal of quality and growth potential. 

	\end{appendices}
\end{document}